\documentclass[iop]{emulateapj}
\pdfoutput=1
\usepackage{natbib}
\usepackage{color}
\usepackage[english]{babel}
\usepackage[normalem]{ulem}
\usepackage{blindtext}
\usepackage{textgreek}
\newcommand{\etal}{et\,al.}

\newcommand{\kms}{km\,s$^{-1}$}

\newcommand{\lsim}{\raise0.3ex\hbox{$<$}\kern-0.75em{\lower0.65ex\hbox{$\sim$}}}
\newcommand{\gsim}{\raise0.3ex\hbox{$>$}\kern-0.75em{\lower0.65ex\hbox{$\sim$}}}

\newcommand{\HI}{H{\sc I}}
\newcommand{\msun}{M$_{\odot}$}

\usepackage{floatrow}
\newfloatcommand{capbtabbox}{table}[][\FBwidth]
\usepackage{gensymb} 
\usepackage{amsmath} 

\begin{document}
\slugcomment{Astrophysical Journal, in press}

\title{SHIELD: Neutral Gas Kinematics and Dynamics}



\author{Andrew T. McNichols}
\affil{Department of Physics \& Astronomy, Macalester College, 1600
  Grand Avenue, Saint Paul, MN 55105}
\affil{National Radio Astronomy Observatory, 520 Edgemont Road, Charlottesville, VA 
22903-2475, USA}
\email{amcnicho@nrao.edu}

\author{Yaron G. Teich}
\affil{Department of Physics \& Astronomy, Macalester College, 1600
  Grand Avenue, Saint Paul, MN 55105}
\affil{School of Education, Boston University, Two Silber Way, Boston, MA 02215}
\email{yateich@gmail.com}

\author{Elise Nims}
\affil{Department of Physics \& Astronomy, Macalester College, 1600
  Grand Avenue, Saint Paul, MN 55105}

\author{John M. Cannon}
\affil{Department of Physics \& Astronomy, Macalester College, 1600
  Grand Avenue, Saint Paul, MN 55105}
\email{jcannon@macalester.edu}


\author{Elizabeth A. K. Adams} \affil{ASTRON, the Netherlands
  Institute for Radio Astronomy, Postbus 2, 7990 AA, Dwingeloo, The
  Netherlands} 

\author{Elijah Z. Bernstein-Cooper} 
\affil{Department of Astronomy, University of Wisconsin, 475 N
  Charter St Madison, WI 53706, USA} 

\author{Riccardo Giovanelli}
\affil{Center for Astrophysics and Planetary Science, Space
  Sciences Building, 122 Sciences Drive, Cornell University,
  Ithaca NY 14853 USA}

\author{Martha P. Haynes}
\affil{Center for Astrophysics and Planetary Science, Space
  Sciences Building, 122 Sciences Drive, Cornell University,
  Ithaca NY 14853 USA}

\author{Gyula I.G. J{\'o}zsa}
\affil{SKA South Africa, Radio Astronomy Research Group, 3rd Floor,
  The Park, Park Road, Pinelands, 7405, South Africa}
\affil{Rhodes University, Department of Physics and Electronics,
  Rhodes Centre for Radio Astronomy Techniques \& Technologies, PO Box
  94, Grahamstown, 6140, South Africa}
\affil{Argelander-Institut f{\"u}r Astronomie, Auf dem H{\"u}gel 71, 
53121 Bonn, Germany}

\author{Kristen B.W. McQuinn}
\affil{Minnesota Institute for Astrophysics, School of Physics and Astronomy, 
116 Church Street, S.E., University of Minnesota, Minneapolis, MN 55455, USA}
\affil{University of Texas at Austin, McDonald Observatory, 2515 Speedway, 
Stop C1402, Austin, TX 78712, USA}

\author{John J. Salzer}  
\affil{Department of Astronomy, Indiana University, 727 East
  Third Street, Bloomington, IN 47405, USA}

\author{Evan D. Skillman}
\affil{Minnesota Institute for Astrophysics, School of Physics and Astronomy, 
116 Church Street, S.E., University of Minnesota, Minneapolis, MN 55455, USA}

\author{Steven R. Warren}
\affil{Cray, Inc.,
380 Jackson Street, Suite 210,
St. Paul, MN 55101, USA}


\author{Andrew Dolphin}
\affil{Raytheon Company, 1151 E. Hermans Road, Tucson, AZ 85756, USA}

\author{E.C. Elson}
\affil{Astrophysics, Cosmology and Gravity Centre (ACGC), Department of Astronomy, University of Cape Town, Private Bag X3, Rondebosch 7701, South Africa}

\author{Nathalie Haurberg} 
\affil{Physics Department, Knox College, 2 East South Street,
  Galesburg, IL 61401, USA}

\author{J{\"u}rgen Ott}
\affil{National Radio Astronomy Observatory, P.O. Box O, 1003
  Lopezville Road, Socorro, NM 87801, USA}

\author{Amelie Saintonge}
\affil{Department of Physics and Astronomy, University College London,
  Gower Place, London WC1E 6BT, UK}


\author{Ian Cave}
\affil{Department of Physics \& Astronomy, Macalester College, 1600
  Grand Avenue, Saint Paul, MN 55105}

\author{Cedric Hagen}
\affil{Department of Physics \& Astronomy, Macalester College, 1600
  Grand Avenue, Saint Paul, MN 55105}

\author{Shan Huang} 
\affil{Academia Sinica, Institute of Astronomy \& Astrophysics,
  P.O. Box 23-141, Taipei 10617, Taiwan}

\author{Steven Janowiecki}  
\affil{Department of Astronomy, Indiana University, 727 East
  Third Street, Bloomington, IN 47405, USA}
\affil{International Centre for Radio Astronomy Research,
University of Western Australia,
35 Stirling Highway,
Crawley, WA 6009, Australia.}

\author{Melissa V. Marshall}
\affil{Department of Physics \& Astronomy, Macalester College, 1600
  Grand Avenue, Saint Paul, MN 55105}

\author{Clara M. Thomann}
\affil{Department of Physics \& Astronomy, Macalester College, 1600
  Grand Avenue, Saint Paul, MN 55105}

\author{Angela Van Sistine}
\affil{Department of Physics
University of Wisconsin-Milwaukee
3135 North Maryland Ave.
Milwaukee, WI 53211}

\begin{abstract}

We present kinematic analyses of the 12 galaxies in the ``Survey of
\HI{} in Extremely Low-mass Dwarfs'' (SHIELD).  We use
multi-configuration interferometric observations of the \HI{} 21cm
emission line from the Karl G.  Jansky Very Large Array
(VLA)\footnote{The National Radio Astronomy Observatory is a facility
  of the National Science Foundation operated under cooperative
  agreement by Associated Universities, Inc.}  to produce image cubes
at a variety of spatial and spectral resolutions.  Both two- and
three-dimensional fitting techniques are employed in an attempt to
derive inclination-corrected rotation curves for each galaxy.  In most
cases, the comparable magnitudes of velocity dispersion and projected
rotation result in degeneracies that prohibit unambiguous circular
velocity solutions. We thus make spatially resolved position-velocity
cuts, corrected for inclination using the stellar components, to
estimate the circular rotation velocities.  We find v$_{\rm circ}$
$\leq$ 30 \kms\ for the entire survey population. Baryonic masses are
calculated using single-dish \HI{} fluxes from Arecibo and stellar
masses derived from HST and Spitzer imaging. Comparison is made with
total dynamical masses estimated from the position-velocity analysis.
The SHIELD galaxies are then placed on the baryonic Tully-Fisher
relation.  There exists an empirical threshold rotational velocity,
V$_{\rm rot}$ $<$ 15 \kms, below which current observations cannot
differentiate coherent rotation from pressure support.  The SHIELD
galaxies are representative of an important population of galaxies
whose properties cannot be described by current models of
rotationally-dominated galaxy dynamics.

\end{abstract}						

\section{Introduction}
\label{S1}

One of the most fundamental correlations in astrophysics is that
rotation velocity is proportional to luminosity.  The Tully-Fisher
relation \citep{tullyfisher77} has been refined over the years (e.g.,
using only the mass of baryons via the ``baryonic Tully-Fisher
relation'', or BTFR; {McGaugh \etal\ 2000}\nocite{mcgaugh00}), and
many investigators have independently verified the remarkably tight
relationship across many orders of magnitude in galaxian mass (see the
recent works by {Lelli \etal\ 2016}\nocite{lelli16}, {Papastergis
  \etal\ 2016}\nocite{papastergis16}, and the references therein).
For massive systems with well-organized and easily-modeled rotation,
the BTFR is well-populated and statistically robust.

How the lowest-mass, gas-rich galaxies populate the BTFR is not yet
well understood.  As the dynamical mass falls, the ratio of bulk
rotation velocity to the magnitude of turbulent motion becomes of
order unity, and current observations become unable to differentiate
between pressure-supported and rotation-dominated galaxies (see, e.g.,
{Tamburro \etal\ 2009}\nocite{tamburro09} and {Stilp
  \etal\ 2013}\nocite{stilp13}).  Empirically, this transition has
been found to occur near a circular velocity of $\sim$20 \kms; for
example, the sample presented in \citet{mcgaugh12} contains no such
galaxies with rotation velocities significantly below this value.
\citet{ezbc14} estimate that the extremely low-mass and metal-poor
galaxy Leo\,P is rotating at 15\,$\pm$\,5 \kms.  For the
slowest-rotating galaxies, the signatures of rotation become
indistinguishable from the random statistical motion of the gaseous
component.

Systems that populate the low end of the BTFR are uniquely important
to our understanding of galaxy evolution.  However, by definition,
these sources are intrinsically faint, physically small, and
technically challenging to study in detail at any significant
distance.  The total number of such galaxies detected to date remains
a significant issue for the $\Lambda$CDM cosmological model, and
discrepancies between simulations and observations still persist (the
``missing satellite problem'' and the ``too-big-to-fail'' problem;
{Kauffmann \etal\ 1993}\nocite{kauffmann93}, {Klypin
  \etal\ 1999}\nocite{klypin99}, {Moore \etal\ 1999}\nocite{moore99},
{Boylan-Kolchin \etal\ 2011}\nocite{boylan11}, {Klypin
  \etal\ 2015}\nocite{klypin15}, {Papastergis
  \etal\ 2015}\nocite{papastergis15}).  Increasing the statistics in
this critical mass range offers an opportunity to better understand
the physical properties of these galaxies via detailed observational
study.

To this end, the ALFALFA survey \citep{giovanelli05} has extended the
faint end of the \HI{} mass function into the 10$^6$ \msun\ $\lsim$
M$_{\rm HI}$ $\lsim$ 10$^7$ \msun\ regime for the first time.  As
discussed in the companion paper by Teich \etal\ (hereafter referred
to as Paper~I), the SHIELD program was designed to identify those
systems from the full ALFALFA catalog with log(M$_{\rm HI}$) $<$ 7.2
and with narrow \HI{} line widths (v$_{\rm 50}$ $<$ 65 \kms, thus
removing massive but \HI{}-deficient galaxies).  In Paper~I and the
present work, 12 of these sources are analyzed extensively in an
effort to understand their physical properties and to contextualize
them among the general population of low-redshift galaxies.  Analysis
continues on the other low-mass galaxies discovered in ALFALFA via the
same criteria.

In this paper, we focus on the dynamical properties of the SHIELD
galaxies to extend the BTFR to the lowest-mass gas-rich galaxies.  We
refer the reader to Paper~I for physical characteristics of the SHIELD
galaxies, for details about the \HI{} data reduction, for details
about the supporting observations used in both works, and for results
specific to the properties of star formation in the SHIELD galaxies
(see also {McQuinn \etal\ 2015a}\nocite{mcquinnsfh}).  Here we only
include discussion of relevant \HI{}-specific data handling.  This is
followed by formal analysis of the data in an effort to determine the
rotation velocities of the SHIELD galaxies.

\section{Observations and Data Handling}
\label{S2}

The SHIELD observational strategy was to observe each galaxy in the D,
C, and B configurations (maximum baseline lengths of 1.03\,km,
3.4\,km, and 11.1\,km, respectively) for 2 hours, 4 hours, and 9
hours, respectively.  The native velocity resolution is 0.824
km\,s$^{-1}$\,ch$^{-1}$.  Data were acquired for programs VLA/10B-187
(legacy identification AC\,990) and VLA/13A-027 (legacy identification
AC\,1115).  As demonstrated in Table~\ref{im.props}, most of these
data were successfully acquired; three sources were not observed in
the B configuration (AGC\,111164, AGC\,111977, AGC\,112521), and two
sources were only observed for 4.5 hours in the B configuration
(AGC\,110482, AGC\,111946).  Paper~I provides complete details about
the calibration and imaging of the 42 independent execution blocks
acquired by the SHIELD programs using the VLA.

Inversion and deconvolution of the visibility data were performed
using the Cotton-Schwab \textsc{clean} algorithm implemented in the
Common Astronomy Software Application (CASA; {McMullin
  \etal\ 2007}\nocite{mcmullin07})\footnote{https://casa.nrao.edu/}.
We produced data cubes using two different values of the Briggs
\textsc{robust} parameter: a ``high resolution'' product with
\textsc{robust}$=$0.5, and a ``low resolution'' product with
\textsc{robust}$=$2.0. Data products that include B configuration data
have 1.5\arcsec\ pixels; the rest of the images have 4\arcsec\ pixels.
Table~\ref{im.props} provides a summary of the multi-configuration
image cube properties. The resulting angular resolutions vary between
$\sim$5\arcsec\ and $\sim$35\arcsec; the corresponding physical
resolutions range from $\sim$140 pc (high resolution images of
AGC\,749241) to $\sim$1 kpc (low resolution images of AGC\,111977).

Paper~I\nocite{teich} presents an exhaustive analysis of the
integrated distribution of the neutral hydrogen in the SHIELD
galaxies.  The spatial distribution and projected mass surface
densities of neutral hydrogen gas allow a detailed comparison with
star formation tracers.  The two-dimensional representation of the
integrated neutral gas surface density, typically referred to as the
``moment zero'' image, was created by manually masking each of the
three-dimensional data cubes.  The moment zero images presented in
Paper~I\nocite{teich} use the ``high resolution'' cubes
(\textsc{robust}$=$0.5) and are corrected for residual flux rescaling
\citep{jorsater95}; here we show the moment zero images from the ``low
resolution'' data products, uncorrected for residual flux rescaling,
in the upper left panels of Figures~\ref{110482.collage} through
\ref{749241.collage}.  The moment zero images are presented in column
density units of 10$^{20}$ cm$^{-2}$.  Channel maps of the full
data cubes from which these moment zero images are derived are
presented in Appendix~\ref{appendix}.

Two important tools used in the kinematic analysis of galaxies are the
first and second moments of the three-dimensional data cube.
Typically these products are respectively referred to as the
``velocity field'' and the ``velocity dispersion'' images.  The first
moment of a typical \HI{} data cube is a two-dimensional image of a
source where each pixel value represents the intensity weighted
average velocity.  The second moment of a data cube likewise
represents the intensity weighted velocity dispersion of the spectral
profile at each sampled position.  

The first and second moments of data cubes are useful, but they do not
always yield the most realistic depiction of the velocity or
dispersion of the gas at a given location within a source, especially
at low ratios of signal to noise (S/N).  This is because moment maps
favor the contributions of the brightest parcels of gas - they are
weighted by intensity.  To mitigate these effects, velocity fields and
dispersion maps can be obtained by fitting (e.g., Gaussians or Hermite
polynomials) through each pixel's velocity profile.  By fitting a
continuous, function to the spectral line profiles of each SHIELD
galaxy, we limit the contribution of high-dispersion spurious noise
that would otherwise potentially skew the weighting of the velocity
fields; this allows more perfect decomposition of the main gas
component from minor additional gas components.

After checking our results for consistency using a variety of fits to
velocity fields produced from data cubes of different resolutions, we
find that single-peaked Gaussian profiles fit to the low-resolution
(\textsc{robust}$=$2.0) data products returned the most continuous and
ordered velocity and dispersion fields.  We fit single Gaussians to
the velocity profiles using the task \textsc{xgaufit} in the software
package \textsc{GIPSY}\footnote{The Groningen Image Processing System
  (\textsc{GIPSY}) is distributed by the Kapteyn Astronomical
  Institute, Groningen, Netherlands.}  \citep{gipsy92}.  The fitting
parameters enforced a lower amplitude bound at twice the measured RMS
in the cubes, a lower dispersion bound equal to the width of a single
channel, and $\sim$60 \kms{} velocity boundaries, using the central
velocity of the data cubes as a prior estimate of systemic velocity.
All velocity information was obtained from the calibrated,
non-blanked, non-residual-flux-rescaled image cubes as in
\citet{ott12}. The amplitude and dispersion of the Gaussian profiles
were then extracted as velocity fields and dispersion maps; these
first and second moments of the data cubes are shown in the upper
middle and in the upper right panels of Figures~\ref{110482.collage}
through \ref{749241.collage}.

The images of the SHIELD galaxies shown in
Figures~\ref{110482.collage} through \ref{749241.collage} allow a
visual comparison of the stellar components with the gaseous
components.  A 2-color Hubble Space Telescope image is shown in the
bottom left panel, while the Spitzer infrared 4.5 $\mu$m image is
shown in the bottom right panel.  These figures also facilitate
comparison of the global spectral profiles of the sources, using both
ALFALFA spectra and the interferometric measurements from the VLA that
are further analyzed in subsequent sections.

\section{Kinematics and Dynamics}
\label{S3}

The primary goal of this work is to determine the rotation velocity of
each SHIELD galaxy, preferably on a spatially resolved basis (i.e., to
extract a rotation curve).  Provided a well-sampled ($u$,$v$) plane,
the deconvolved three-dimensional image cube is a faithful
representation of the gas kinematics of a particular source.  As
discussed in detail above, the collapse of the three-dimensional
velocity structure into a two-dimensional velocity field
representation is inherently limited: the output image is weighted by
intensity and thus offers an incomplete perspective of the retrieved
velocity structure. Nonetheless, it is common to attempt modeling a
galaxy's rotational dynamics directly from the velocity field
\citep{fraternali02, oh11, davis13, jadams14,elson14, oh15}.

In this work, we explore the gas kinematics of the SHIELD galaxies
using three approaches.  In \S~\ref{S3.1} we attempt traditional
tilted ring fitting using the two-dimensional velocity fields as input.  In
\S~\ref{S3.2} we apply multiple three-dimensional fitting techniques
to the image cubes.  In \S~\ref{S3.3} we perform a spatially resolved
position-velocity analysis.

\subsection{Two-Dimensional Modeling:  Tilted Ring Analysis}
\label{S3.1}

One of the standard analytical methods used to derive a rotation curve
from a 2-dimensional velocity field is tilted ring modeling
\citep{trm}.  Tilted ring models (TRM) attempt to reconstruct the {\em
  three-dimensional} structure and dynamics of sources from {\em
  two-dimensional} velocity fields and velocity dispersion maps.  For
systems with ordered disk rotation, TRM are a proven diagnostic of
galaxy dynamics \citep{cannon12, schmidt14, salak16}.

Prior to fitting the velocity fields of the SHIELD galaxies, they were
blanked using a Boolean mask admitting high S/N emission in the moment
zero maps.  This threshold masking eliminates noise and unphysical
velocities (which usually manifest as single isolated pixels) from the
edges of the velocity fields; the regions of emission above the
Gaussian fitting threshold parameters fall well within the footprint
of high S/N emission in the moment zero maps and thus are not
affected.  As noted above, the blanked velocity fields of each source
are presented as the top center panel of Figures \ref{110482.collage}
through \ref{749241.collage}; note that the color scale bar at the top
of each center panel represents source recessional velocity in
km\,s$^{-1}$.

A TRM attempts to fit concentric ellipses of known inclination to a
velocity field and thus provides a best-fit model solutions for the free
kinematic parameters of major axis position angle (PA) and inclination
($i$) as a function of radial distance from the dynamical center.
This model of nested tilted rings assumes rotational support and gas
coherence, and enables deprojection of the velocity contributions of
different parcels of gas into the deprojected circular velocity,
hereafter referred as V$_{\rm rot}$.  There are a variety of software
packages commonly used to perform this deprojection (\textsc{reswri},
\textsc{ringfit}, \textsc{kinemetry}); we employ the \textsc{GIPSY}
task \textsc{rotcur}, which performs a least-squares-fitting algorithm
to the following function:

\begin{equation}
\begin{aligned}
  v(x,y) ={}
  V_{sys} + V_{rot} \cdot \cos(\theta) \cdot \sin(i)
  + V_{exp} \cdot \sin(\theta) \cdot \sin(i)
\end{aligned}
\label{rceq}
\end{equation}
where
\begin{equation}
\begin{aligned}
  \cos(\theta) ={}
  - \bigg(\frac{(x-XPOS) \cdot \sin(PA)}{r} \\
  + \frac{(y-YPOS) \cdot \cos(PA)}{r}\bigg)
\end{aligned}
\label{expand}
\end{equation}

\noindent In Equations \ref {rceq} and \ref{expand}, $v(x,y)$ is the
radial velocity in rectangular coordinates, $V_{sys}$ is the systemic
recessional velocity of the Doppler-shifted emission, $V_{rot}$ is the
rotational component of the projected velocity $V_{max}$, $i$ is the
inclination of a given ring (positive increase defined along the line
of sight, out of the plane of the sky), $V_{exp}$ is the radial
component of the projected velocity (i.e., the expansion velocity),
XPOS and YPOS are the right ascension and declination of the kinematic
center with respect to the center of the imaged field, and $PA$ is the
position angle of the receding side of the major axis of rotation
defined with north$=$0$\degree{}$ and increasing to the east.  To
reduce the number of free parameters in our model and explicitly
determine the rotation-supported component of the circular velocities
at each ring, we assume negligible asymmetric drift (see
{Bernstein-Cooper \etal\ 2014}\nocite{ezbc14}); that is, we assume
zero radial component to the motions of the rings.

\textsc{rotcur}'s Levenberg-Marquardt solver fits kinematic parameters
within concentric rings of finite thickness (typically half of the
width of the resolving beam major axis).  There is less possibility of
finding degenerate solutions with this algorithm when the number of
points inside of each ring is maximized, and when the tilted rings can
fit to significant emission to the maximum radial extent.  For the
faint galaxies of this sample, the highest sensitivity to extended
emission comes from fitting to the \textsc{robust}=2.0 (i.e.,
``natural'' image weight) moment maps.  However, for those sources
without VLA B configuration data, we used ROBUST=0.5 to achieve the
higher resolution models.

In order to explore the effects of varying spectral and angular
resolution, \textsc{rotcur} was run on velocity fields at four
different resolutions for each galaxy: 1) natural weighting and native
spectral resolution; 2) natural weighting and spectral resolution
Hanning smoothed by a factor of three; 3) natural image weighting with
a Gaussian taper on the ($u$,$v$) plane and native spectral
resolution; 4) natural image weighting with a Gaussian taper on the
($u$,$v$) plane and spectral resolution Hanning smoothed by a factor
of three.  Each masked velocity field was used as input into
\textsc{rotcur} and the program was allowed to run iteratively.  At
first, every kinematic parameter of the fit was left free for each
ring, and then parameters were constrained and held constant for all
subsequent rounds of the fitting process at the rings' mean value,
weighted by the residual error. The order in which the parameters were
constrained does not produce a statistically significant difference in
the fit for any parameter except in the most extreme cases where
\textsc{rotcur} had difficulty fitting the rotation curve altogether.
Thus the order in which the parameters was constrained followed the
same pattern for each velocity field: $V_{sys}$, $PA$, $XPOS$, $YPOS$,
and then $i$.  The expansion velocity of each ring $V_{exp}$ was
explicitly held at zero for all model fits; under the assumption of
zero net expansion velocity, the minimization procedure describes only
the rotation support of the gas for each ring.

This iterative tilted ring fitting process was attempted for each of
the SHIELD sources.  However, only five galaxies (AGC\, 110482,
AGC\,112521, AGC\, 175605, AGC\,731457, and AGC\, 749237) had
convergent models.  The resulting rotation curves for these galaxies
are shown in Figure~\ref{rotation.curves}.

AGC\,749237 has the rotation curve most closely resembling those of
larger dwarf and spiral galaxies (i.e., steeply rising with radius,
then flat).  There appears to be a sharp velocity increase across the
innermost $|sim$10\arcsec\ ($\sim$560) parsecs of the galaxy, giving
way to a flattened rotation curve at greater radial distance.  The
tapered and smoothed data have higher modeled deprojected rotation
velocities because of a markedly different fitted inclination angle
(by more than 20$\degree$) than the other fits. It is important to
note that AGC\,749237 has the largest single-dish \HI{} line width of
any of the SHIELD galaxies.  And yet, the various TRMs still show
ambiguity in the final value of the circular rotation velocity of the
source. This result is characteristic of the rest of the sample; the
lower (and the more diffuse) a source's cumulative \HI{} flux, the
more difficult it is to create spatially resolved kinematic models
with high significance.

The solid-body rotation that characterizes many well-studied dwarf
galaxies \citep{spekkens05, deblok10} is evident in the rotation curve
model solutions for AGC\,110482, AGC\,112521, and AGC\,174605.  The
fits rise relatively smoothly to V$_{\rm rot}$ $\simeq$ 15\,$\pm$\,5
\kms\ (AGC\,110482 and AGC\,11252); the result for AGC\,174605 favors
an even lower V$_{\rm rot}$, although the uncertainties are
significant.  The dispersion of the four different fits for a given
galaxy provides an indication of the systematic uncertainties.  The
model constructed from the tapered data for AGC\,112521 shows marginal
evidence for a downturn at large radii, but this interpretation is
tenuous because of the relatively high resulting uncertainties and low
S/N of the gas at large radial distance.


The rotation curve of AGC\,731457 is difficult to interpret.  The
tapered data favor a rising rotation curve at all radii, but the full
resolution images are consistent with projected velocities equal to
the velocity dispersion ($\sim$10-15 \kms).  This source is the
second-most distant SHIELD galaxy except for AGC\,749237 (see
Table~\ref{kin.props}). The solutions that we derive using this method
should be interpreted with caution for each of the sources. However,
unlike AGC\,749237 (whose rotation curve is spatially resolved to a
degree comparable with studies of more massive, closer galaxies), the
rotation curve solutions for AGC\,731457 appear to disagree even
between maps of different resolution.

From this analysis we conclude that only AGC\,749237 is adequately fit
by a simple two-dimensional TRM. This is perhaps expected, given that
inclination and rotation velocity are completely degenerate for
velocity-field fits if the rotation curve is solid-body.  The low S/N,
diffuse interstellar media, and comparable magnitudes of projected
rotation and velocity dispersion found in the SHIELD galaxies require
characterization using a model that brings in external constraints in
an effective manner, thus decreasing degeneracies in the solution.

\subsection{Three-Dimensional Modeling}
\label{S3.2}

Based on the limitations encountered by the two-dimensional methods
described above, we next attempted to model the dynamics of the SHIELD
galaxies using the full three-dimensional information in the data
cubes.  As shown in the channel maps presented in the
Appendix~\ref{appendix}, there is movement of the \HI{} gas through
many of the three-dimensional data cubes that is visible to the eye.
The primary limitation in accessing the rotational information is the
size of the resolution element (the synthesized restoring beam).  For
the SHIELD galaxies, only a few disks are resolved at the Nyquist
limit ($\sim$3.4 synthesized beam elements across the rotation axis).

We explored three modeling packages to attempt this analysis: the
\textsc{GIPSY} task \textsc{GALMOD} \citep{gipsy92}; the Tilted Ring
Fitting Code \textsc{TiRiFiC} \citep{jozsa07}; and the 3D-Based
Analysis of Rotating Objects from Line Observations code
\textsc{3dBAROLO} \citep{diteodoro15}.  Each of these software
packages uses numerical methods to construct TRMs from
three-dimensional intensity and velocity information.

Three dimensional modeling of the SHIELD sample has proved
inconclusive for even the most massive galaxies in the sample.  These
systems are the most amenable to dynamical modeling: they have the
highest column densities and the largest projected rotation
velocities. These results highlight the significant degeneracies in
the kinematic parameters of the SHIELD galaxies using the \HI{} data
alone.  Some of these degeneracies can be constrained by using
optically derived properties. However, in order to provide an
un-biased examination of the observations, we explore the \HI{}
observations through position-velocity mapping (see next section).
Note that we will later rely on optical observations for some
properties (e.g., inclination) in order to derive inherent properties
(e.g., dynamical masses).

\subsection{Position-Velocity Mapping}
\label{S3.3}

In the absence of convergent three-dimensional model fitting
procedures, we explore the three-dimensional velocity information in
each cube by undertaking a spatially resolved position-velocity (P-V)
analysis.  Here we leverage the well-known capability of traditional
P-V analysis to identify two important maxima in a given data cube.
The first is the maximum projected rotation velocity along a given
slice; this occurs when that slice is drawn along the kinematic major
axis of a galaxy.  The second is the intrinsic projected velocity
width; this is the velocity extent of the gas along a slice that is
orthogonal to the major axis slice.  In the presence of ordered
rotation, this analysis provides a reliable estimate of the kinematic
major axis.  The inclination of the disk remains poorly constrained,
and needs an additional prior.

As implemented in \citet{most11a} and \citet{ezbc14}, we employ a
spatially resolved P-V analysis using the low-resolution
(robust$=$2.0) data cubes.  The kinematic major axis is identified
through inspection.  Once achieved, we then create a series of minor
axis slices that span the length of the galaxy's gas disk.  The
central minor axis slice intersects the major axis slice at the
dynamical center of the source.  The other minor axis slices are
offset by the synthesized beam width along the major axis slice.
Examining the velocity centroids of these minor axis cuts as a
function of position along the major axis serve as a diagnostic of the
magnitude of the projected rotation of the source.

For each SHIELD galaxy, we manually identified the position angle of
the major axis (positive moving east of north) using the
\textsc{KPVSLICE} tool in the \textsc{KARMA} package. The position of
the major axis slice was chosen to produce the maximum spatial and
velocity extent; a secondary requirement was that the slice passes
through an \HI{} surface density maximum if evident.  For systems with
well-defined major axes from the velocity fields, this position is
obvious.  However, for sources without signatures of strong rotation,
the position of the major axis slice effectively attempts to maximize
S/N.  The location of the major axis slice through the AGC\,748778
data cube (see Figure~\ref{748778.collage}) is a useful example; there
is no obvious center of rotation in the velocity field image, and so
the major axis slice passes along the extent of the bulk of the \HI{}
gas.  The major axis position angle of each source is given in
Table~\ref{kin.props}, along with other kinematic properties.

The locations of the major and minor axis slice locations are shown in
the velocity field panels of Figures~\ref{110482.collage} -
\ref{749241.collage}.  In all the maps, position angle is defined as
positive east of north. Positive offset in the major axis frame is
defined with respect to the receding half of the galaxy in the cases
where disk-like rotation was readily identifiable from the rotation
cubes, or towards the more northerly direction for sources without a
strong rotation gradient.

The resulting spatially resolved P-V diagrams are presented in
Figures~\ref{110482.slices} through \ref{749241.slices}.  Contours
overlain shown levels of increasing surface brightness in the cube.
The solid and dashed lines show the full velocity extent of \HI{} gas
from each source, while the dotted line shows the geometric midpoint
of those boundary values; the dotted line can be considered a P-V
based estimate of the systemic velocity of each source.  Half of the
sample members show some evidence for solid-body rotation in the major
axis slices: AGC\,110482, AGC\,111164, AGC\,111977, AGC\,112521,
AGC\,174605, and AGC\,749237.  The other sample members (AGC\,111946,
AGC\,174585, AGC\,182595, AGC\,731457, AGC\,748778, and AGC\,749241)
do not show a perceptible velocity gradient along the major axis
slice.

AGC\,731457 presents an especially difficult dynamical case (see
Figures~\ref{731457.collage} and \ref{731457.slices}).  The
moment-zero map shows a centrally concentrated \HI{} distribution,
with low surface brightness structure in the outer disk.  The stellar
component is compact compared to the neutral gas.  Regardless of which
value was used for the kinematic major axis, the velocity extents of
the major and minor axis P-V slices was essentially unchanged.  The
location of the dynamical center thus passes through the \HI{} surface
density maximum, including gas below the 10$^{20}$ cm$^{-2}$ level;
the orientation also carries the slice through the highest surface
brightness \HI{} gas. This orientation appears to be in conflict with
the (very weak) velocity gradient apparent in the upper middle panel
of Figure~\ref{731457.collage}.  However, we stress that the maximum
velocity extent seen by this major axis slice is essentially
indistinguishable from any others that pass through the \HI{} surface
density maximum.
 
The advantage of this spatially resolved P-V analysis is clear:
signatures of projected rotation can be quantified, even for some
systems where the two-dimensional (\S~\ref{S3.1}) and
three-dimensional (\S~\ref{S3.2}) modeling fail.  AGC\,111977 (see
Figures \ref{111977.collage} and \ref{111977.slices}) is a good
example; the velocity field image suggests rotation along a clear
major axis.  The major axis P-V slice suggests that \HI{} gas is
moving at projected velocities between 180 \kms\ and 210 \kms, over an
angular region spanning $\pm$45\arcsec.  The projected rotation is
apparent as a gradient in the velocity of the centroids of the \HI{}
gas along the minor axis slices; the same $\pm$15 \kms\ of projected
rotation is apparent.

These PV diagrams provide robust measurements of the projected gas
velocity (V$_{max}$) for each of the 12 SHIELD galaxies.  This comes
with the added benefits that P-V slice mapping does not suffer from
the effects of beam smearing due to collapse to two dimensions.
Further, P-V slice mapping does not depend on the potentially
ambiguous geometrical parametrizations inherent to three-dimensional
modeling.  Crucially, however, by adopting maximum projected rotation
values from the P-V diagrams, we have not fit a convergent tilted ring
model to these sources. Consequently, the inclinations of their gas
disk components remains unconstrained.

\subsection{Dynamical Masses}
\label{S3.4}

Given that the SHIELD galaxies are not amenable to resolved rotation
curve analysis, the next most important physical parameter that we can
determine is the total dynamical mass of each galaxy.  By estimating
the rotational velocity at the largest reliable distance from the
dynamical center of each source, we can make an estimate of the total
depth of the gravitational potential well.  By comparing to previous
measurements of the luminous components (stars, gas, dust), we can
infer global dark matter fractions.  Finally, with a reliable estimate
of the rotational velocity and the sum of the baryonic masses, we can
contextualize the SHIELD galaxies on the BTFR.

An important physical parameter in determining the maximum rotational
velocity of the SHIELD galaxies is the inclination of the disk with
respect to the line of sight.  Ideally this parameter is determined
from the gas kinematics, and is allowed to vary as a function of
position within the disk (e.g., to account for warps).  However, as
discussed in \S~\ref{S3.3}, we are unable to achieve unambiguous
rotational models using either two or three-dimensional analysis
techniques. 

Without a kinematic measure of inclination from the \HI{}, we thus
turn to the stellar component for a determination of the
inclination. We note that the stellar and gaseous inclinations are
often evidently different, especially in the case of gas-dominated
dwarfs whose neutral hydrogen reservoirs are significantly more
extended than the stellar component.  In the most extreme examples
(e.g., AGC\,749241; see Figure~\ref{749241.collage}), there is very
little resemblance between the \HI{} and optical morphologies.  The
inclination derived from the gas component is only reliable in those
cases where coherent rotation is obvious (see
Figure~\ref{rotation.curves}).  Nonetheless, an estimate of the
stellar disk inclination offers a meaningful substitute; importantly,
it is one that can be applied in a uniform and reproducible way for
all members of the SHIELD sample.

As discussed and shown in the companion {Paper~I}\nocite{teich}, the
inclination used for deprojecting V$_{max}$ into tangential velocity
was determined using the axial ratios of ellipses fit to the stellar
population in masked I-band Hubble Space Telescope images using the
\textsc{CleanGalaxy} isophote-fitting code ({Hagen
  \etal\ 2014}\nocite{hagen14}; see also FITGALAXY {Marshall
  2013}\nocite{marshall13}).  \textsc{CleanGalaxy} allows removal of
foreground and background contaminants, and then automatically fits
elliptical surface brightness contours as a function of galactocentric
radius.  Note that these inclination measurements describe a different
underlying galactic population (the stars), and are constrained from
observations of higher spatial resolution. The adopted inclination
values are listed in Table~\ref{kin.props} and shown in Figure 1 of
{Paper~I}\nocite{teich}.

The compilation of our derived kinematic parameters for the SHIELD
galaxies is presented in Table~\ref{kin.props}.  The largest angular
extent to which we confidently measure \HI{} gas in projected rotation
is listed as R$_{\rm max}$.  The inclination-corrected circular
velocity at this location is then given as V$_{\rm max}$; note that
the maximum velocity of significant emission along the ``major'' axis
of each galaxy's PV diagram was halved under the simplifying
assumption of axisymmetric gas disks.  Note that R$_{\rm max}$ V$_{\rm
  max}$ as defined here are not the host halo's maximum circular
velocity and the radius at which the circular velocity curve peaks;
comparison with simulations should bear this in mind.

We determine the baryonic mass of each source by adding the total gas
mass to the total stellar mass.  As tabulated in
{Paper~I}\nocite{teich}, the total \HI{} mass (using the Arecibo flux
integral) is corrected by a factor of 1.35 to account for other gas
species.  We do not correct the gas masses for a contribution from
molecular gas or from dust.  However, we expect these components to be
less massive than the \HI{} component; the galaxies are metal-poor
\citep{haurberg15} and therefore do not have a significant amount of
dust, and the paucity of molecular gas in low-mass galaxies is
well-documented (see, e.g., {Warren \etal\ 2015}\nocite{warren15},
{Rubio \etal\ 2015}\nocite{rubio15} and references therein).

\begin{figure}
  \includegraphics[width=\textwidth]{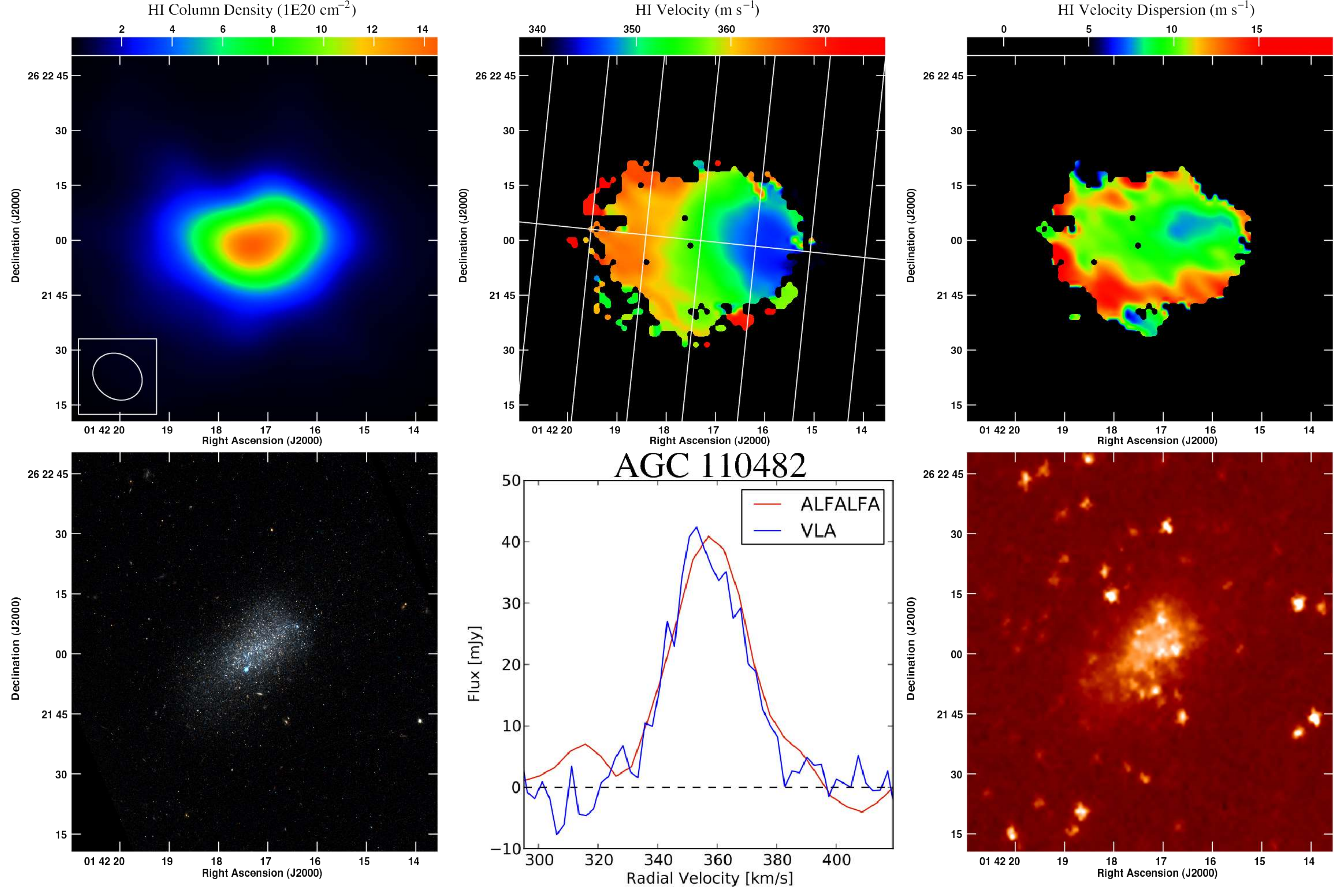}
  \caption{The gaseous and stellar components of the SHIELD galaxy
    AGC\,110482.
{\it Upper left:} moment zero image generated from the naturally weighted image
cube, manually blanked using the method described in  Paper~I,
with the synthesized beam overlaid; the scale bar shows column density in units of 
10$^{20}$ cm$^{-2}$.
{\it Upper middle:} intensity-weighted velocity field, produced by fitting a 
single Gaussian function and blanking at the 10$^{20}$ atoms\,cm$^{-2}$ level from the 
moment zero image; the scale bar shows velocity in units of 
\kms.
The singly-oriented white line represents the major axis position
angle used to produce the top panel of Figure \ref{110482.slices}. The
seven perpendicular white lines indicate the minor axis slices used to
produce the bottom panels of Figure \ref{110482.slices}.  The
intersection of the central minor axis slice line with the major axis
slice line is centered at the determined dynamical center of the
galaxy.
{\it Upper right:} velocity dispersion image, produced by fitting a 
single Gaussian function and blanking at the 10$^{20}$ atoms\,cm$^{-2}$ level from the 
moment zero image; the scale bar shows velocity dispersion in units of 
\kms.
{\it Lower left:} three-color HST image of AGC\,110482, as presented in
McQuinn \etal\ (2014). 
{\it Lower middle:} global \HI{} spectra using the VLA data (blue) and using 
the ALFALFA data (red).
{\it Lower right:} Spitzer 4.5 \textmu{}m image of AGC\,110482.}
\label{110482.collage}
\end{figure}

\begin{figure}
\includegraphics[width=\textwidth]{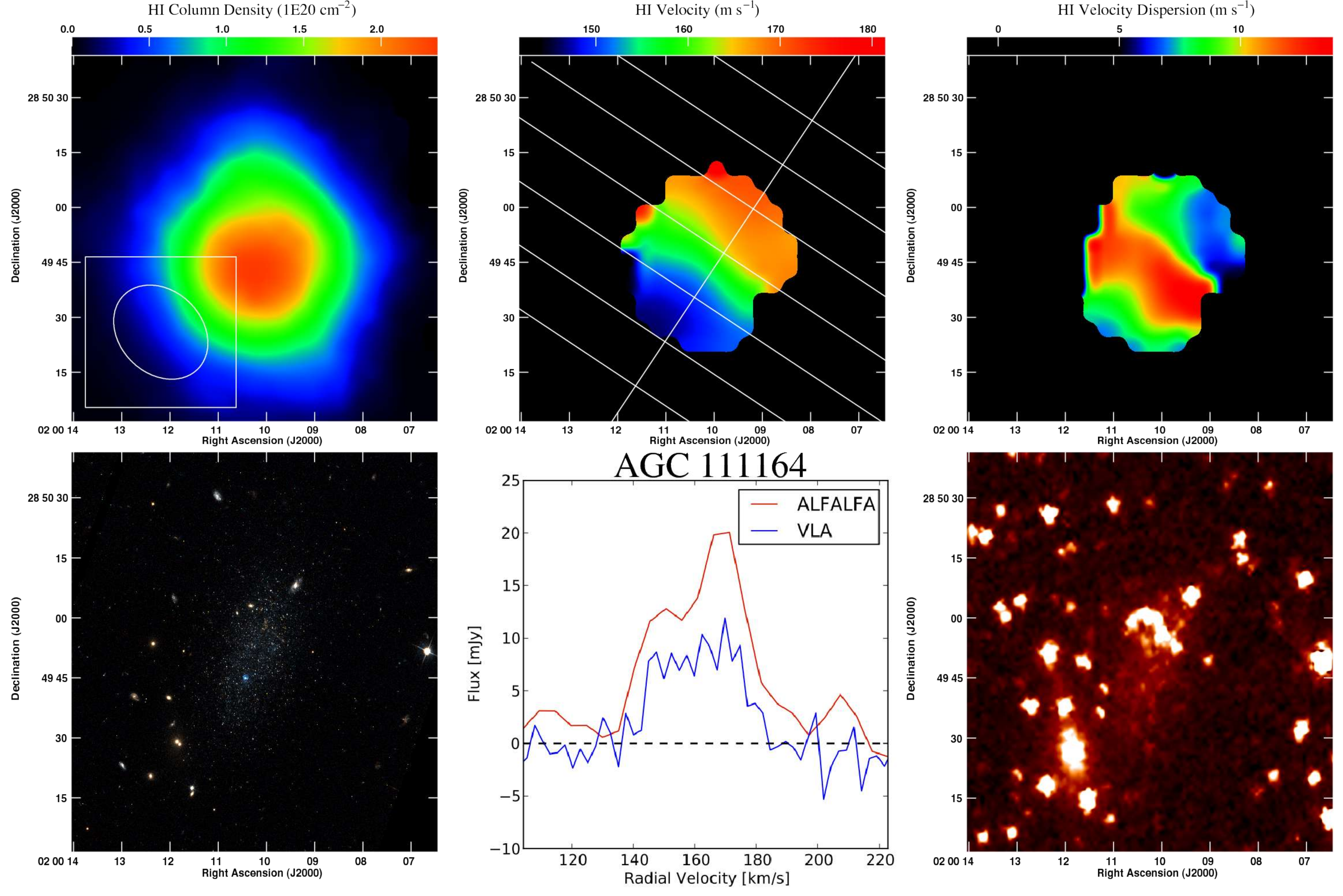}
  \caption{Same as Figure 1, for AGC\,111164}.
\label{111164.collage}
\end{figure}

\begin{figure}
\includegraphics[width=\textwidth]{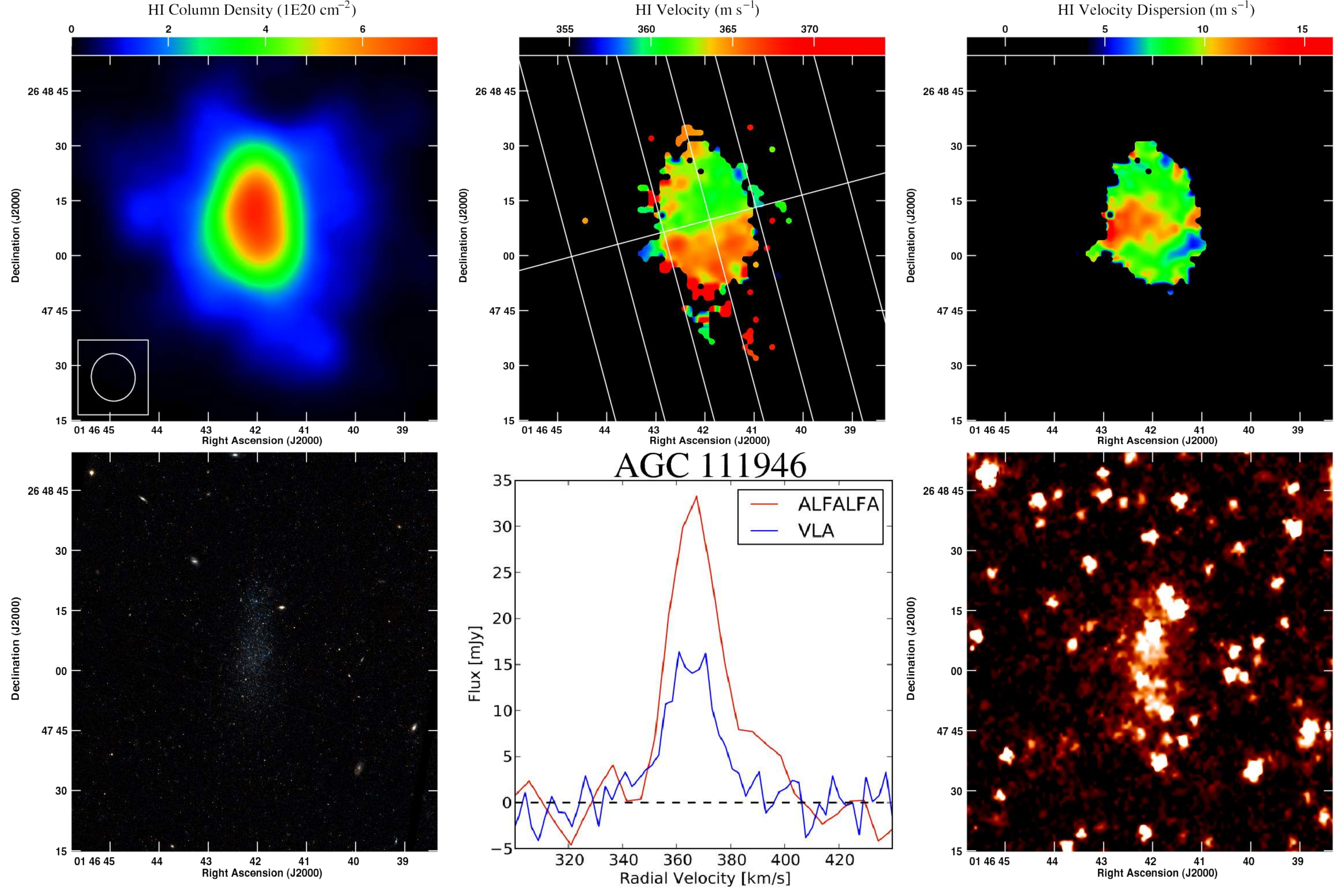}
  \caption{Same as Figure 1, for AGC\,111946.}
\label{111946.collage}
\end{figure}

\begin{figure}
\includegraphics[width=\textwidth]{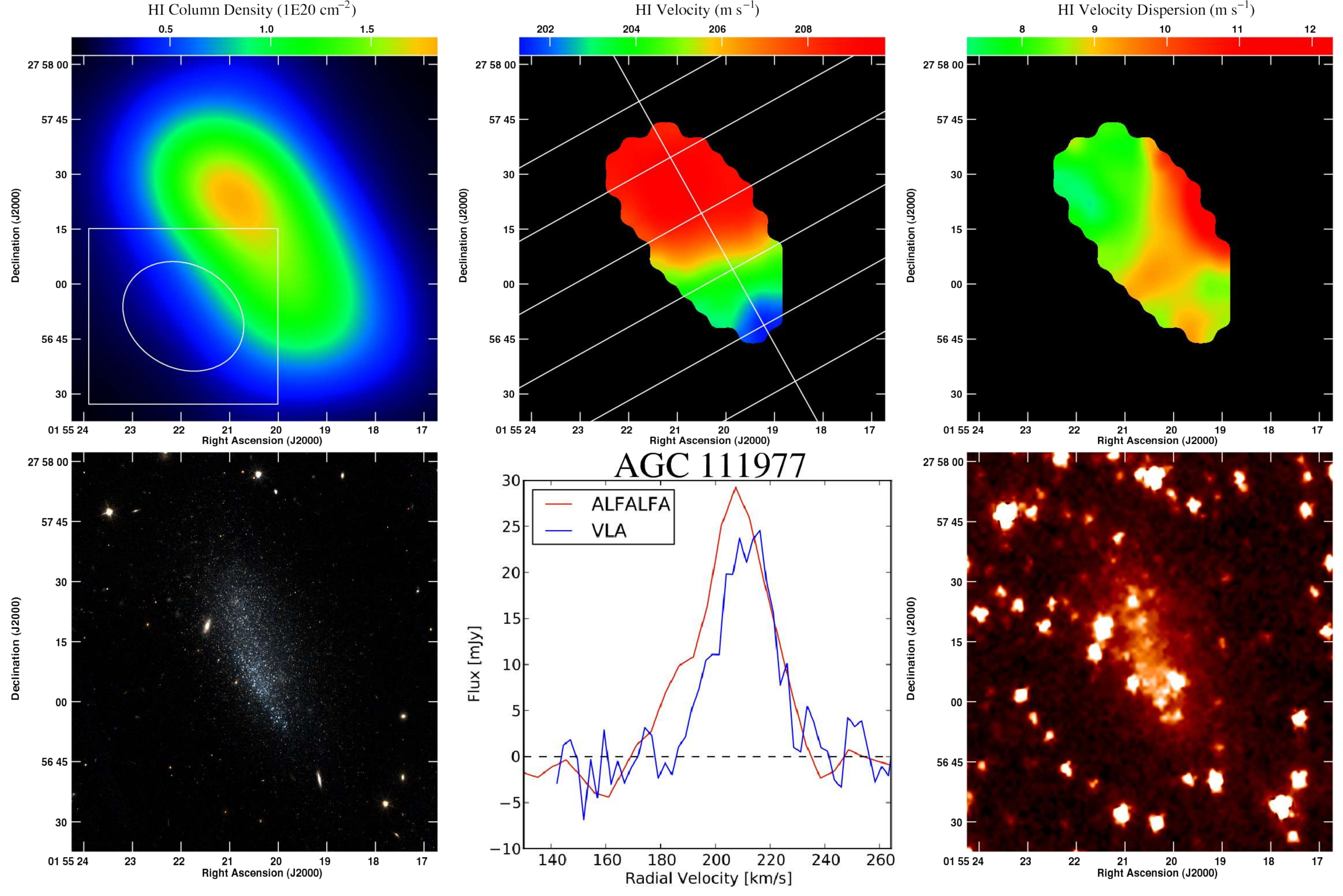}
  \caption{Same as Figure 1, for AGC\,111977.}
\label{111977.collage}
\end{figure}

\begin{figure}
\includegraphics[width=\textwidth]{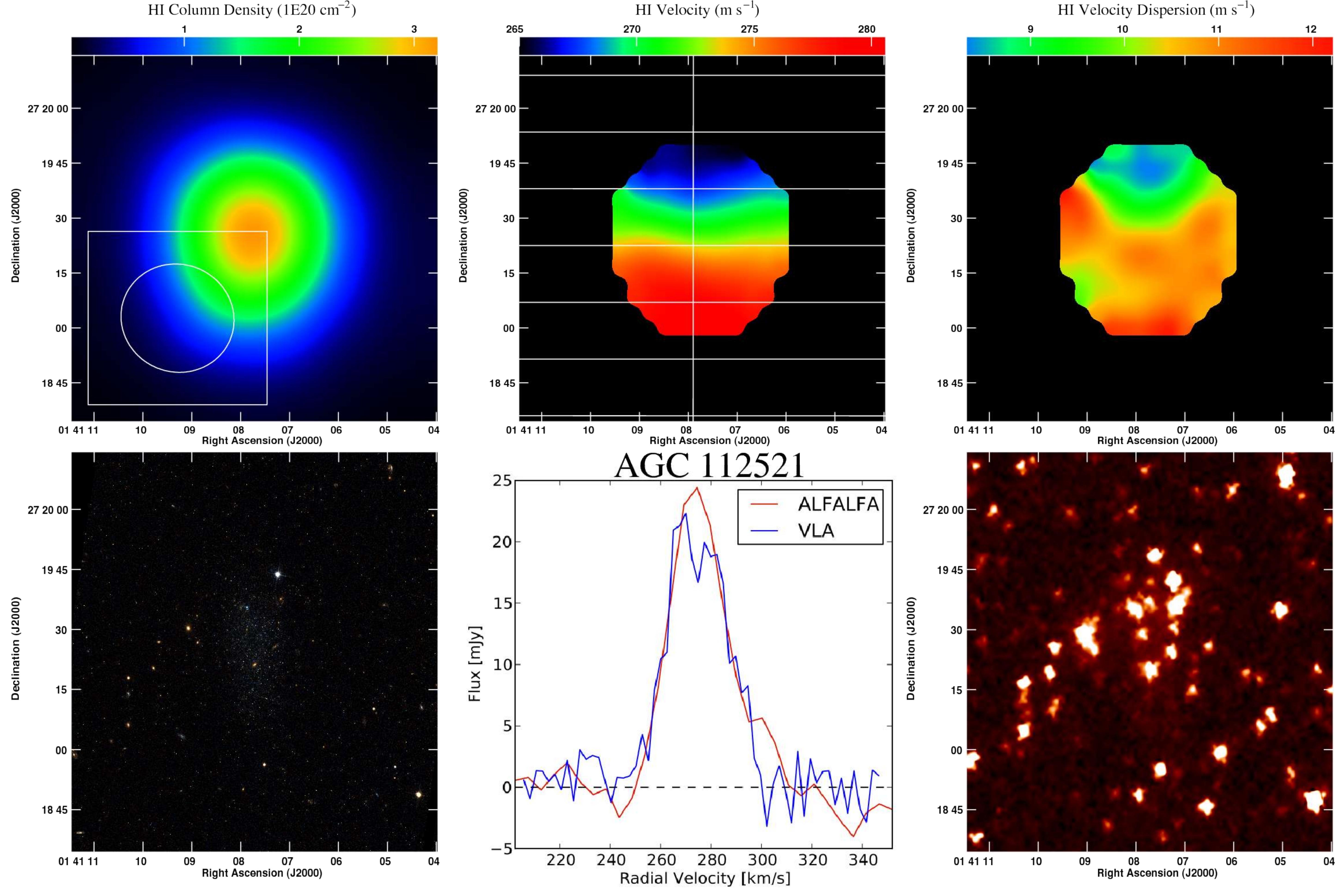}
  \caption{Same as Figure 1, for AGC\,112521.}
\label{112521.collage}
\end{figure}

\begin{figure}
\includegraphics[width=\textwidth]{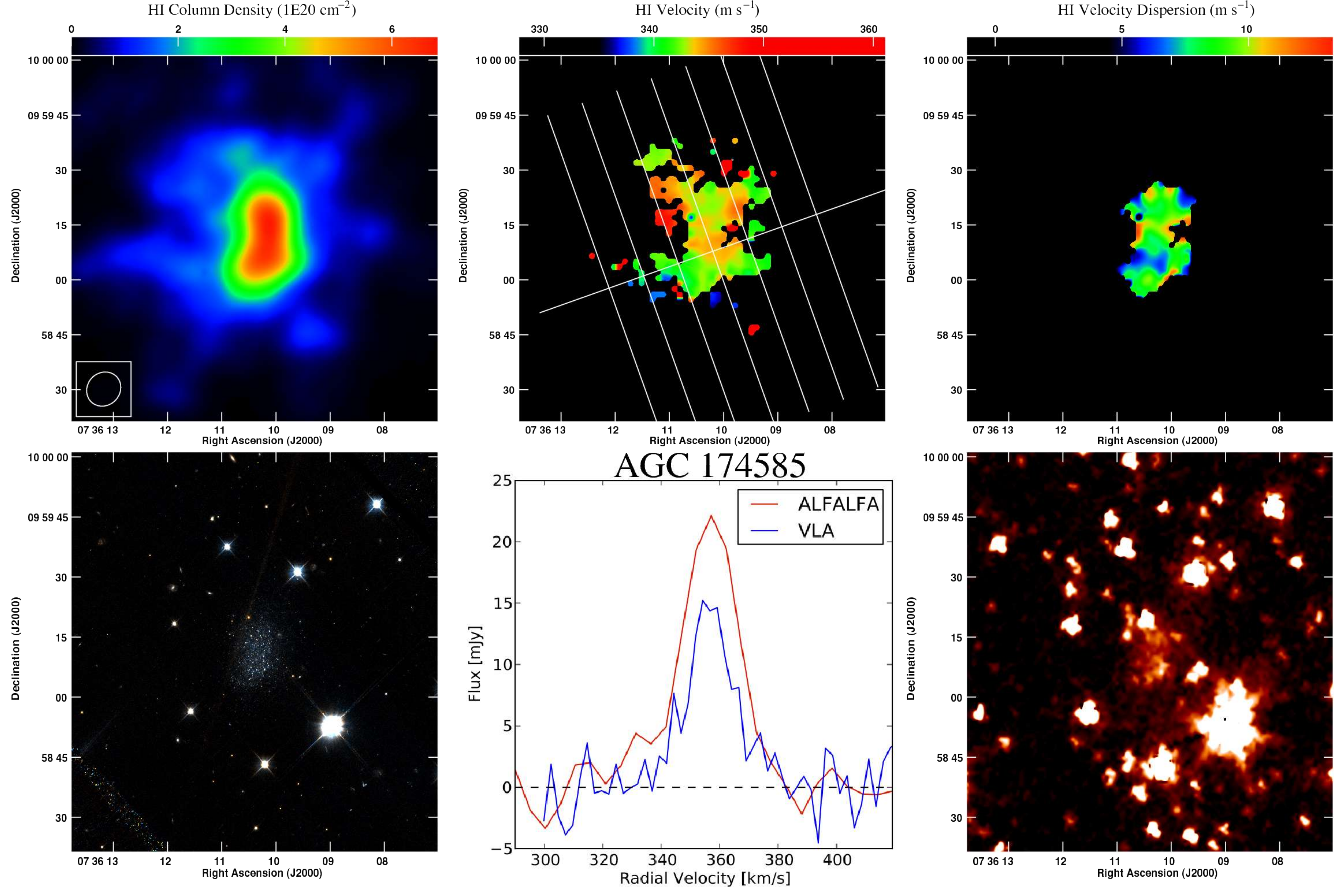}
  \caption{Same as Figure 1, for AGC\,174585.}
\label{174585.collage}
\end{figure}

\begin{figure}
\includegraphics[width=\textwidth]{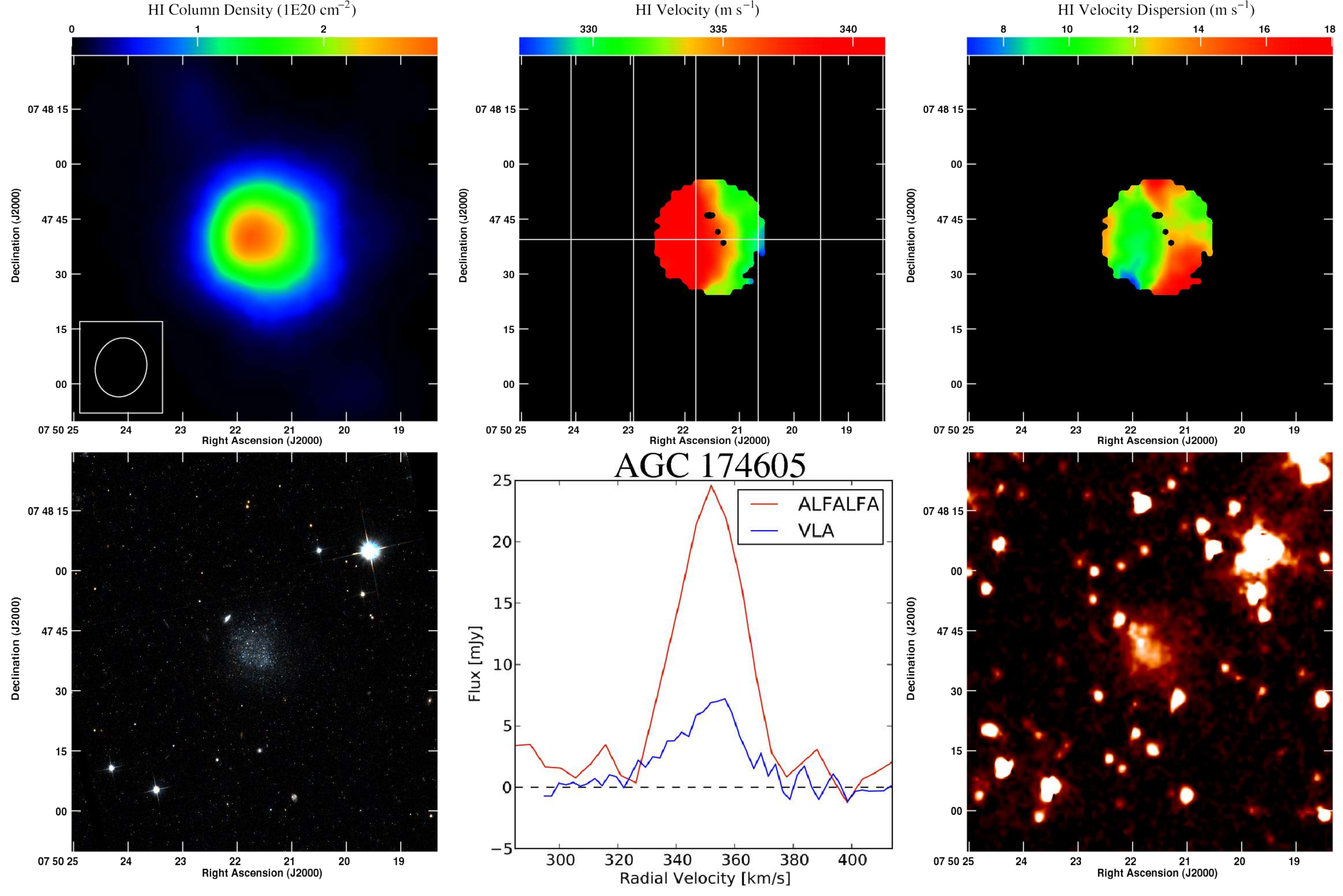}
  \caption{Same as Figure 1, for AGC\,174605.}
\label{174605.collage}
\end{figure}

\begin{figure}
\includegraphics[width=\textwidth]{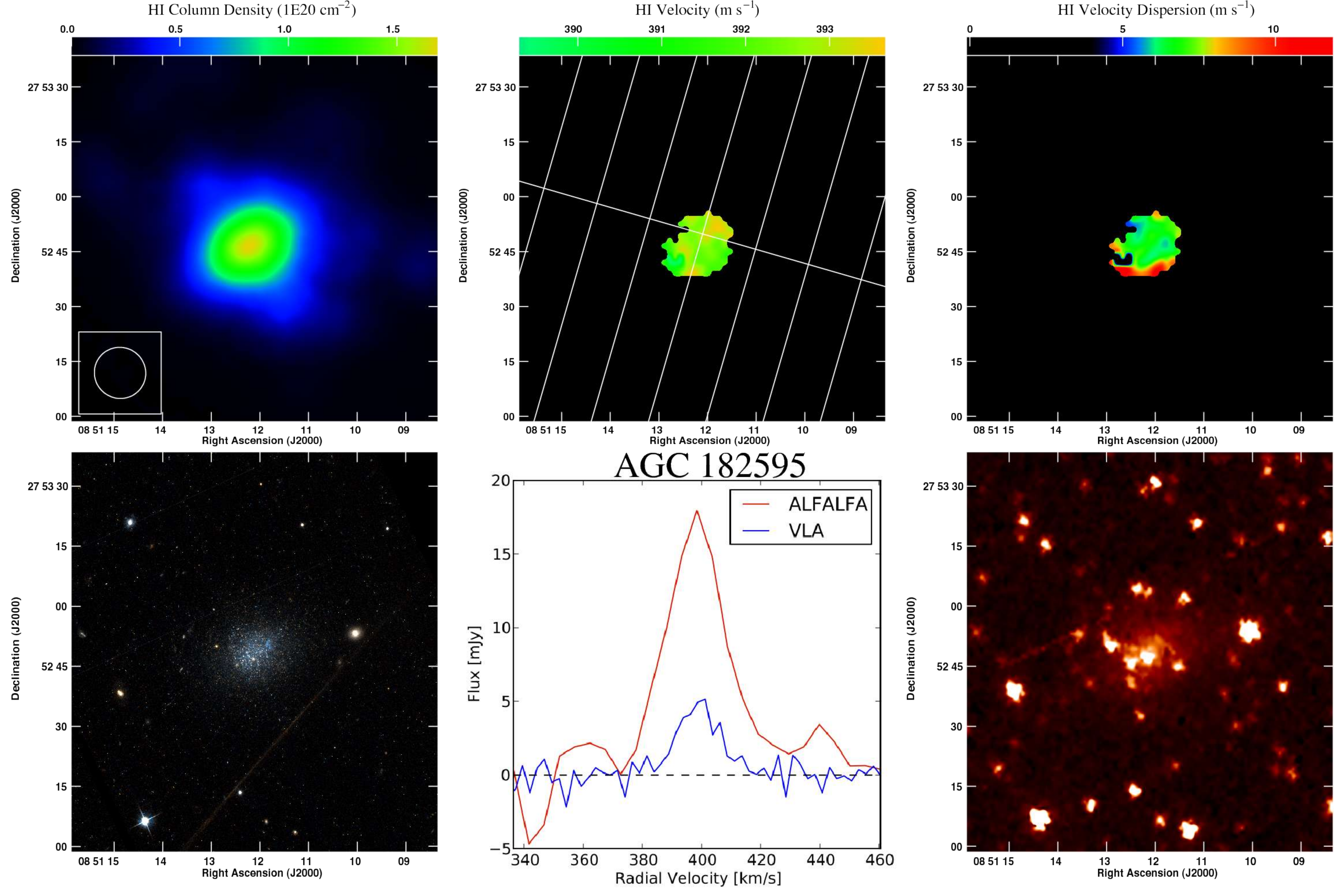}
  \caption{Same as Figure 1, for AGC\,182595.}
\label{182595.collage}
\end{figure}

\begin{figure}
\includegraphics[width=\textwidth]{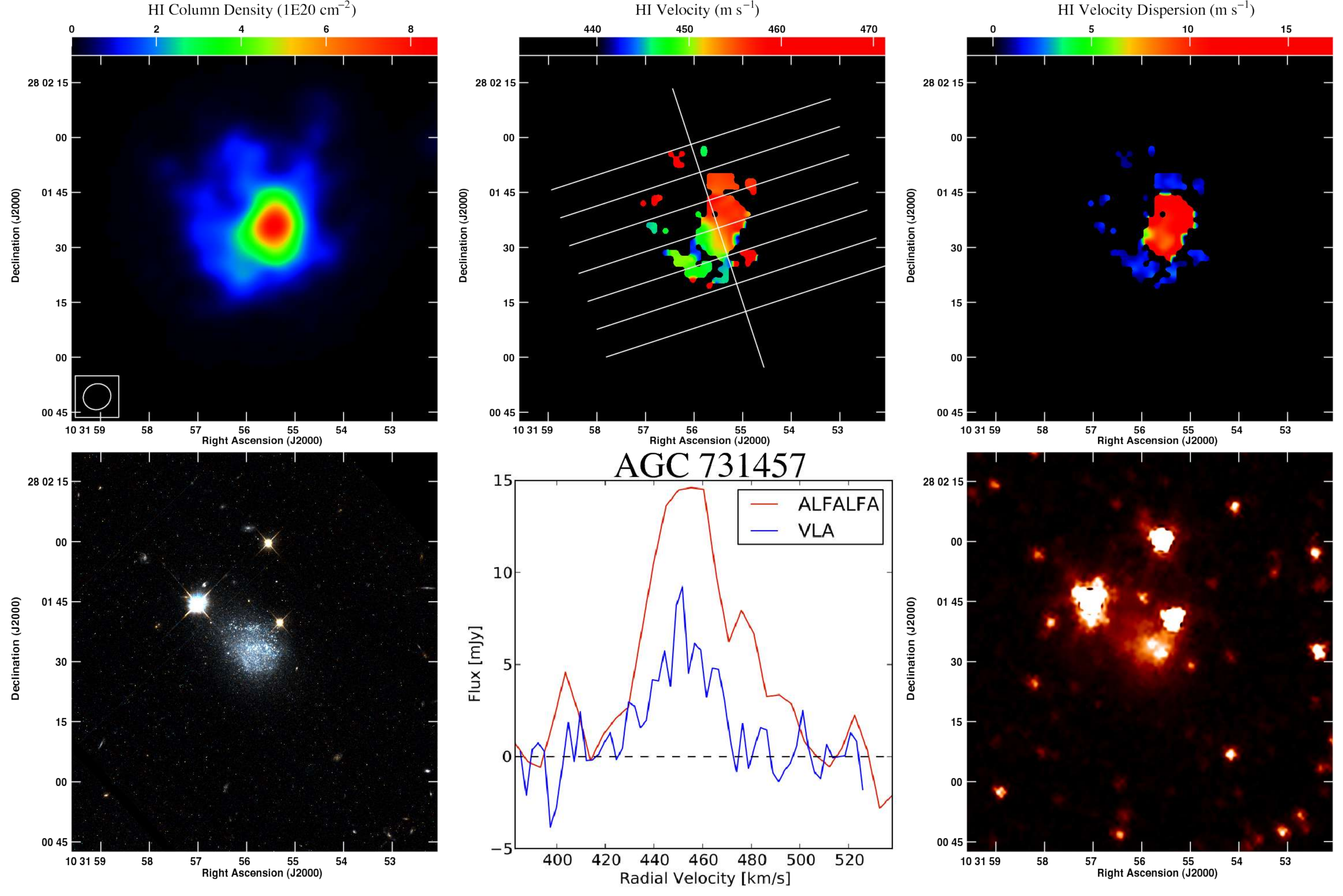}
  \caption{Same as Figure 1, for AGC\,731457.}
\label{731457.collage}
\end{figure}

\begin{figure}
\includegraphics[width=\textwidth]{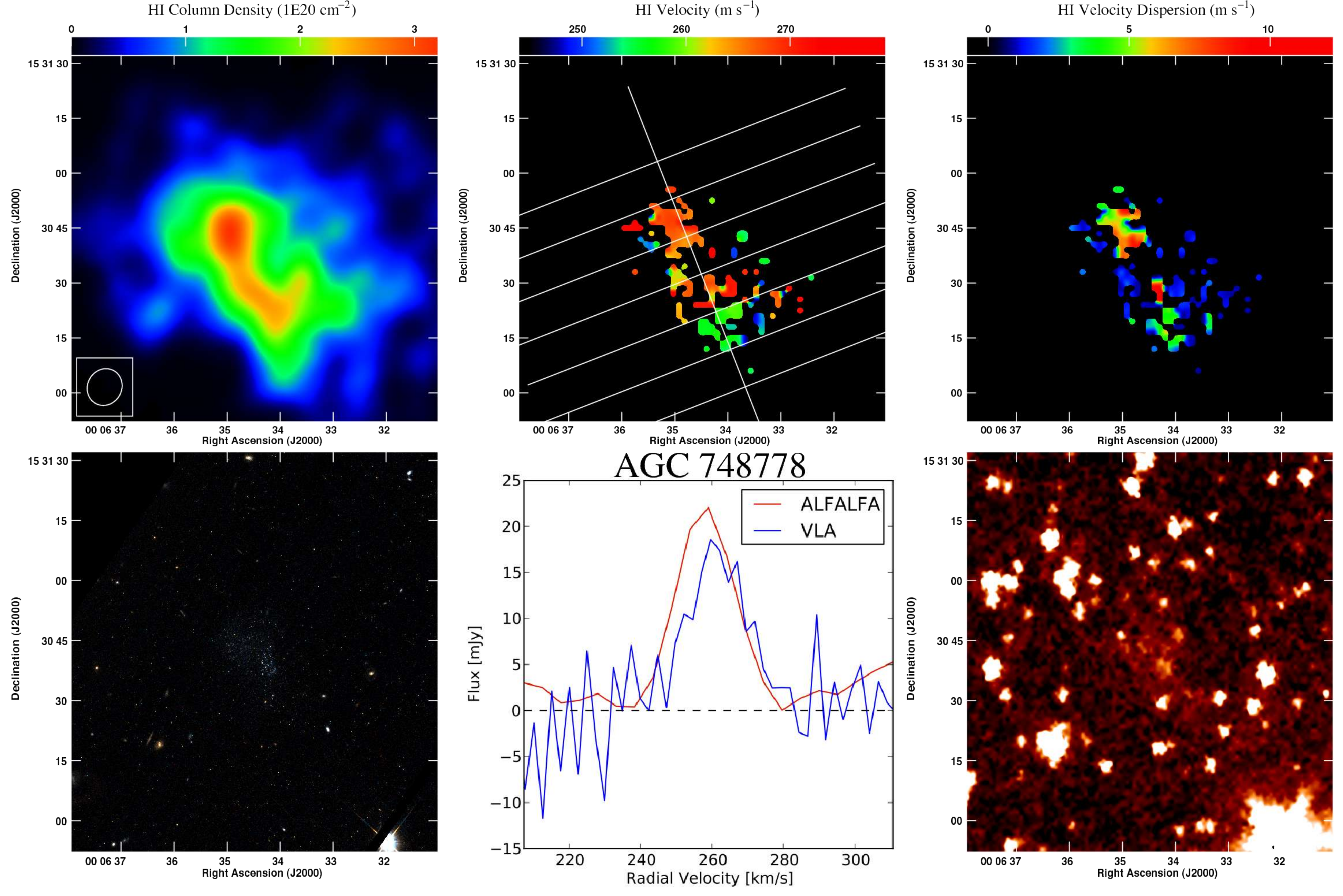}
  \caption{Same as Figure 1, for AGC\,748778.}
\label{748778.collage}
\end{figure}

\begin{figure}
\includegraphics[width=\textwidth]{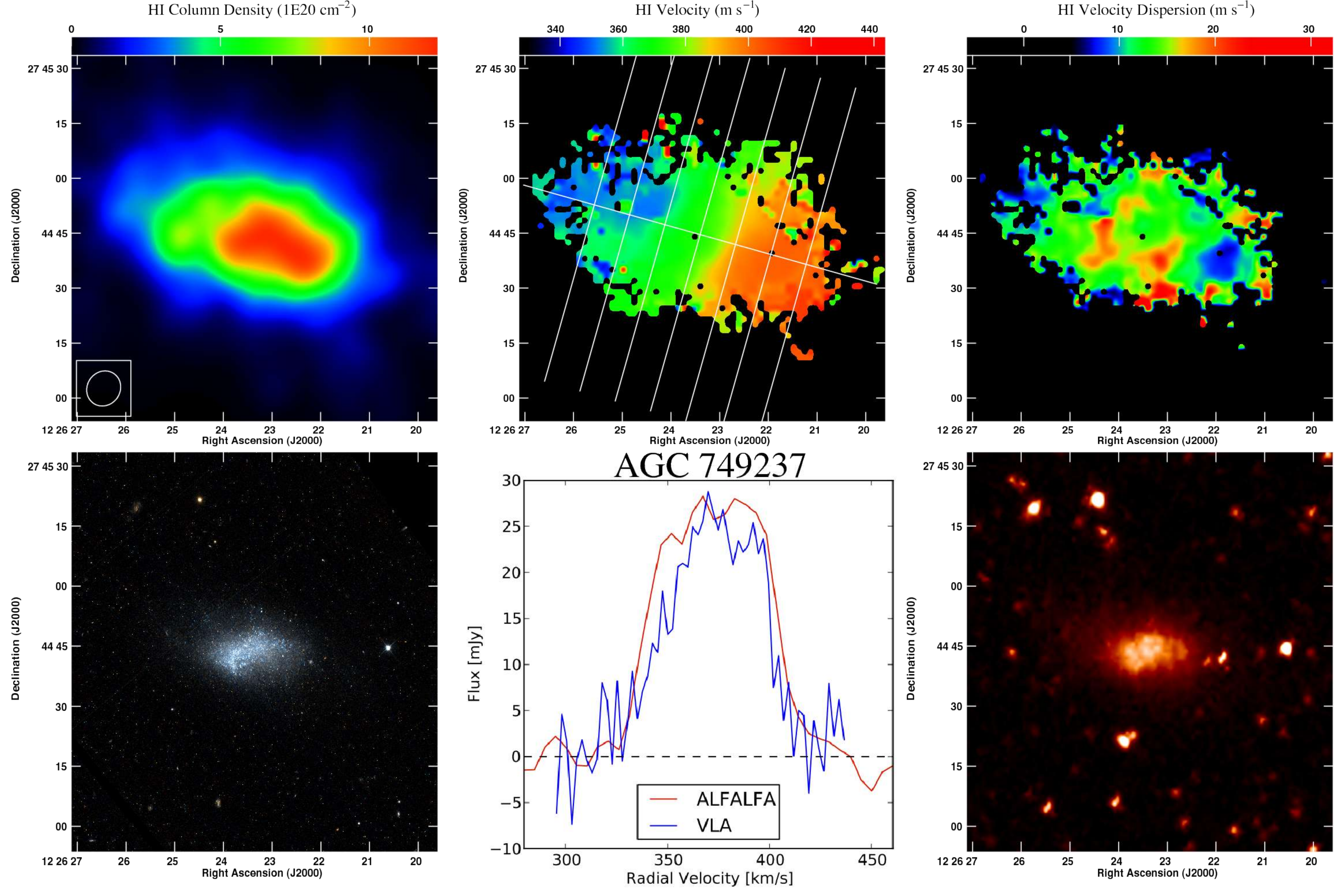}
  \caption{Same as Figure 1, for AGC\,749237.}
\label{749237.collage}
\end{figure}

\begin{figure}
\includegraphics[width=\textwidth]{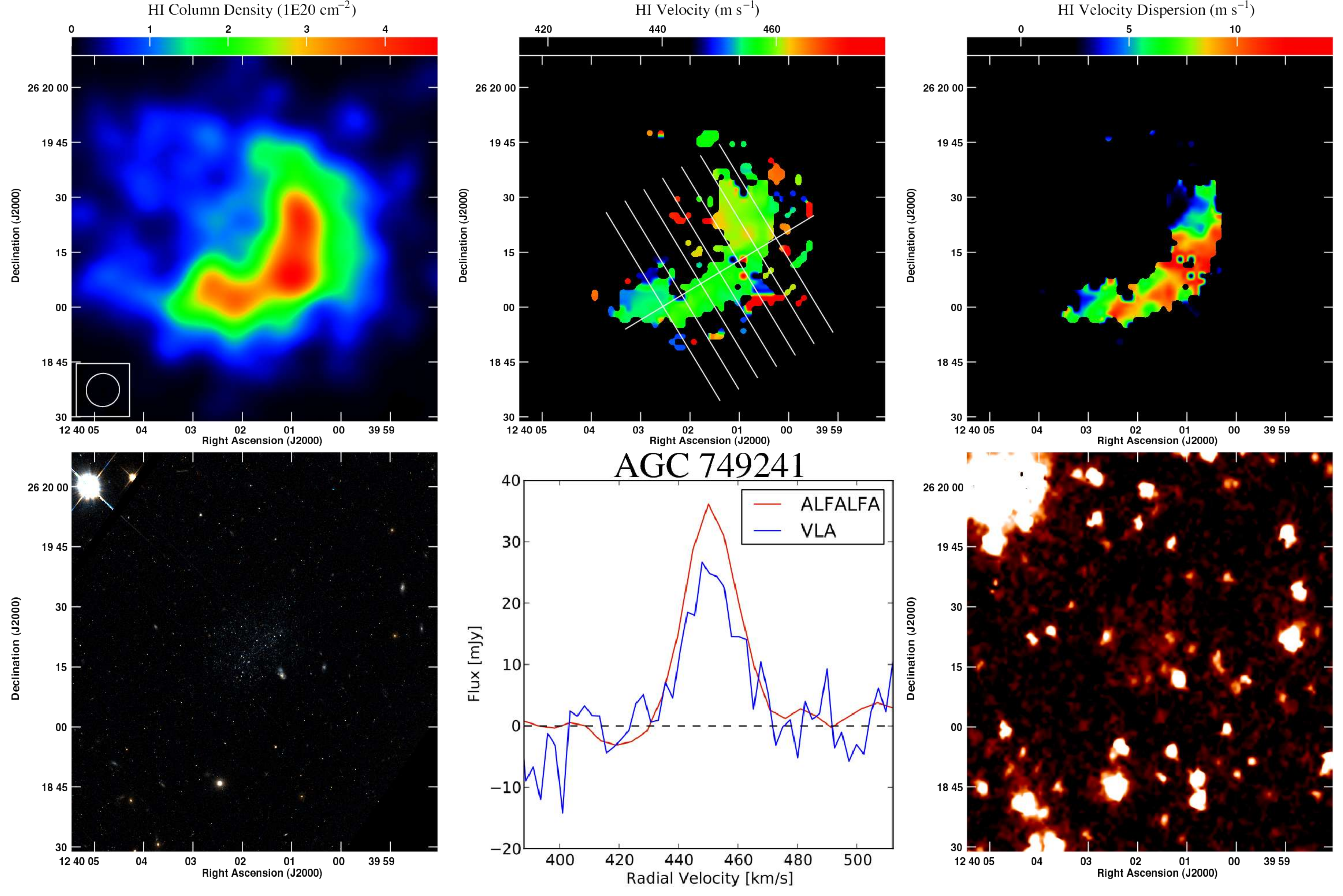}
  \caption{Same as Figure 1, for AGC\,749241.}
\label{749241.collage}
\end{figure}

\begin{figure}
\includegraphics[width=\textwidth]{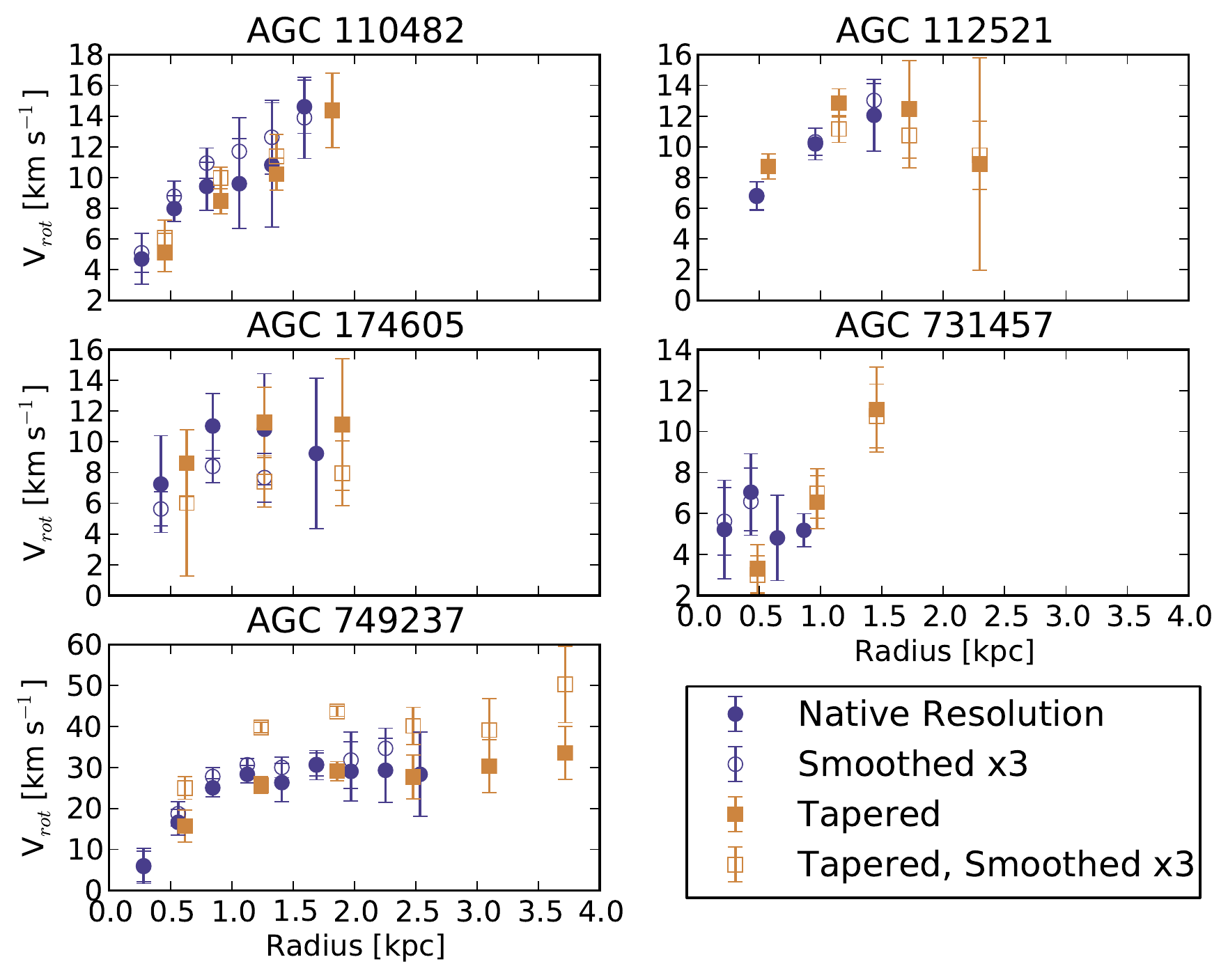}
\caption{Rotation curves of five SHIELD galaxies, as labeled; all
  results were obtained using the GIPSY task \textsc{rotcur}.  The
  filled purple circles correspond to rotation velocities derived from
  velocity fields created with the full spatial and spectral
  resolution cubes; the filled gold squares correspond to rotation
  velocities derived from tapered cubes at full spectral resolution.
  The open purple circles signify full spatial resolution with velocity
  resolution decreased by a factor of three; the open gold squares
  show tapered data that have been likewise smoothed.}
\label{rotation.curves}
\end{figure}

\begin{figure}
\includegraphics[width=\textwidth]{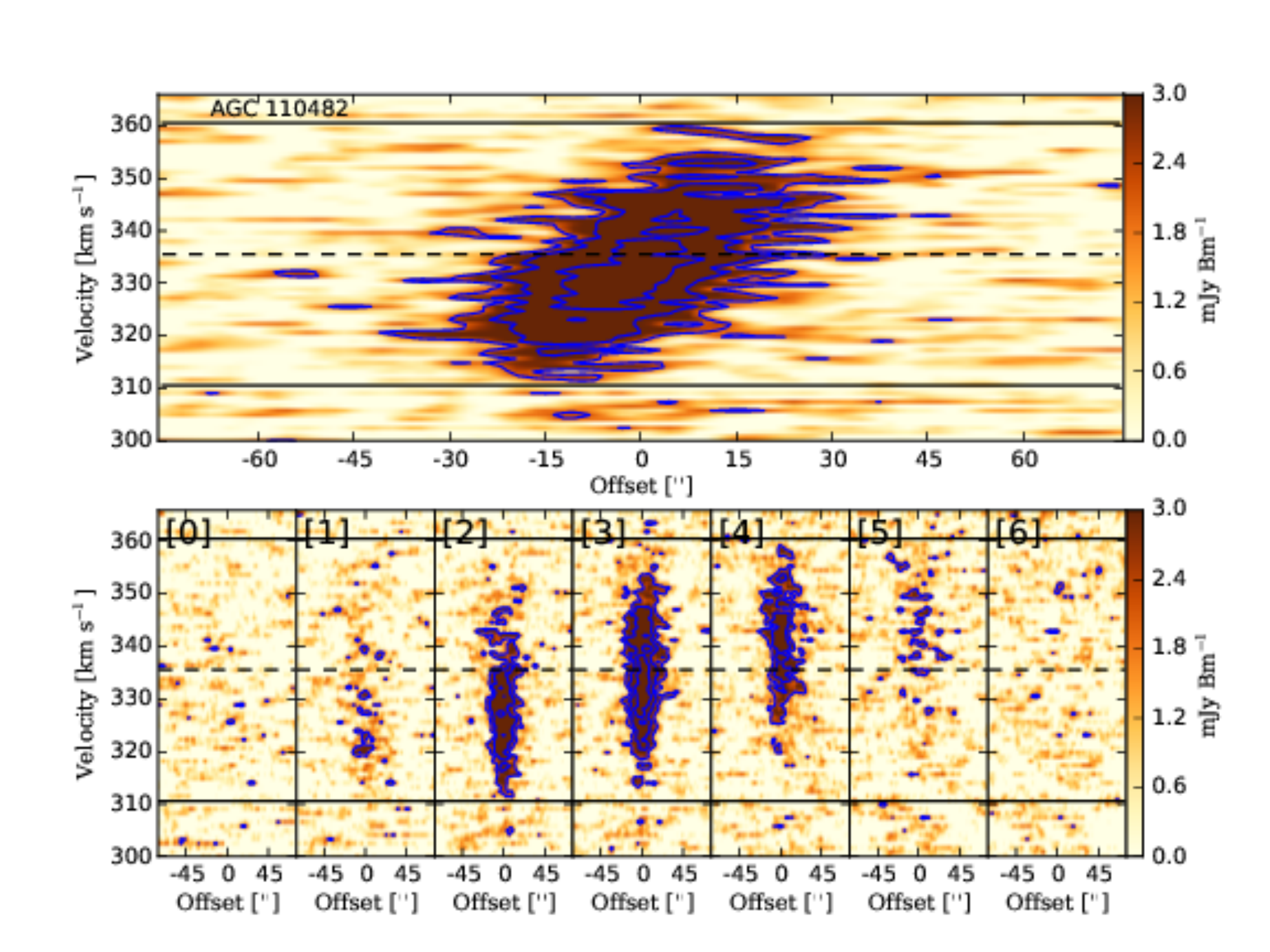}
\caption{Spatially resolved P-V diagrams across the major and minor
  axes of AGC\,110482. The upper panel shows the slice taken across
  what was identified as the ``major axis'' of rotation (see Table
  \ref{kin.props}) and that passes through the kinematic center. The
  lower panels show minor axis P-V cuts, spaced evenly by one beam
  width along the major axis; the central panel intersects the major
  axis slice at the dynamical center position.  The slices used to
  generate these P-V maps are overlaid on the upper middle panel of Figure
  \ref{110482.collage}.}
\label{110482.slices}
\end{figure}

\begin{figure}
\includegraphics[width=\textwidth]{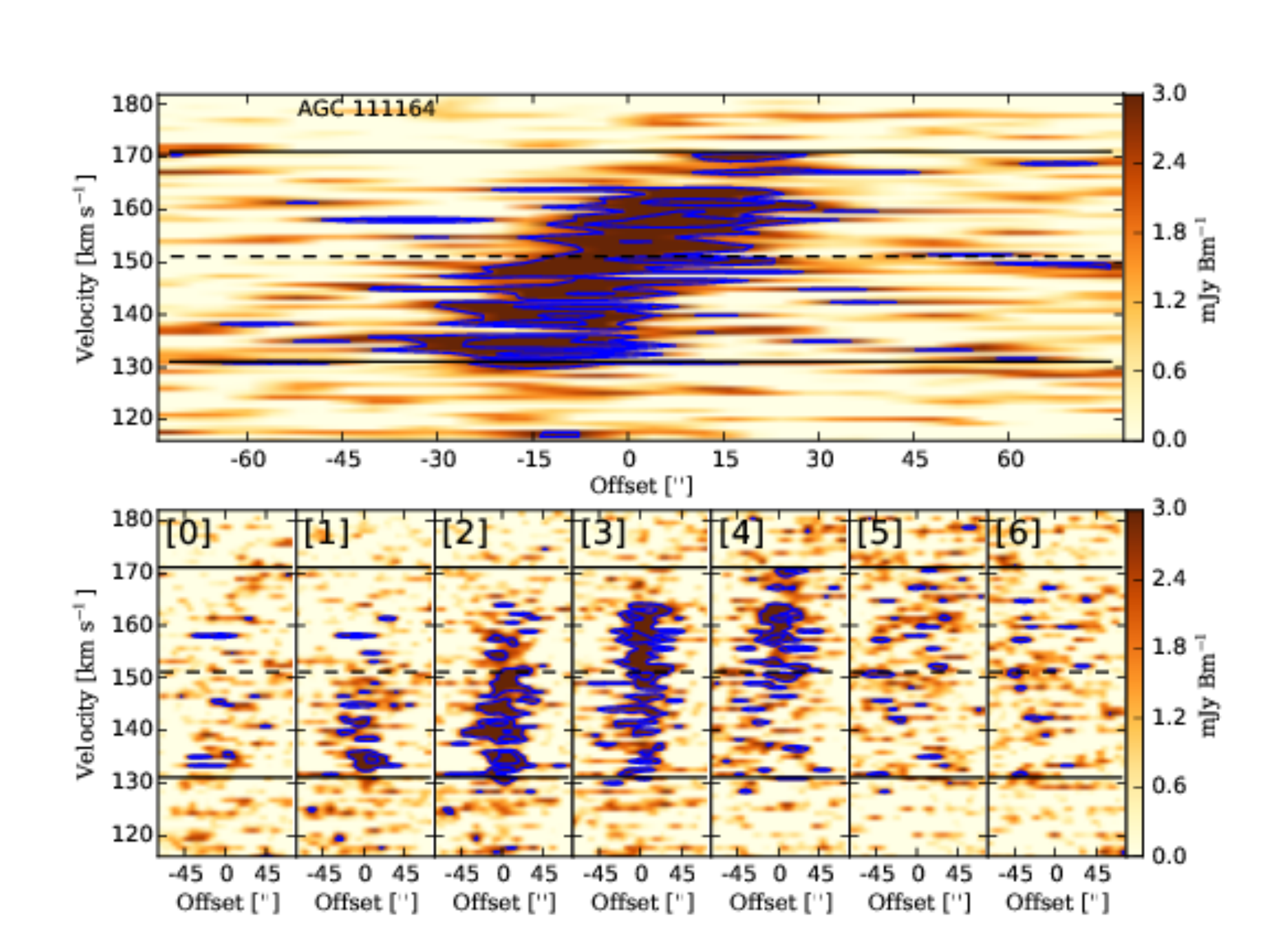}
\caption{Spatially resolved P-V diagrams across the major and minor
  axes of AGC\,111164. The upper panel shows the slice taken across
  what was identified as the ``major axis'' of rotation (see Table
  \ref{kin.props}) and that passes through the kinematic center. The
  lower panels show minor axis P-V cuts, spaced evenly by one beam
  width along the major axis; the central panel intersects the major
  axis slice at the dynamical center position.  The slices used to
  generate these P-V maps are overlaid on the upper middle panel of Figure
  \ref{111164.collage}.}
\label{111164.slices}
\end{figure}

\begin{figure}
\includegraphics[width=\textwidth]{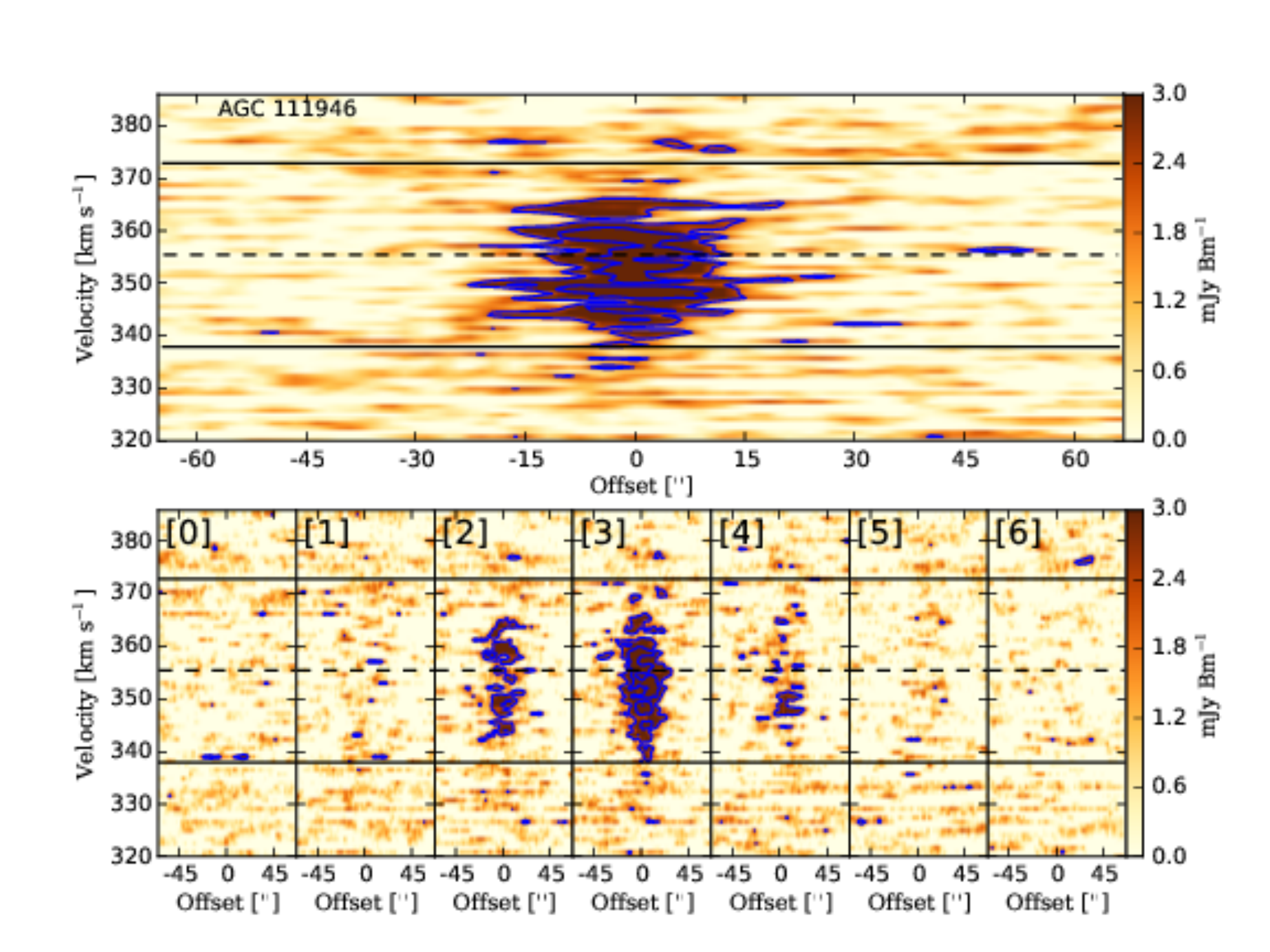}
\caption{Spatially resolved P-V diagrams across the major and minor
  axes of AGC\,111946. The upper panel shows the slice taken across
  what was identified as the ``major axis'' of rotation (see Table
  \ref{kin.props}) and that passes through the kinematic center.  The
  major axis was defined to pass through the largest rotation gradient
  of the source, in spite of the $\sim$370\kms{} outlying points to
  the northwest.  The lower panels show minor axis P-V cuts, spaced
  evenly by one beam width along the major axis; the central panel
  intersects the major axis slice at the dynamical center position.
  The slices used to generate these P-V maps are overlaid on the upper
  middle panel of  Figure \ref{111946.collage}.}
\label{111946.slices}
\end{figure}

\begin{figure}
\includegraphics[width=\textwidth]{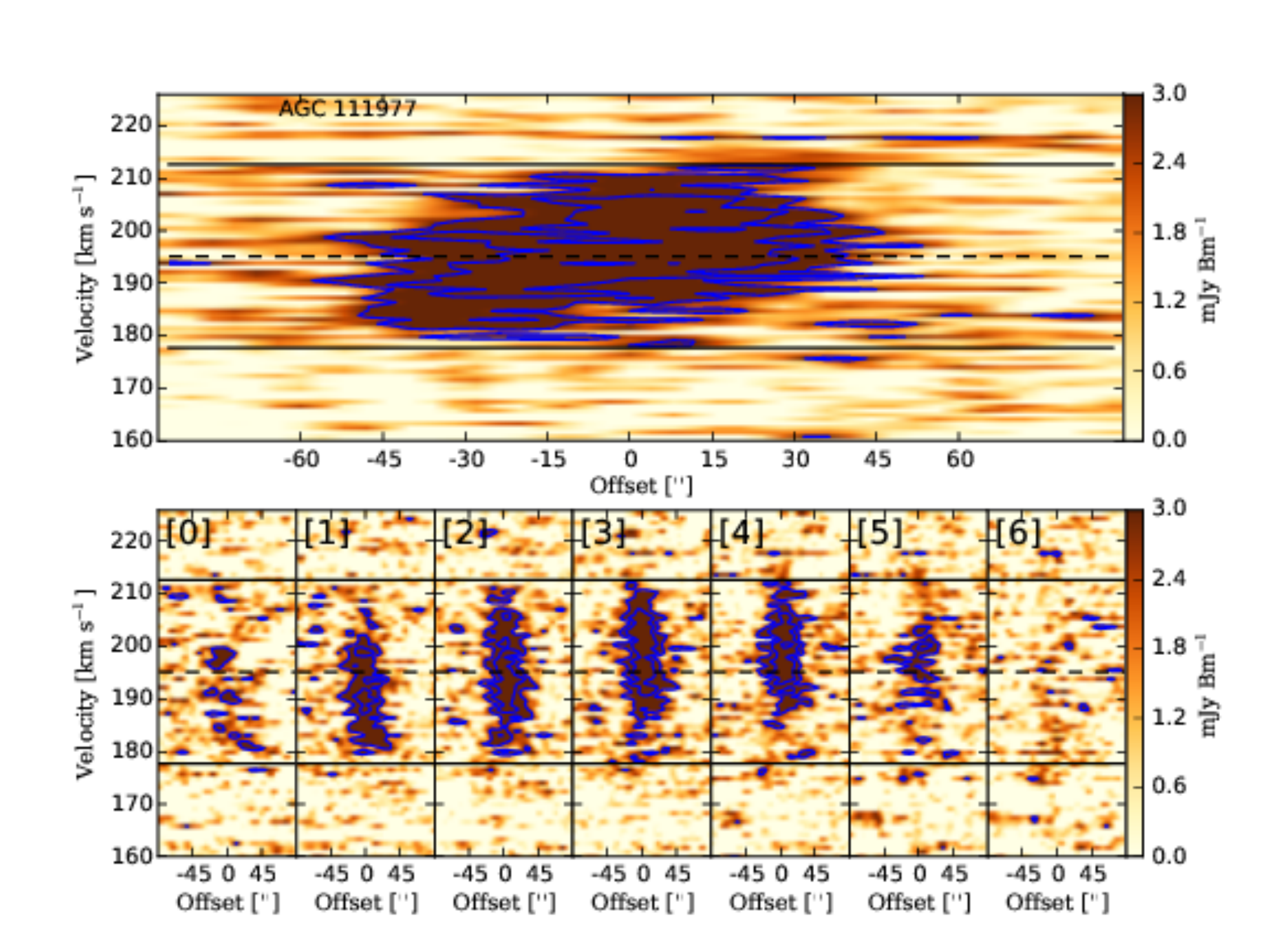}
\caption{Spatially resolved P-V diagrams across the major and minor
  axes of AGC\,111977. The upper panel shows the slice taken across
  what was identified as the ``major axis'' of rotation (see Table
  \ref{kin.props}) and that passes through the kinematic center. The
  lower panels show minor axis P-V cuts, spaced evenly by one beam
  width along the major axis; the central panel intersects the major
  axis slice at the dynamical center position.  The slices used to
  generate these P-V maps are overlaid on the upper middle panel of
  Figure \ref{111977.collage}.}
\label{111977.slices}
\end{figure}

\begin{figure}
\includegraphics[width=\textwidth]{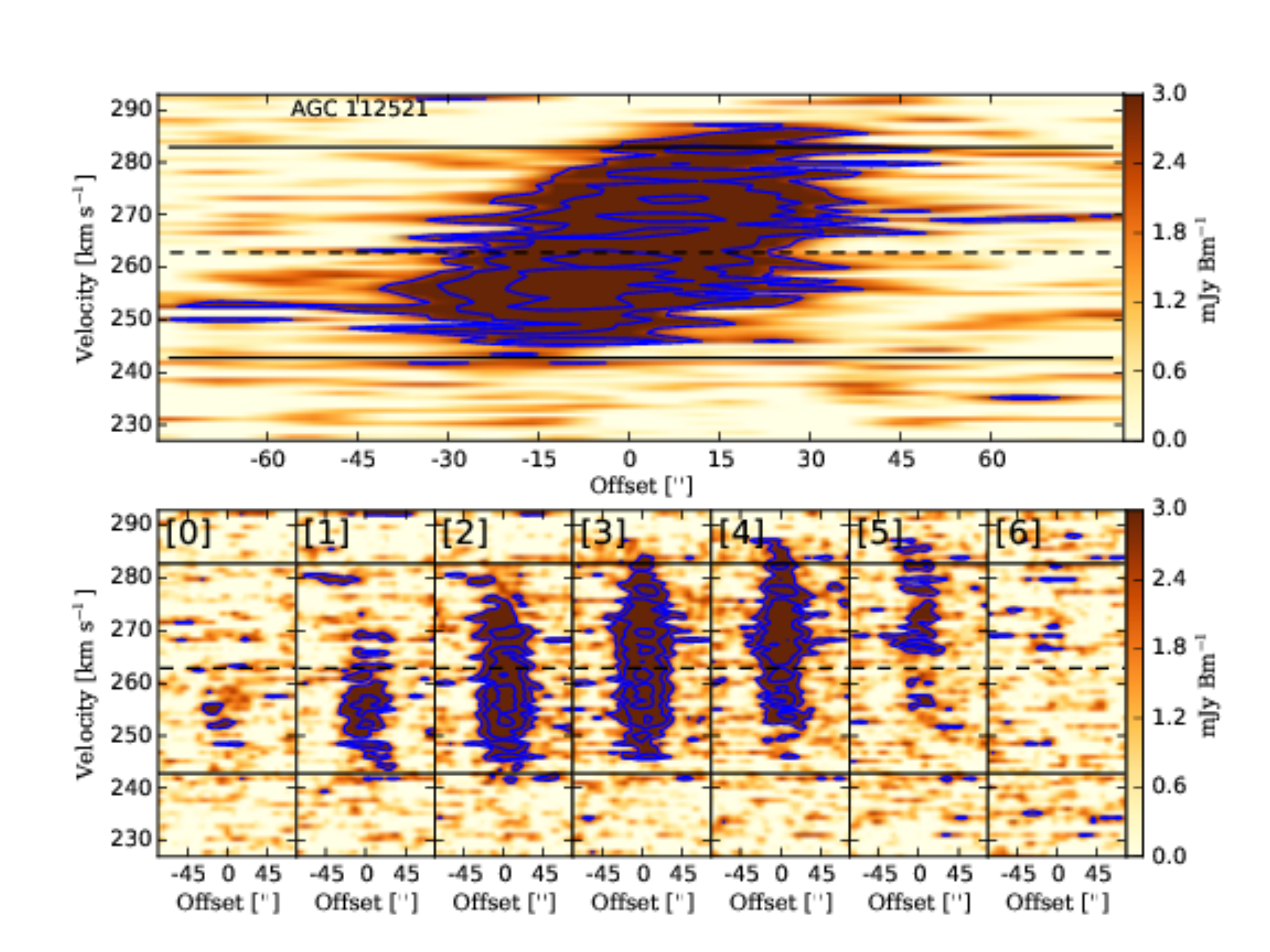}
\caption{Spatially resolved P-V diagrams across the major and minor
  axes of AGC\,112521. The upper panel shows the slice taken across
  what was identified as the ``major axis'' of rotation (see Table
  \ref{kin.props}) and that passes through the kinematic center. The
  lower panels show minor axis P-V cuts, spaced evenly by one beam
  width along the major axis; the central panel intersects the major
  axis slice at the dynamical center position.  The slices used to
  generate these P-V maps are overlaid on the upper middle panel of  Figure
  \ref{112521.collage}.}
\label{112521.slices}
\end{figure}

\begin{figure}
\includegraphics[width=\textwidth]{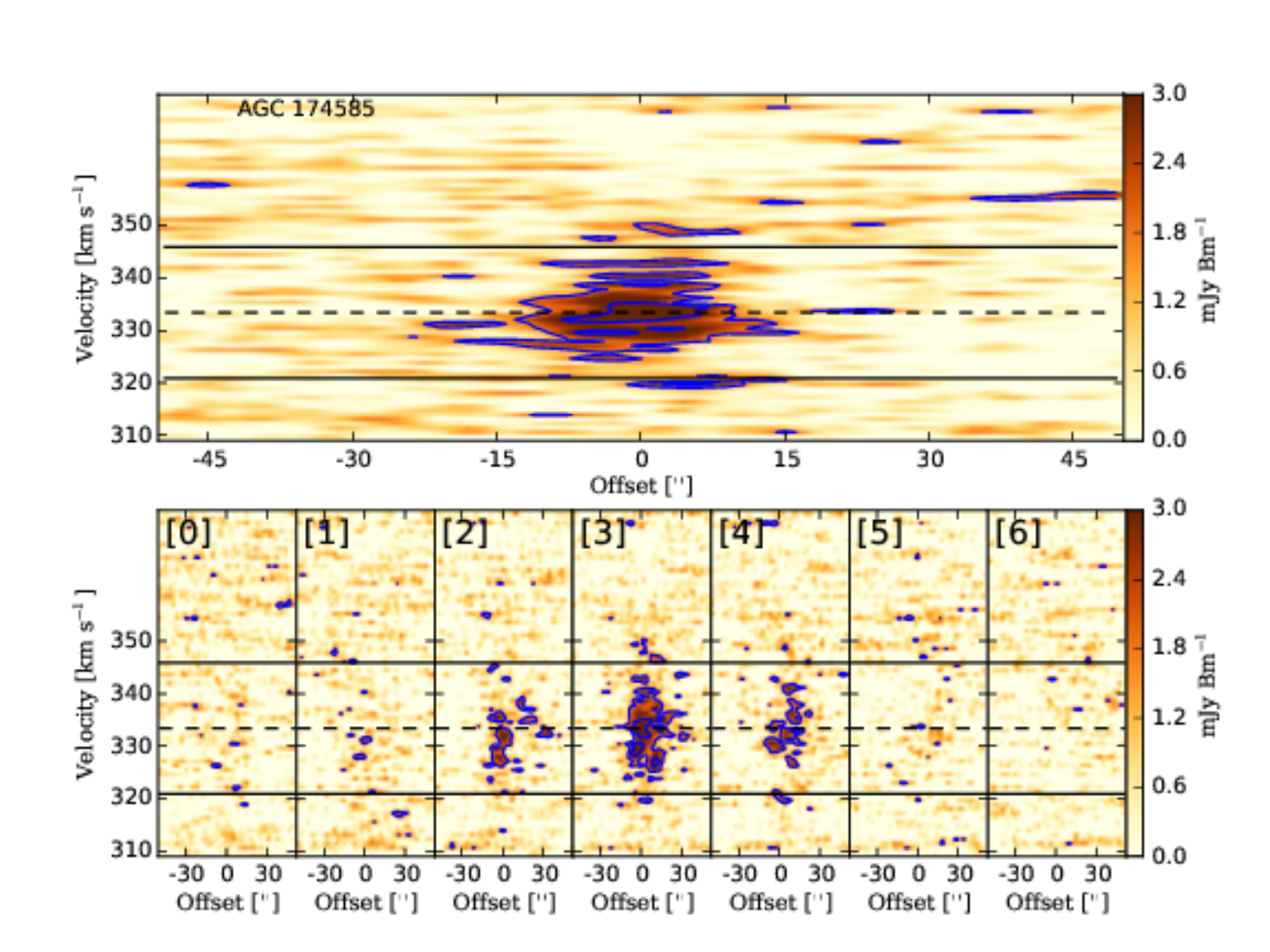}
\caption{Spatially resolved P-V diagrams across the major and minor
  axes of AGC\,174585. The upper panel shows the slice taken across
  what was identified as the ``major axis'' of rotation (see Table
  \ref{kin.props}) and that passes through the kinematic center. The
  lower panels show minor axis P-V cuts, spaced evenly by one beam
  width along the major axis; the central panel intersects the major
  axis slice at the dynamical center position.  The slices used to
  generate these P-V maps are overlaid on the upper middle panel of  Figure 
  \ref{174585.collage}.  Note that the major axis of AGC\,174585 could
  be defined in two directions. There is an apparent rotation gradient
  across the southern lobe of the galaxy, but higher sensitivity
  images (the naturally-weighted data cube) appear to reveal stronger
  emission with higher velocity to the northwest, so the PV slice
  major axis was defined to trace through both lobes instead of across
  the bottom one for a few pixels of faint slow gas. Either low
  surface brightness gas- unresolved at high angular resolution- and
  the high surface brightness gas detected in our highest resolution
  maps have different rotation axes, or there is no preferential axis
  of rotation in this source.}
\label{174585.slices}
\end{figure}

\begin{figure}
 \includegraphics[width=\textwidth]{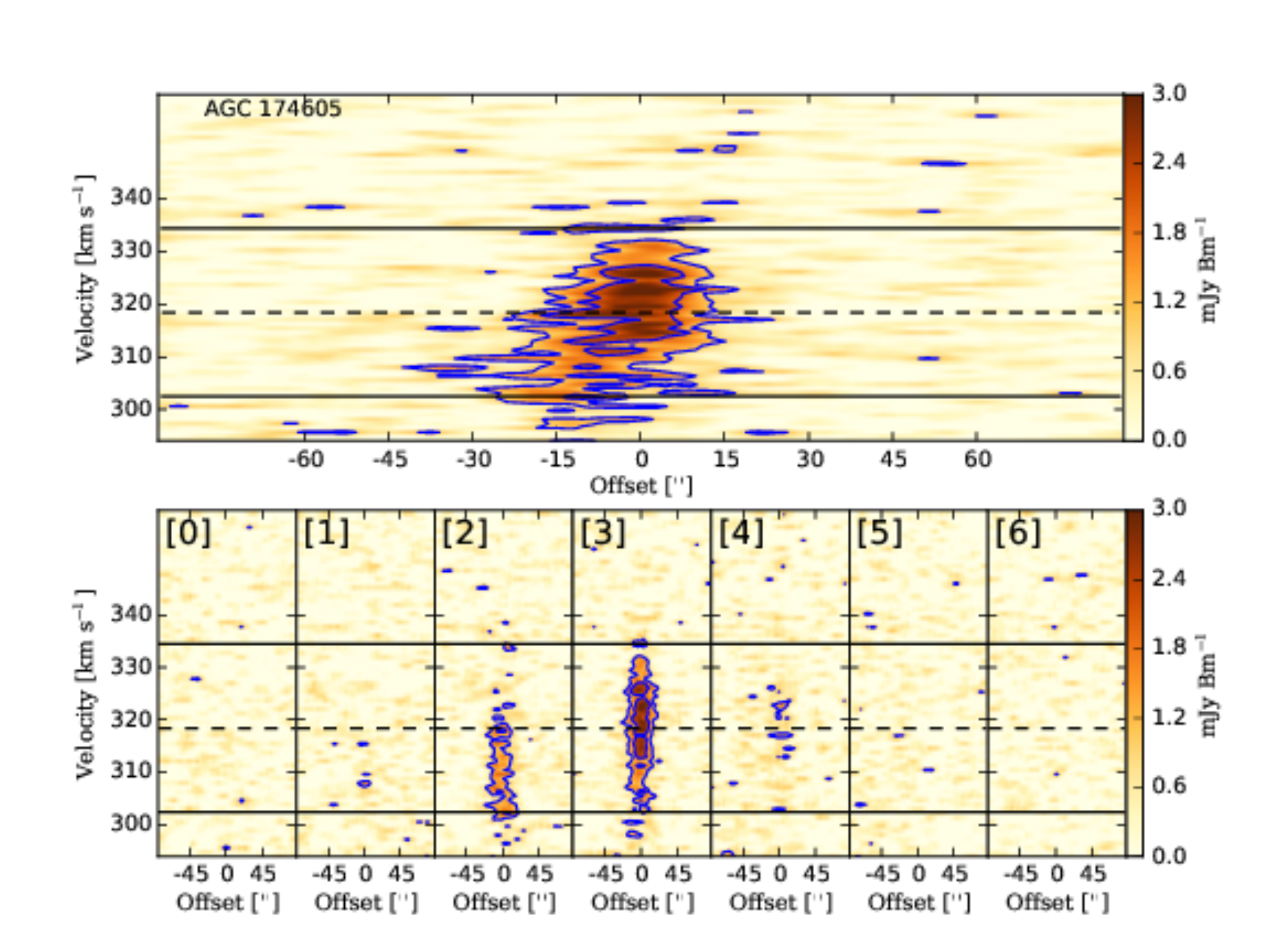}
\caption{Spatially resolved P-V diagrams across the major and minor
  axes of AGC\,174605. The upper panel shows the slice taken across
  what was identified as the ``major axis'' of rotation (see Table
  \ref{kin.props}) and that passes through the kinematic center. The
  lower panels show minor axis P-V cuts, spaced evenly by one beam
  width along the major axis; the central panel intersects the major
  axis slice at the dynamical center position.  The slices used to
  generate these P-V maps are overlaid on the upper middle panel of  Figure
  \ref{174605.collage}.}
\label{174605.slices}
\end{figure}

\begin{figure}
\includegraphics[width=\textwidth]{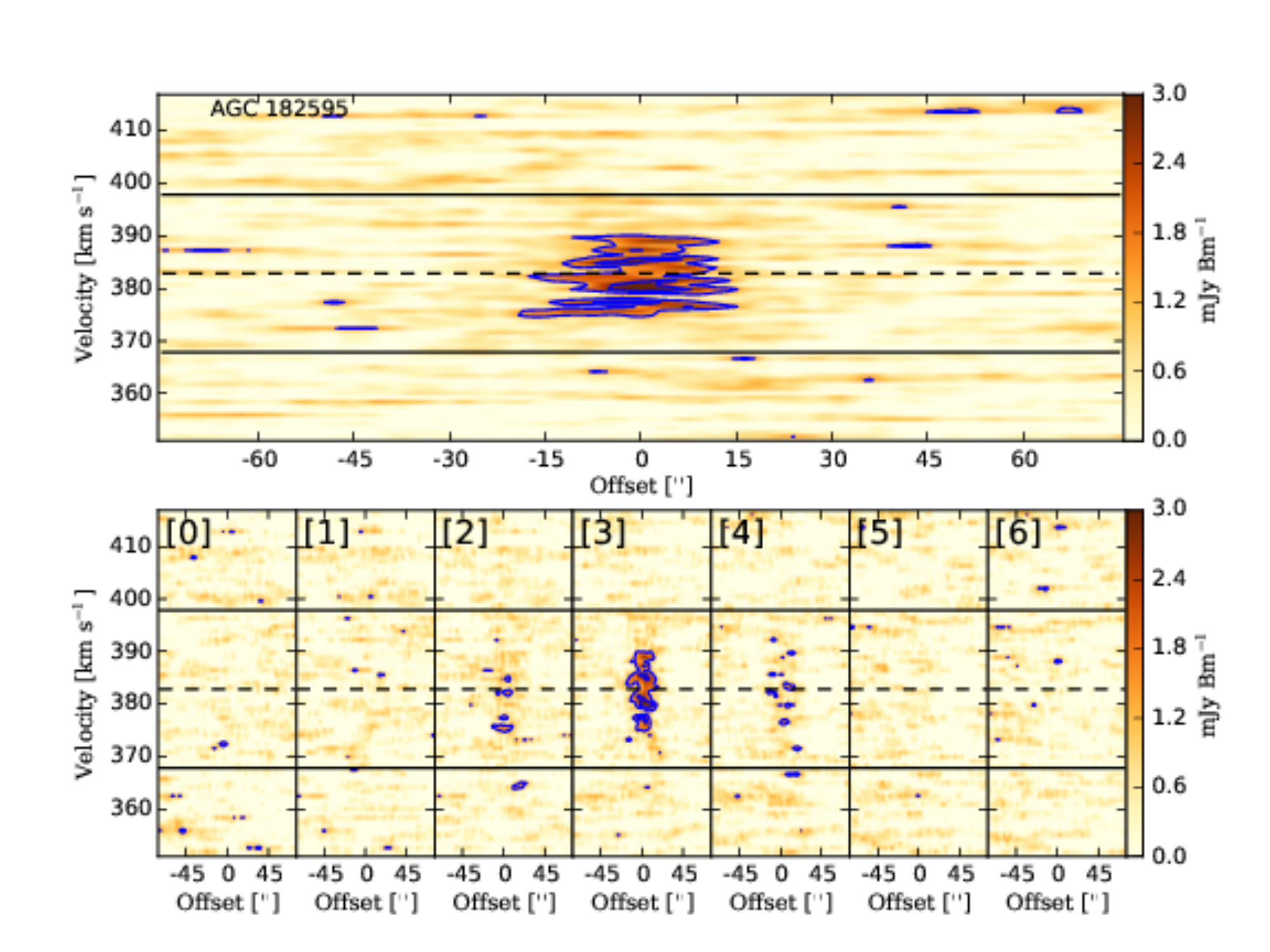}
\caption{Spatially resolved P-V diagrams across the major and minor
  axes of AGC\,182595. The upper panel shows the slice taken across
  what was identified as the ``major axis'' of rotation (see Table
  \ref{kin.props}) and that passes through the kinematic center. The
  lower panels show minor axis P-V cuts, spaced evenly by one beam
  width along the major axis; the central panel intersects the major
  axis slice at the dynamical center position.  The slices used to
  generate these P-V maps are overlaid on the upper middle panel of Figure
  \ref{182595.collage}.  Note that AGC\,182595 is a source whose P-V
  diagrams appear to indicate no ordered rotation: the position angle
  of the P-V slice in this case has practically no effect on the
  resulting maps, and it was chosen such that by eye the {\em radial}
  offset (in position space) was maximized.  Since the velocity ranges
  of the P-V maps were practically identical for all values of
  position angle, a few pixels' change in radial offset was taken as a
  proxy for highest dispersive motion, the closest we could come to
  defining a ``major'' axis of rotation for this galaxy.}
\label{182595.slices}
\end{figure}

\begin{figure}
\includegraphics[width=\textwidth]{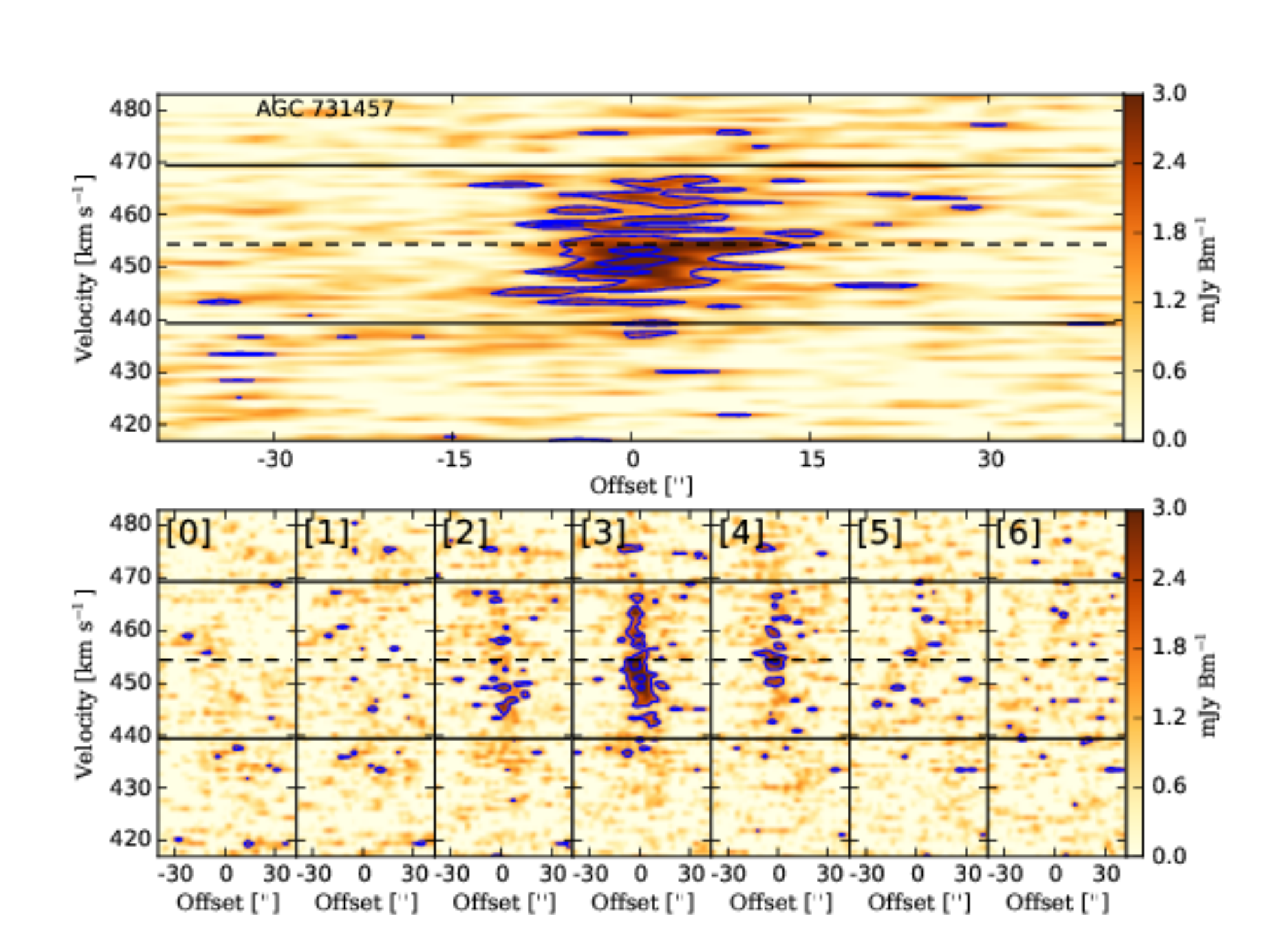}
\caption{Spatially resolved P-V diagrams across the major and minor
  axes of AGC\,731457. The upper panel shows the slice taken across
  what was identified as the ``major axis'' of rotation (see Table
  \ref{kin.props}) and that passes through the kinematic center. The
  lower panels show minor axis P-V cuts, spaced evenly by one beam
  width along the major axis; the central panel intersects the major
  axis slice at the dynamical center position.  The slices used to
  generate these P-V maps are overlaid on the upper middle panel of Figure
  \ref{731457.collage}.}
\label{731457.slices}
\end{figure}

\begin{figure}
\includegraphics[width=\textwidth]{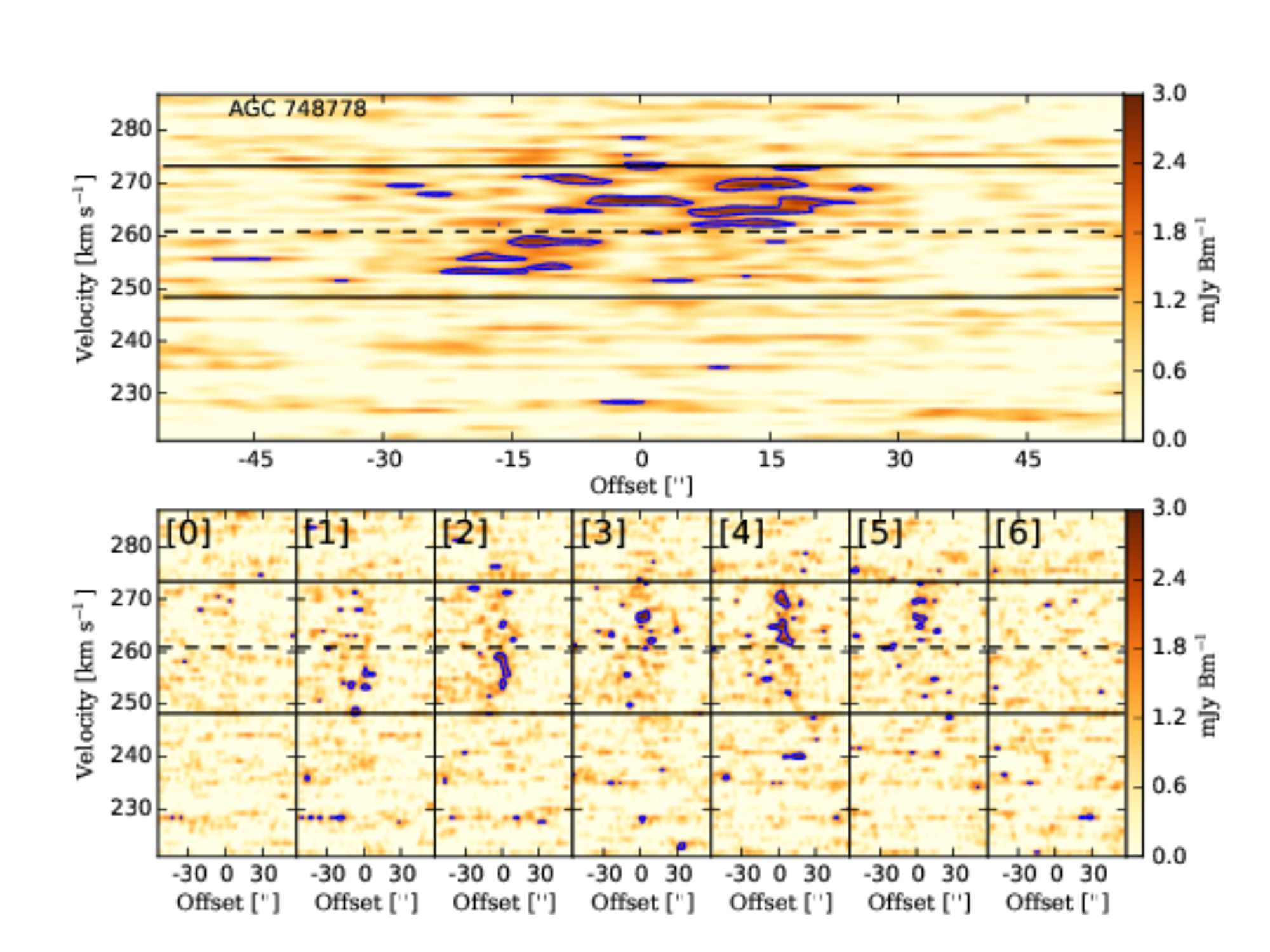}
\caption{Spatially resolved P-V diagrams across the major and minor
  axes of AGC\,748778. The upper panel shows the slice taken across
  what was identified as the ``major axis'' of rotation (see Table
  \ref{kin.props}) and that passes through the kinematic center. The
  lower panels show minor axis P-V cuts, spaced evenly by one beam
  width along the major axis; the central panel intersects the major
  axis slice at the dynamical center position.  The slices used to
  generate these P-V maps are overlaid on the upper middle panel of  Figure
  \ref{748778.collage}.  Note that at the angular resolution of these
  data, the velocity structure that we detect in AGC\,748778 is
  attributable to only the parcels of neutral gas with highest surface
  brightness. Therefore, the position angle used to define a ``major''
  axis was determined to be across the region of highest significant
  emission in the moment 0 map, which also happens to correspond to
  what looks like a weak velocity gradient from the southeast to the
  northwest.}
\label{748778.slices}
\end{figure}

\begin{figure}
\includegraphics[width=\textwidth]{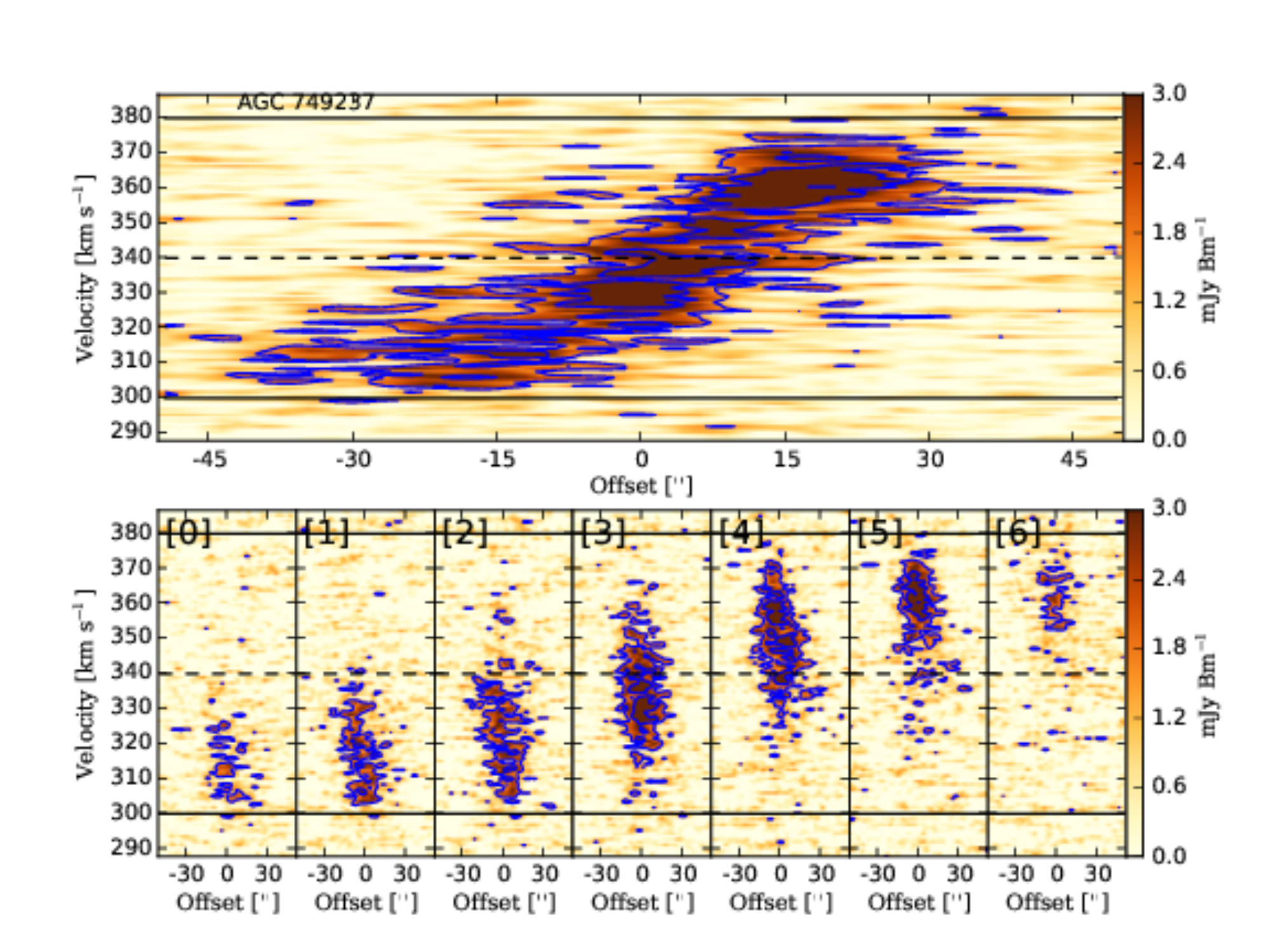}
\caption{Spatially resolved P-V diagrams across the major and minor
  axes of AGC\,749237. The upper panel shows the slice taken across
  what was identified as the ``major axis'' of rotation (see Table
  \ref{kin.props}) and that passes through the kinematic center. The
  lower panels show minor axis P-V cuts, spaced evenly by one beam
  width along the major axis; the central panel intersects the major
  axis slice at the dynamical center position.  The slices used to
  generate these P-V maps are overlaid on the upper middle panel of
  Figure \ref{749237.collage}.}
\label{749237.slices}
\end{figure}

\begin{figure}
\includegraphics[width=\textwidth]{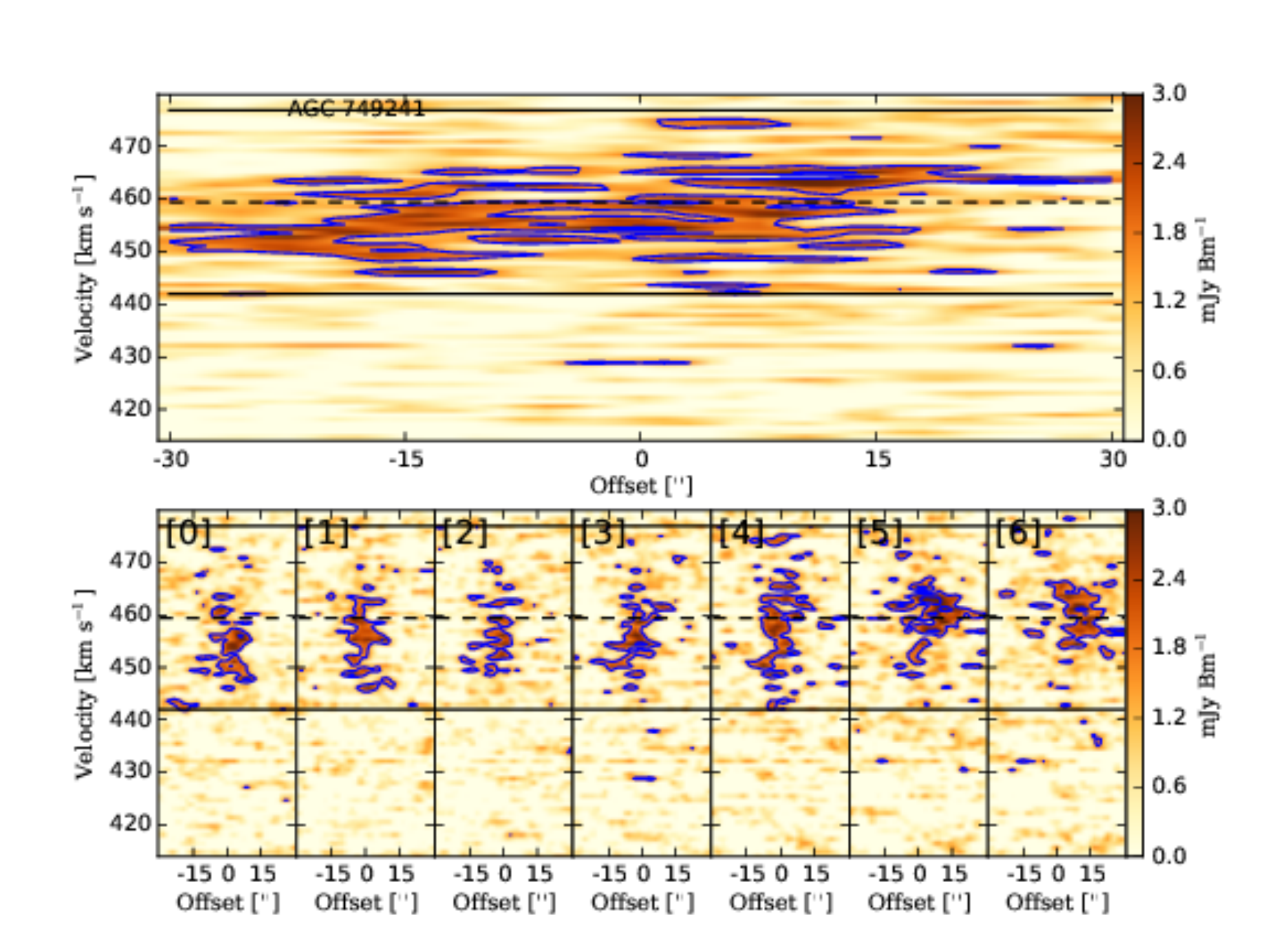}
\caption{Spatially resolved P-V diagrams across the major and minor
  axes of AGC\,749241. The upper panel shows the slice taken across
  what was identified as the ``major axis'' of rotation (see Table
  \ref{kin.props}) and that passes through the kinematic center. The
  lower panels show minor axis P-V cuts, spaced evenly by one beam
  width along the major axis; the central panel intersects the major
  axis slice at the dynamical center position.  The slices used to
  generate these P-V maps are overlaid on the upper middle panel of
  \ref{749241.collage}.  Note that the highly irregular
  ``crescent-shaped'' distribution of neutral gas in this source makes
  the selection of an unambiguous position angle extremely subjective.
  It was chosen to match the arcing structure of the highest column
  density gas, which also shows a very weak gradient from outer
  southwest to the inner northeast edges.}
\label{749241.slices}
\end{figure}

For the stellar mass of the each SHIELD galaxy, we follow
{Paper~I}\nocite{teich} in using the stellar masses derived from
Hubble Space Telescope images \citep{mcquinnsfh}.  Note that we have
dedicated Spitzer imaging of the SHIELD galaxies, and that these
images are shown in Figures~\ref{110482.collage} -
\ref{749241.collage}.  Ideally a radial luminosity profile derived
from these images can be converted to a mass profile via adoption of a
(usually constant) infrared mass-to-light ratio.  Regrettably, the
small physical sizes, faintness, and significant distances of the
SHIELD galaxies result in some systems being significantly
contaminated by foreground and background sources that preclude clean
surface brightness profiles.  The resulting stellar masses and stellar
mass profiles are presented in \citet{cannon13}, to which we refer the
interested reader for details.  The total baryonic masses of the
SHIELD galaxies are tabulated by summing the \HI{} gas mass and the
stellar mass as presented in column 8 of Table~1 of
{Paper~I}\nocite{teich}.

As is evident from the velocity fields shown in
Figures~\ref{110482.collage} through \ref{749241.collage}, the
amplitude of projected rotation is comparable to the average velocity
dispersion of the \HI{} gas in many of the SHIELD galaxies.  This
strongly suggests that the SHIELD galaxies populate the mass regime
where galaxies transition from rotationally supported to pressure
supported systems.  Disentangling rotation from velocity dispersion
may represent a fundamental and limiting challenge for the least
massive, gas-rich galaxies in the local volume.

We seek to quantify the magnitude of pressure support that the \HI{}
velocity dispersion provides in the SHIELD galaxies.  As discussed in
\citet{staveleysmith92}, this component can be significant for
low-mass systems.  Following the formalism presented in
\citet{hoffman96}, we correct the enclosed dynamical mass for the
contribution from the \HI{} velocity dispersion via the relation
\begin{equation}
\begin{aligned}
  M_{dyn}(r) = \frac{(V_{rot}(r)^{2} + 3\sigma_{z}(r)^{2}) \cdot r}{G} \\
  = 2.325 \times 10^{5}M_{\odot}\left(\frac{V_{rot}^{2} + 3\sigma_{z}^{2}}{km^{2} s^{-2}}\right)\left(\frac{r}{kpc}\right)
\end{aligned}
\label{hoff}
\end{equation}

\noindent where M$_{dyn}(r)$ represents the radially-dependent
enclosed dynamical mass in solar mass units, V$_{rot}(r)$ is the
projected rotation velocity corrected for disk inclination,
$\sigma_{z}(r)$ is the gas dispersion along the line of sight, and r
is the distance from the dynamical center, and $G$ is the Universal
constant of gravitation.  The dynamical masses of the SHIELD galaxies,
corrected for pressure support within the disk, are tabulated in
column (9) of Table \ref{dyn.props}.  Comparing these values to the
baryonic masses, we arrive at the global ratio of total mass to
luminous mass (M$_{\rm dyn}$/M$_{\rm bary}$) as tabulated in columns
10 and 11 of Table \ref{dyn.props}.

Using these data, we can now contextualize the SHIELD galaxies by
placing them on the BTFR.  In Figure \ref{btfr}, the SHIELD galaxies
are each plotted on the BTFR alongside the galaxy populations from the
comprehensive review of \citet{mcgaugh12}: gas-dominated spirals,
gas-rich dwarfs, and gas-poor dwarf spheroidal galaxies, along with
the least massive known \HI{}-bearing galaxy in the local univsere,
Leo P \citep{ezbc14}.  In agreement with recent results (e.g., {Lelli
  \etal\ 2016}\nocite{lelli16}; {Papastergis
  \etal\ 2016}\nocite{papastergis16}, although note that the method
used to determine rotational velocity there uses W$_{\rm 50}$), the
SHIELD galaxies fall on the BTFR within measurement error.  Note that
in the sparsely-sampled portion of parameter space at low rotational
velocities (v$_{\rm circ}$ $<$ 30 \kms), the SHIELD galaxies make an
important contribution toward improving the statistics (more than
doubling the number of systems plotted in Figure~\ref{btfr}.  While
the dispersion appears to increase at these low velocities, we suspect
that observational uncertainty and model degeneracies play important
roles.

\section{Discussion and Conclusions}
\label{S4}

The study of the SHIELD galaxies represents a significant legacy of
the ALFALFA survey: those sources that populate the faint end of the
\HI{} mass function and which also harbor an easily-detectable stellar
component.  In this work, we have presented a detailed examination of
the neutral gas dynamics of 12 systems.  The discussion in previous
sections tells a clear story: the contributions from rotational and
pressure support are effectively equal in the SHIELD galaxies.

\begin{figure}
\includegraphics[width=\textwidth]{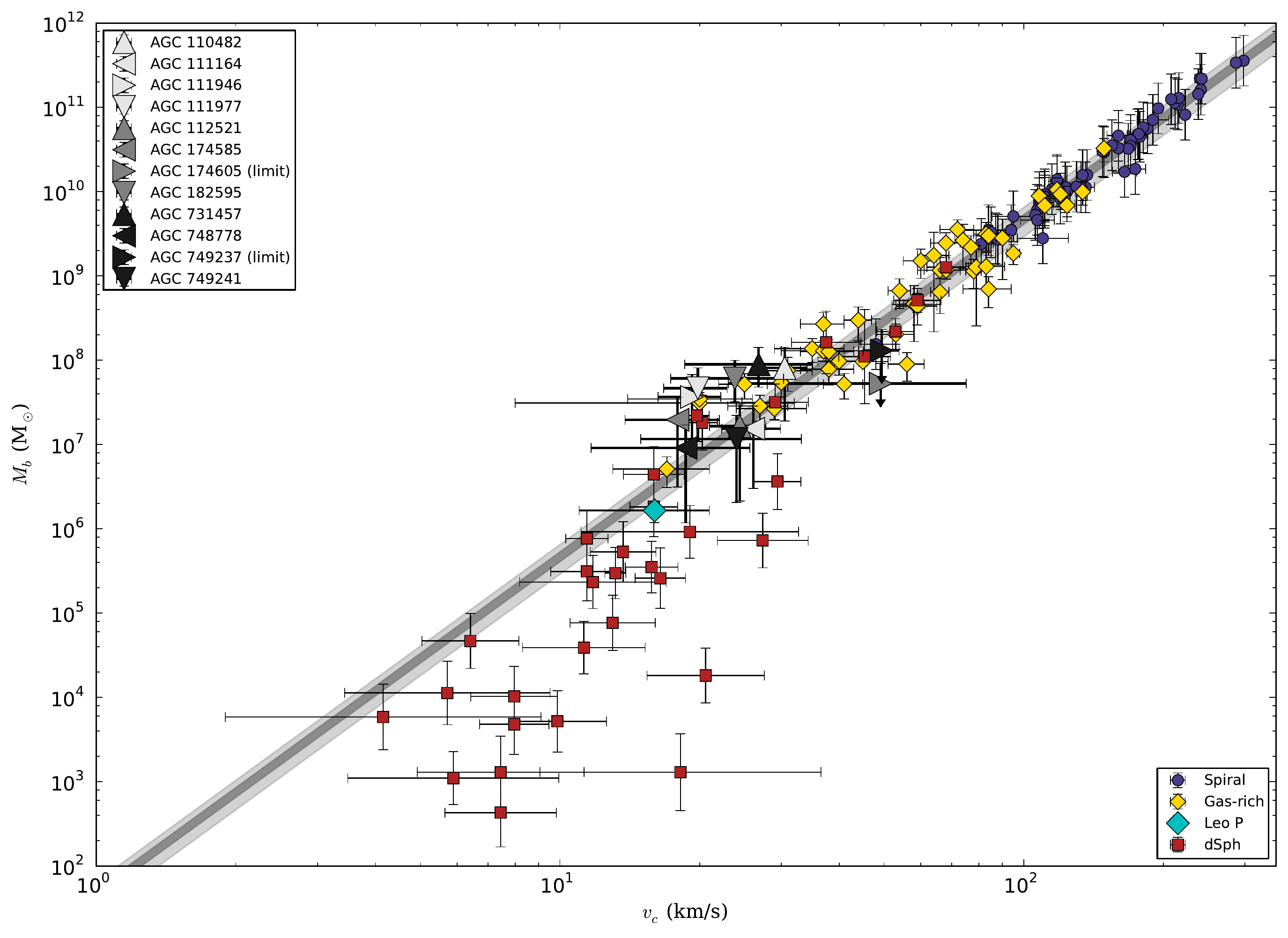}
\caption{The fundamental plane of the mass-velocity scaling relation,
  commonly referred to as the Baryonic Tully-Fisher relation
  (BTFR). The smaller points correspond to data from the literature
  (see the compilation by McGaugh (2012) and references therein). The
  purple circles correspond to spiral galaxies with available \HI{}
  line-width data whose baryonic mass is dominated by the stellar
  component. The gold diamonds represent the less massive gas-rich
  galaxies used to calibrate the model, and the red squares represent
  spheroidal dwarf galaxies with no detectable \HI{}. The larger cyan
  diamond represents Leo P, the slowest rotating and lowest-mass
  galaxy known to still be relatively rich with interstellar gas
  (Bernstein-Cooper \etal\ 2014). The gray bars represent 1 and 3
  $\sigma$ deviations from a fit of the BTFR to the gas-rich galaxy
  sample. The large greyscale triangles represent the SHIELD galaxies.
  The SHIELD galaxy sample is fit significantly by the model; ten
  galaxies agree within model uncertainty of 1 $\sigma$ and all 12
  agree within 3 $\sigma$ model uncertainty.}
\label{btfr}
\end{figure}

Using Figure~\ref{btfr} as an interpretive guide, we see that the
primary contribution of the SHIELD program to our understanding of the
dynamics of low-mass galaxies comes in the form of improved statistics
in the lowest mass bins.  This sample of low-HI mass galaxies
effectively doubles the number of points (with v$_{\rm c}$ $\lsim$
30 \kms\ ) that can be placed on the BTFR.  The gas-rich SHIELD
galaxies have higher baryon fractions and are less dark matter
dominated than dSph galaxies with similar rotational velocities.

All of the SHIELD galaxies agree within 3 $\sigma$ model uncertainty
to the BTFR presented in Figure~\ref{btfr}.  The most massive dSph
systems can be considered to be rough analogs of the SHIELD galaxies
with stripped \HI{} components.  dSphs with v$_{\rm c}$ $\gsim$ 20
\kms\ can be made to lie on the BTFR if an amount of gas which would
be appropriate to bring the dSph to a typical M$_{\rm HI}$/M$_{\star}$
($\sim$10$^7$ \msun\ for systems in this range of circular velocities)
were added to their baryonic mass budgets.  However, the less massive
dSph galaxies are fundamentally different; they are less massive in
total, likely a result of significant tidal stripping that has
affected both their baryonic and dark matter components.

This gain in low-mass systems on the BTFR comes with a significant
caveat: for most SHIELD galaxies, the rotational velocities are
estimated from methods without the benefit of close constraints on the
gas inclination. In comparison with studies of larger dwarfs using
similar observational strategies, the rotational dynamics in the
SHIELD galaxies are not resolved at high spatial resolution.  For
example, the recent dynamical modeling of the LITTLE THINGS galaxies
by \citet{oh15} performs a full radial mass decomposition for most of
these marginally closer, brighter, and more massive sources.

There are two empirical limitations that preclude such detailed
analysis in the SHIELD galaxies.  The first is the simple and perhaps
predictable issue of the distance of the sources: the nearest SHIELD
galaxy, AGC\,111164, lies at D $=$ 5.11\,$\pm$\,0.07 Mpc; the most
distant systems lie beyond 10 Mpc (AGC\,174605, AGC\,731457,
AGC\,749237).  At these distances, even B configuration resolution VLA
data presents a beam smearing of hundreds of parsecs.  The second
limitation is that the SHIELD galaxies have small total \HI{} flux
integrals.  These limitations are in agreement with those found in 
similar studies of low-mass galaxies \citep[e.g., ][]{mcgaugh12}.

By way of comparison, the \citet{oh15} sample contains multiple
systems with \HI{} masses in the same range as those of the SHIELD
galaxies, and in fact some that are less massive still.  However,
importantly, all three of the \citet{oh15} systems whose rotational
velocities are lower than 20 \kms\ are in or just outside the Local
Group (DDO\,210, DDO\,216, IC\,1613).  The gain in angular resolution
and in \HI{} flux from these sources facilitates a depth of analysis
that is simply unavailable with current observational capabilities
outside of the Local Group.  Note that the observational strategies
used in this work are very similar to those used in \citet{oh15}.

An interesting comparison can be found in Leo\,P, a nearby
(D$=$1.62\,$\pm$0.15 Mpc; {McQuinn \etal\ 2015}\nocite{mcquinnleop}),
extremely low-mass (log(M$_{\rm HI}$) $=$ 8.1\,$\times$\,10$^{5}$
\msun) galaxy that was discovered by ALFALFA
\citep{giovanelli13,rhode13}. In a detailed \HI{} study by
\citet{ezbc14}, the authors examine deep VLA \HI{} 21\,cm data that
are very similar to the data presented here for the SHIELD galaxies.
The conclusion is the same as that in the present work: extracting a
meaningful and non-degenerate model of the gas kinematics is extremely
challenging at rotation velocities lower than 20 \kms and without well
constrained gas inclination.

Based on the multiple lines of evidence outlined above, we conclude
that there exists an empirical lower threshold rotational velocity,
below which current observations cannot differentiate coherent
rotation from pressure support.  Using the SHIELD galaxies, and the
systems from the aforementioned studies, this threshold appears below
V$_{\rm rot}$ $\sim$15 \kms.  Our observations demand models which can
reproduce the kinematics of low-mass galaxies whose gas is dominated
by both pressure and rotational dynamics.

It is interesting to note that that the ALFALFA survey has discovered
many candidate objects whose \HI{} properties are galaxy-like, but
that lack an obvious stellar population in survey-depth optical data
products.  These systems can broadly be categorized as ``ultra compact
high velocity clouds'' (UCHVCs; {Adams \etal\ 2013}\nocite{adams13})
and ``Almost Dark'' galaxy candidates \citep{cannon15,janowiecki15}.
Further comparisons of all of the SHIELD-class galaxies with members
of these ALFALFA sub-samples promise to populate the continuum of
sources at the lowest and most extreme masses.

\clearpage
\acknowledgements

The authors acknowledge the work of the entire ALFALFA collaboration
team in observing, flagging, and extracting the catalog of galaxies
used to identify the SHIELD sample. The ALFALFA team at Cornell is
supported by NSF grants AST-0607007 and AST-1107390 to RG and MPH and
by grants from the Brinson Foundation.  Y.G.T, A.T.M., and J.M.C. are
supported by NSF grant AST-1211683.  E.A.K.A. is supported by
TOP1EW.14.105, which is financed by the Netherlands Organisation for
Scientific Research (NWO).  ATM would like to gratefully acknowledge
Robert Pipes, Marianne Takamiya, Ramprasad Rao, Edward Molter,
Charlotte Martinkus, the members of the NRAO/UVa galaxy discussion
group, and the Akamai Workforce Initiative.  This research made use of
\textsc{Astropy}, a community-developed core \textsc{Python} package
for Astronomy \citep{astropy13}.

Support for Hubble Space Telescope data in this work was provided by
NASA through grant GO-12658 from the Space Telescope Institute, which
is operated by Aura, Inc., under NASA contract NAS5-26555.  The
Arecibo Observatory is operated by SRI International under a
cooperative agreement with the National Science Foundation
(AST-1100968), and in alliance with Ana G. M{\'e}ndez-Universidad
Metropolitana, and the Universities Space Research Association.  This
research made use of NASA's Astrophysical Data System, the NASA/IPAC
Extragalactic Database which is operated by the Jet Propulsion
Laboratory, California Institute of Technology, under contract with
the National Aeronautics and Space Administration, and Montage, funded
by the NASA's Earth Science Technology Office, Computation
Technologies Project, under Cooperative Agreement Number NCC5-626
between NASA and the California Institute of Technology.  Montage is
maintained by the NASA/IPAC Infrared Science Archive.

\textit{Facilities}: \facility{HST}, \facility{GALEX},
\facility{WIYN:3.5m}, \facility{WIYN:0.9m}, \facility{Spitzer},
\facility{VLA}


\bibliographystyle{apj}                                                 

\begin{thebibliography}{}
\bibitem[Adams et al.(2013)]{adams13} Adams, E.~A.~K., Giovanelli, R.,
  \& Haynes, M.~P.\ 2013, \apj, 768, 77
  
\bibitem[Adams et al.(2014)]{jadams14} Adams, J.~J., Simon, 
J.~D., Fabricius, M.~H., et al.\ 2014, \apj, 789, 63 

\bibitem[Astropy Collaboration et al.(2013)]{astropy13}
  Astropy Collaboration, Robitaille, T.~P., Tollerud, E.~J., et
  al.\ 2013, \aap, 558, A33

\bibitem[Begum et al.(2008)]{begum08} Begum, A., Chengalur, J.~N.,
  Karachentsev, I.~D., Sharina, M.~E., \& Kaisin, S.~S.\ 2008, \mnras,
  386, 1667

\bibitem[Bernstein-Cooper et al.(2014)]{ezbc14} Bernstein-Cooper,
  E.~Z., Cannon, J.~M., Elson, E.~C., et al.\ 2014, \aj, 148, 35

\bibitem[Boylan-Kolchin et al.(2011)]{boylan11}
  Boylan-Kolchin, M., Bullock, J.~S., \& Kaplinghat, M.\ 2011, \mnras,
  415, L40

\bibitem[(1995)]{briggs95} Briggs, D.~S.\ 1995, Bulletin 
of the American Astronomical Society, 27, \#112.02 

\bibitem[Cannon et al.(2011a)]{most11a} Cannon, J.~M., Most,
  H.~P., Skillman, E.~D., et al.\ 2011, \apj, 735, 35

\bibitem[Cannon et al.(2012)]{cannon12} Cannon, J.~M.,
  Bernstein-Cooper, E.~Z., Cave, I.~M., et al.\ 2012, \aj, 144, 82
  
\bibitem[Cannon \etal(2013)]{cannon13} Cannon, J.~M., Marshall, 
M., Cave, I., et al.\ 2013, American Astronomical Society Meeting Abstracts 
\#221, 221, 352.04

\bibitem[Cannon et al.(2015)]{cannon15} Cannon, J.~M., Martinkus,
  C.~P., Leisman, L., et al.\ 2015, \aj, 149, 72

\bibitem[Carignan(2015)]{2015IAUGA..2301455C} Carignan, C.\ 2015, IAU
  General Assembly, 23, 1455

\bibitem[Corbelli \& Schneider(1997)]{corbelli97} Corbelli,
  E., \& Schneider, S.~E.\ 1997, \apj, 479, 244

\bibitem[Davis et al.(2013)]{davis13} Davis, T.~A., Alatalo, 
K., Bureau, M., et al.\ 2013, \mnras, 429, 534 

\bibitem[de Blok(2010)]{deblok10} de Blok, W.~J.~G.\ 2010, Advances in
  Astronomy, 2010, 789293

\bibitem[Di Cintio et al.(2014)]{dc14} Di Cintio, A., Brook, C.~B.,
  Dutton, A.~A., et al.\ 2014, \mnras, 441, 2986
  
\bibitem[Di Teodoro \& Fraternali(2015)]{diteodoro15} Di Teodoro,
  E.~M., \& Fraternali, F.\ 2015, \mnras, 451, 3021

\bibitem[Elson et al.(2013)]{elson13} Elson, E.~C., de Blok, W.~J.~G.,
  \& Kraan-Korteweg, R.~C.\ 2013, \mnras, 429, 2550
  
\bibitem[Elson(2014)]{elson14} Elson, E.~C.\ 2014, \mnras, 437, 3736 

\bibitem[Eskew et al.(2012)]{eskew12} Eskew, M., Zaritsky, D., 
\& Meidt, S.\ 2012, \aj, 143, 139 

\bibitem[Ewen \& Purcell(1951)]{ewen51} Ewen, H.~I., \&
  Purcell, E.~M.\ 1951, \nat, 168, 356

\bibitem[Fraternali et al.(2002)]{fraternali02} Fraternali, F., van 
Moorsel, G., Sancisi, R., \& Oosterloo, T.\ 2002, \aj, 123, 3124 

\bibitem[Frusciante et al.(2012)]{frusciante12} Frusciante, N.,
  Salucci, P., Vernieri, D., Cannon, J.~M., \& Elson, E.~C.\ 2012,
  \mnras, 426, 751

\bibitem[Garrison-Kimmel et al.(2014)]{garrison14}
  Garrison-Kimmel, S., Boylan-Kolchin, M., Bullock, J.~S., \& Kirby,
  E.~N.\ 2014, \mnras, 444, 222
  
\bibitem[Giovanelli et al.(2005)]{giovanelli05} Giovanelli, R., 
  Haynes, M.~P., Kent, B.~R., et al.\ 2005, \aj, 130, 2598

\bibitem[Giovanelli \etal(2013)]{giovanelli13} Giovanelli, R., Haynes,
  M.~P., Adams, E.~A.~K., et al.\ 2013, \aj, 146, 15

\bibitem[Gooch(1996)]{gooch96} Gooch, R.\ 1996, Astronomical Data
  Analysis Software and Systems V, 101, 80

\bibitem[Gordon  et  al.(2016)]{gordon16}  Gordon,  A.~J.~R.,  Cannon,
  J.~M., Adams, E.~A.,  \& SHIELD II Team  2016, American Astronomical
  Society Meeting Abstracts, 227, 136.04

\bibitem[Hagen et al.(2014)]{hagen14} Hagen, C., Cannon,
  J.~M., Cave, I., et al.\ 2014, American Astronomical Society Meeting
  Abstracts \#223, 223, \#355.16

\bibitem[Haurberg et al.(2015)]{haurberg15} Haurberg, N.~C., Salzer,
  J.~J., Cannon, J.~M., \& Marshall, M.~V.\ 2015, \apj, 800, 121
  
\bibitem[Hoffman et al.(1996)]{hoffman96} Hoffman, G.~L., 
Salpeter, E.~E., Farhat, B., et al.\ 1996, \apjs, 105, 269 

\bibitem[Huang et al.(2012a)]{huang12pop} Huang, S., Haynes, M.~P.,
  Giovanelli, R., \& Brinchmann, J.\ 2012, \apj, 756, 113

\bibitem[Huang et al.(2012b)]{huang12gas} Huang, S., Haynes, M.~P.,
  Giovanelli, R., et al.\ 2012, \aj, 143, 133

\bibitem[Hunter et al.(2012)]{hunter12} Hunter, D.~A., Ficut-Vicas,
  D., Ashley, T., et al.\ 2012, \aj, 144, 134

\bibitem[Janowiecki et al.(2015)]{janowiecki15} Janowiecki, S.,
  Leisman, L., J{\'o}zsa, G., et al.\ 2015, \apj, 801, 96

\bibitem[Jorsater \& van Moorsel(1995)]{jorsater95} Jorsater, S., \&
  van Moorsel, G.~A.\ 1995, \aj, 110, 2037

\bibitem[J{\'o}zsa et al.(2007)]{jozsa07} J{\'o}zsa,
  G.~I.~G., Kenn, F., Klein, U., \& Oosterloo, T.~A.\ 2007, \aap, 468,
  731

\bibitem[Kauffmann et al.(1993)]{kauffmann93} Kauffmann, G.,
  White, S.~D.~M., \& Guiderdoni, B.\ 1993, \mnras, 264, 201

\bibitem[Kerr et al.(1954)]{kerr54} Kerr, F.~J., Hindman, 
  J.~F., \& Robinson, B.~J.\ 1954, Australian Journal of Physics, 7, 297

\bibitem[Kirby et al.(2012)]{kirby12} Kirby, E.~M., Koribalski, B.,
  Jerjen, H., \& L{\'o}pez-S{\'a}nchez, {\'A}.\ 2012, \mnras, 420,
  2924

\bibitem[Klypin et al.(2015)]{klypin15} Klypin, A., 
Karachentsev, I., Makarov, D., \& Nasonova, O.\ 2015, \mnras, 454, 1798 

\bibitem[Klypin et al.(1999)]{klypin99} Klypin, A., Kravtsov, A.~V.,
  Valenzuela, O., \& Prada, F.\ 1999, \apj, 522, 82

\bibitem[Lee et  al.(2016)]{lee16} Lee, E., Cannon,  J.~M., McNichols,
  A., Teich, Y., \& SHIELD II Team 2016, American Astronomical Society
  Meeting Abstracts, 227, 136.07

\bibitem[Lelli et al.(2016)]{lelli16} Lelli, F., McGaugh, 
S.~S., \& Schombert, J.~M.\ 2016, \apjl, 816, L14 

\bibitem[Marshall(2013)]{marshall13} Marshall, M.\ 2013, American
  Astronomical Society Meeting Abstracts \#221, 221, 352.05

\bibitem[Masters(2005)]{masters05} Masters, K.~L.\ 2005, Ph.D.~Thesis,
  
\bibitem[Mateo(1998)]{mateo98} Mateo, M.~L.\ 1998, \araa, 36, 435

\bibitem[McGaugh et al.(2000)]{mcgaugh00} McGaugh, S.~S., Schombert,
  J.~M., Bothun, G.~D., \& de Blok, W.~J.~G.\ 2000, \apjl, 533, L9

\bibitem[McGaugh(2012)]{mcgaugh12} McGaugh, S.~S.\ 2012,
  \aj, 143, 40

\bibitem[McMullin et al.(2007)]{mcmullin07} McMullin, J.~P.,
  Waters, B., Schiebel, D., Young, W., \& Golap, K.\ 2007,
  Astronomical Data Analysis Software and Systems XVI, 376, 127
  
\bibitem[McNichols et al.(2015)]{AAS} McNichols, A., Teich, Y.,
  Cannon, J.~M., et al.\ 2015, American Astronomical Society Meeting
  Abstracts, 225, \#248.20

\bibitem[McQuinn  et  al.(2014)]{mcquinntrgb}  McQuinn,  K.~B.~W.,
  Cannon, J.~M., Dolphin, A.~E., et al.\ 2014, \apj, 785, 3

\bibitem[McQuinn et al.(2015a)]{mcquinnsfh} McQuinn, K.~B.~W., 
  Cannon, J.~M., Dolphin, A.~E., et al.\ 2015a, \apj, 802, 66

\bibitem[McQuinn et al.(2015b)]{mcquinnleop} McQuinn, K.~B.~W., 
  Skillman, E.~D., Dolphin, A.~E., et al.\ 2015b, \apj, 812, 156

\bibitem[Meyer  et  al.(2004)]{meyer04}  Meyer, M.~J.,  Zwaan,  M.~A.,
  Webster, R.~L., et al.\ 2004, \mnras, 350, 1195
  
\bibitem[Moore et al.(1999)]{moore99} Moore, B., Ghigna,
  S., Governato, F., et al.\ 1999, \apjl, 524, L19

\bibitem[Navarro et al.(1997)]{nfw97} Navarro, J.~F., Frenk, C.~S., \&
  White, S.~D.~M.\ 1997, \apj, 490, 493

\bibitem[Oh et al.(2011)]{oh11} Oh, S.-H., de Blok, W.~J.~G., Brinks,
  E., Walter, F., \& Kennicutt, R.~C., Jr.\ 2011, \aj, 141, 193

\bibitem[Oh et al.(2015)]{oh15} Oh, S.-H., Hunter,
  D.~A., Brinks, E., et al.\ 2015, \aj, 149, 180

\bibitem[Ott et al.(2012)]{ott12} Ott, J., Stilp, A.~M., Warren,
  S.~R., et al.\ 2012, \aj, 144, 123

\bibitem[Papastergis et al.(2015)]{papastergis15} Papastergis,
  E., Giovanelli, R., Haynes, M.~P., \& Shankar, F.\ 2015, \aap, 574,
  A113

\bibitem[Papastergis et al.(2016)]{papastergis16} Papastergis, E.,
  Adams, E.A.K., \& van der Hulst, J.M.\ 2016, \aap, submitted
  (ArXiV/1602.09087)

\bibitem[Perley \& Butler(2013)]{pb12} Perley, R.~A., \& Butler,
  B.~J.\ 2013, \apjs, 204, 19

\bibitem[Rhode \etal(2013)]{rhode13} Rhode, K.~L., Salzer, J.~J.,
  Haurberg, N.~C., et al.\ 2013, \aj, 145, 149
  
\bibitem[Robles et al.(2015)]{2015ApJ...810...99R} Robles, V.~H.,
  Lora, V., Matos, T., \& S{\'a}nchez-Salcedo, F.~J.\ 2015, \apj, 810,
  99

\bibitem[Rogstad et al.(1974)]{trm} Rogstad, D.~H., 
Lockhart, I.~A., \& Wright, M.~C.~H.\ 1974, \apj, 193, 309 

\bibitem[Rubio et al.(2015)]{rubio15} Rubio, M.,
  Elmegreen, B.~G., Hunter, D.~A., et al.\ 2015, \nat, 525, 218

\bibitem[Salak et al.(2016)]{salak16} Salak, D., Nakai, N.,
  Hatakeyama, T., \& Miyamoto, Y.\ 2016, arXiv:1603.05881

\bibitem[Schmidt et al.(2014)]{schmidt14} Schmidt, P., J{\'o}zsa,
  G.~I.~G., Gentile, G., et al.\ 2014, \aap, 561, A28
  
\bibitem[Schwab(1984)]{schwab84} Schwab, F.~R.\ 1984, \aj, 89, 1076

\bibitem[Spekkens et al.(2005)]{spekkens05} Spekkens, K., Giovanelli,
  R., \& Haynes, M.~P.\ 2005, \aj, 129, 2119

\bibitem[Stark et al.(2009)]{stark09} Stark, D.~V., McGaugh, S.~S., \&
  Swaters, R.~A.\ 2009, \aj, 138, 392
  
\bibitem[Staveley-Smith et al.(1992)]{staveleysmith92} Staveley-Smith,
  L., Davies, R.~D., \& Kinman, T.~D.\ 1992, \mnras, 258, 334

\bibitem[Stevens et al.(2016)]{stevens16} Stevens, K., Cannon, J.~M.,
  McNichols, A., et al.\ 2016, American Astronomical Society Meeting
  Abstracts, 227, 136.05

\bibitem[Stilp et al.(2013)]{stilp13} Stilp, A.~M.,
  Dalcanton, J.~J., Skillman, E., et al.\ 2013, \apj, 773, 88
  
\bibitem[Swaters et al.(2002)]{swaters02}  Swaters, R.~A., van Albada,
  T.~S., van der Hulst, J.~M., \& Sancisi, R.\ 2002, \aap, 390, 829
  
\bibitem[Tamburro et al.(2009)]{tamburro09} Tamburro, D., Rix, H.-W.,
  Leroy, A.~K., et al.\ 2009, \aj, 137, 4424

\bibitem[Teich et al.(2016)]{teich} Teich, Y.~G., Cannon, J.~M.,
  McNichols, A.~T., et al.\ 2016, \apj, submitted

\bibitem[Thomann et al.(2012)]{thomann12} Thomann, C., Cannon, J.~M.,
  Elson, E.~C., et al.\ 2012, American Astronomical Society Meeting
  Abstracts \#219, 219, \#244.03

\bibitem[Trachternach et al.(2009)]{trachternach09} Trachternach, C.,
  de Blok, W.~J.~G., McGaugh, S.~S., van der Hulst, J.~M., \& Dettmar,
  R.-J.\ 2009, \aap, 505, 577

\bibitem[Tully \& Fisher(1977)]{tullyfisher77} Tully, R.~B., \&
  Fisher, J.~R.\ 1977, \aap, 54, 661

\bibitem[van der Hulst et al.(1992)]{gipsy92} van der Hulst, J.~M.,
  Terlouw, J.~P., Begeman, K.~G., Zwitser, W., \& Roelfsema,
  P.~R.\ 1992, Astronomical Data Analysis Software and Systems I, 25,
  131

\bibitem[Verbeke et al.(2015)]{verbekeIAU} Verbeke, R., Vandenbroucke,
  B., De Rijcke, S., \& Koleva, M.\ 2015, IAU General Assembly, 22,
  2232938
  

\bibitem[Walker \& Pe{\~n}arrubia(2011)]{penarrubia11} Walker, M.~G.,
  \& Pe{\~n}arrubia, J.\ 2011, \apj, 742, 20

\bibitem[Warren \etal(2015)]{warren15} Warren, S.~R., Molter, E.,
  Cannon, J.~M., et al.\ 2015, \apj, 814, 30
  
\bibitem[Wiegert et al.(2006)]{wiegertAAS06} Wiegert, T.,
  English, J., \& Fiege, J.\ 2006, Bulletin of the American
  Astronomical Society, 38, 96

\bibitem[Wiegert(2011)]{wiegert_thesis11} Wiegert, T.~B.~V.\ 2011, 
  Ph.D.~Thesis

\bibitem[Wiegert \& English(2014)]{wiegert14} Wiegert, T.,
  \& English, J.\ 2014, New Astronomy, 26, 40
  
\bibitem[Wendt et al.(2008)]{wendt08} Wendt, H.,
  Orchiston, W., \& Slee, B.\ 2008, Journal of Astronomical History
  and Heritage, 11, 185

\end{thebibliography}


\clearpage
\thispagestyle{empty}
\tabletypesize{\small}
\begin{deluxetable}{lcc}
\tablecaption{Combined Imaging Properties} 
\tablewidth{0pt} 
\tablecolumns{3} 
\tablehead{   \colhead{AGC} & \colhead{Beam Dimensions}                             & \colhead{RMS Noise per Channel}\\
              \colhead{\#}  & \colhead{(B$_{\rm Maj}$ $\times$ B$_{\rm Min}$ $@$ BPA)} & \colhead{(Jy~bm$^{-1}$)}
}
\startdata 
\cutinhead{Briggs' Weighting $R = 0.5$} 
110482\tablenotemark{a} & 11.98\arcsec\ $\times$ 9.04\arcsec\ $@$ 49.4$\degree$ & 1.0$\times$10$^{-3}$       \\
111164\tablenotemark{b} & 21.56\arcsec\ $\times$ 21.24\arcsec\ $@$ 28.8$\degree$ & 1.4$\times$10$^{-3}$    \\ 
111946\tablenotemark{a} & 10.30\arcsec\ $\times$ 8.86\arcsec\ $@$ $-$169.1$\degree$ & 1.1$\times$10$^{-3}$   \\ 
111977\tablenotemark{b} & 24.01\arcsec\ $\times$ 19.95\arcsec\ $@$ 56.5$\degree$ & 1.5$\times$10$^{-3}$      \\ 
112521\tablenotemark{b} & 22.03\arcsec\ $\times$ 19.51\arcsec\ $@$ $-$42.9$\degree$ & 1.3$\times$10$^{-3}$   \\ 
174585                  & 6.19\arcsec\ $\times$ 5.52\arcsec\ $@$ $-$45.9$\degree$ & 8.8$\times$10$^{-4}$      \\
174605                  & 11.81\arcsec\ $\times$ 9.99\arcsec\ $@$ 2.8$\degree$ & 5.1$\times$10$^{-4}$        \\ 
182595                  & 10.05\arcsec\ $\times$ 9.93\arcsec\ $@$ 74.7$\degree$ & 6.9$\times$10$^{-4}$       \\ 
731457                  & 6.04\arcsec\ $\times$ 5.53\arcsec\ $@$ $-$55.1$\degree$ & 8.8$\times$10$^{-4}$     \\ 
748778                  & 5.91\arcsec\ $\times$ 5.23\arcsec\ $@$ $-$29.2$\degree$ & 9.3$\times$10$^{-4}$     \\
749237                  & 6.21\arcsec\ $\times$ 5.59\arcsec\ $@$ $-$24.1$\degree$ & 7.9$\times$10$^{-4}$     \\  
749241                  & 6.06\arcsec\ $\times$ 5.82\arcsec\ $@$ 51.5$\degree$ & 7.9$\times$10$^{-4}$     \\ 
\cutinhead{Briggs' Weighting $R = 2.0$} 
110482\tablenotemark{a} & 14.16\arcsec\ $\times$ 12.02\arcsec\ $@$ 53.0$\degree$ & 1.0$\times$10$^{-3}$  \\
111164\tablenotemark{b} & 28.50\arcsec\ $\times$ 22.51\arcsec\ $@$ $-$45.0$\degree$ & 1.4$\times$10$^{-3}$  \\ 
111946\tablenotemark{a} & 12.99\arcsec\ $\times$ 11.91\arcsec\ $@$ 8.3$\degree$ & 1.1$\times$10$^{-3}$  \\ 
111977\tablenotemark{b} & 34.47\arcsec\ $\times$ 28.15\arcsec\ $@$ 59.4$\degree$ & 1.4$\times$10$^{-3}$  \\ 
112521\tablenotemark{b} & 31.00\arcsec\ $\times$ 29.38\arcsec\ $@$ 69.7$\degree$ & 1.2$\times$10$^{-3}$  \\ 
174585                  & 9.76\arcsec\ $\times$ 8.85\arcsec\ $@$ $-$44.2$\degree$ & 8.1$\times$10$^{-4}$  \\
174605                  & 16.28\arcsec\ $\times$ 13.87\arcsec\ $@$ $-$15.9$\degree$ & 4.7$\times$10$^{-4}$  \\ 
182595                  & 14.09\arcsec\ $\times$ 13.88\arcsec\ $@$ 50.7$\degree$ & 6.0$\times$10$^{-4}$  \\ 
731457                  & 7.61\arcsec\ $\times$ 6.96\arcsec\ $@$ $-$64.8$\degree$ & 8.4$\times$10$^{-4}$  \\ 
748778                  & 10.23\arcsec\ $\times$ 9.31\arcsec\ $@$ $-$28.3$\degree$ & 8.5$\times$10$^{-4}$  \\
749237                  & 9.84\arcsec\ $\times$ 8.99\arcsec\ $@$ $-$34.3$\degree$ & 7.3$\times$10$^{-4}$  \\  
749241                  & 5.45\arcsec\ $\times$ 4.73\arcsec\ $@$ $-$51.5$\degree$ & 7.3$\times$10$^{-4}$  \\ 
\enddata
\tablenotetext{a}{4.5 hours of B configuration observaton.}
\tablenotetext{b}{Not observed in the B configuration.}
\label{im.props}
\end{deluxetable}\clearpage

\thispagestyle{empty}
\tabletypesize{\small}
\begin{deluxetable}{ccccccccccc}
\tablecaption{Combined Kinematic Properties} 
\tablewidth{0pt}
\tablecolumns{9}
\tablehead{
  \colhead{AGC} &
  \colhead{RA} &
  \colhead{Dec} &
  \colhead{PA} &
  \colhead{R$_{max}$} &
  \colhead{V$_{max}$} &
  \colhead{V$_{rot}$} &
  \colhead{$\sigma_{max}$} &
  \colhead{$i$} &\\
  \colhead{\#} &
  \colhead{(J2000)} &
  \colhead{(J2000)} &
  \colhead{[$\degree$]} &
  \colhead{[\arcsec]} &
  \colhead{[\kms{}]} &
  \colhead{[\kms{}]} &
  \colhead{[\kms{}]} &
  \colhead{[$\degree$]} &\\
  \colhead{(1)} &
  \colhead{(2)} &
  \colhead{(3)} &
  \colhead{(4)} &
  \colhead{(5)} &
  \colhead{(6)} &
  \colhead{(7)} &
  \colhead{(8)} &
  \colhead{(9)} }
\startdata
110482 & 01:42:17 & 26:21:60 & 84  & 30 & 50 & 31 & 13   & 55 $\pm$ 5 \\ 
111164 & 02:00:10 & 28:49:48 & 326 & 30 & 40 & 26 & 9    & 50 $\pm$ 5 \\
111946 & 01:46:42 & 26:48:10 & 285 & 15 & 35 & 20 & 15   & 62 $\pm$ 5 \\ 
111977 & 01:55:21 & 27:57:19 & 29  & 45 & 35 & 20 & 10   & 59 $\pm$ 5 \\ 
112521 & 01:41:08 & 27:19:23 & 180 & 40 & 40 & 24 & 10   & 55 $\pm$ 5 \\ 
174585 & 07:36:10 & 09:59:08 & 290 & 15 & 25 & 19 & 13.5 & 42 $\pm$ 5 \\
174605 & 07:50:22 & 07:47:39 & 90  & 20 & 30 & 49 & 12   & 19 $\pm$ 10\\ 
182595 & 08:51:12 & 27:52:50 & 74  & 15 & 30 & 24 & 4    & 39 $\pm$ 10\\ 
731457 & 10:31:56 & 28:01:35 & 18  & 10 & 30 & 29 & 12   & 34 $\pm$ 10\\ 
748778 & 00:06:35 & 15:30:32 & 21  & 25 & 25 & 19 & 7    & 40 $\pm$ 15\\
749237 & 12:26:23 & 27:44:45 & 254 & 30 & 80 & 49 & 10   & 54 $\pm$ 5 \\  
749241 & 12:40:01 & 26:19:10 & 301 & 30 & 35 & 25 & 6.5 & 45 $\pm$ 20\\ 
\enddata
\tablecomments{
Column 1 - AGC catalog name; 
Columns 2 and 3 - RA and DEC of kinematic centers derived from PV slicing analysis;
Column 4 - position angle of receding side of major axis, measured east of north, derived from PV slicing analysis; 
Column 5 - furthest projected radius at which significant gas emission is detected;
Column 6 - difference between the largest and smallest velocities associated with emission in the PV slice maps;
Column 7 -  V$_{rot}$ is V$_{max}$ projected by the $i$ using the method of Papastergis \etal\ (2015) assuming a constant value of q=0.13;
Column 8 - average \HI{} velocity dispersion at R$_{max}$;
Column 9 - galaxy inclination, derived from the stellar component.}
\label{kin.props}
\end{deluxetable}
\clearpage

\thispagestyle{empty}
\tabletypesize{\small}
\begin{deluxetable}{ccccccc}
\tablecaption{Derived Kinematic Properties} 
\tablewidth{0pt}
\tablecolumns{11}
\tablehead{
\colhead{AGC} &
\colhead{Distance} &
\colhead{M$_{\star}$} &
\colhead{M$_{\rm HI}$} &
\colhead{M$_{\rm bary}$} &
\colhead{M$_{\rm dyn}$} &
\colhead{M$_{\rm dyn}$/M$_{\rm bary}$}\\
\colhead{\#} &
\colhead{[Mpc]} &
\colhead{[10$^{7}$\,\msun]} &
\colhead{[10$^{7}$\,\msun]} &
\colhead{[10$^{7}$\,\msun]} &
\colhead{[10$^{8}$\,\msun]} &
\colhead{}\\
\colhead{(1)} &
\colhead{(2)} &
\colhead{(3)} &
\colhead{(4)} &
\colhead{(5)} &
\colhead{(6)} &
\colhead{(7)} 
}
\startdata
110482 & 7.82 $\pm$ 0.21        & 5.5 $\pm$ 1.9         & 1.92 $\pm$ 0.12        & 8.1  & 3.8  & 4.7 \\
111164 & 5.11 $\pm$ 0.07        & 1.0$^{+0.20}_{-0.30}$   & 0.40 $\pm$ 0.03        & 1.5  & 1.6  & 10.6 \\
111946 & 9.02$^{+0.20}_{-0.29}$ & 1.7$^{+0.60}_{-0.70}$     & 1.46$^{+0.09}_{-0.11}$   & 3.7  & 1.6  & 4.3 \\ 
111977 & 5.96$^{+0.11}_{-0.09}$ & 3.7$^{+1.2}_{-1.1}$      & 0.71$^{+0.05}_{-0.05}$    & 4.7  & 2.1  & 4.5 \\ 
112521 & 6.58 $\pm$ 0.18        & 0.70$^{+0.30}_{-0.20}$  & 0.71 $\pm$ 0.06        & 1.7  & 2.7  & 15.9 \\ 
174585 & 7.89$^{+0.21}_{-0.17}$ & 0.90 $\pm$ 0.30         & 0.79$^{+0.07}_{-0.07}$   & 2.0  & 1.5  & 7.5  \\
174605 &10.89 $\pm$ 0.28        & 2.8$^{+1.4}_{-2.8}$    & 1.85 $\pm$ 0.15         & 5.3  & 7.0  & 13.2 \\ 
182595 & 9.02 $\pm$ 0.28        & 5.0$^{+2.2}_{-3.2}$    & 0.81 $\pm$ 0.08         & 6.1  & 0.9 & 1.5 \\ 
731457 &11.13$^{+0.20}_{-0.16}$ & 6.5$^{+3.7}_{-4.8}$      & 1.81$^{+0.13}_{-0.13}$    & 8.9  & 1.4  & 1.6 \\ 
748778 & 6.46$^{+0.14}_{-0.17}$ & 0.3 $\pm$ 0.10         & 0.45$^{+0.04}_{-0.05}$    & 0.91  & 0.9 & 9.9 \\
749237 &11.62$^{+0.20}_{-0.16}$ & 5.3$^{+2.9}_{-5.3}$      & 5.74$^{+0.25}_{-0.22}$    & 13  & 10.8  & 8.3 \\  
749241 & 5.62$^{+0.17}_{-0.14}$ & 4.0$^{+0.10}_{-0.20}$    & 0.57$^{+0.04}_{-0.04}$    & 4.8  & 1.3  & 2.7 \\ 
\enddata\vspace{0.5 cm}
\tablecomments{
Column 1 - AGC catalog name;
Column 2 - TRGB distance derived from McQuinn \etal\ (2014);
Column 3 - Stellar mass derived from McQuinn \etal\ (2015a);
Column 4 - \HI{} mass calculated from the ALFALFA flux integrals \citep{giovanelli05} and the distances of column 2;
Column 5 - cumulative baryonic mass;
Column 6 - dynamical mass;
Column 7 - ratio of dynamical mass to baryonic mass.}
\label{dyn.props}
\end{deluxetable}
\clearpage


\clearpage

\clearpage

\clearpage

\clearpage

\clearpage

\clearpage

\clearpage

\clearpage

\clearpage

\clearpage

\clearpage

\clearpage

\clearpage

\clearpage

\clearpage

\clearpage


\appendix
\section{Channel Maps of the SHIELD Galaxies}
\label{appendix}

We present channel maps for all 12 SHIELD galaxies.  These data cubes
were used to produce the moment images shown along the top row in
Figures~\ref{110482.collage} through \ref{749241.collage}.

\begin{figure}
\includegraphics[width=\textwidth]{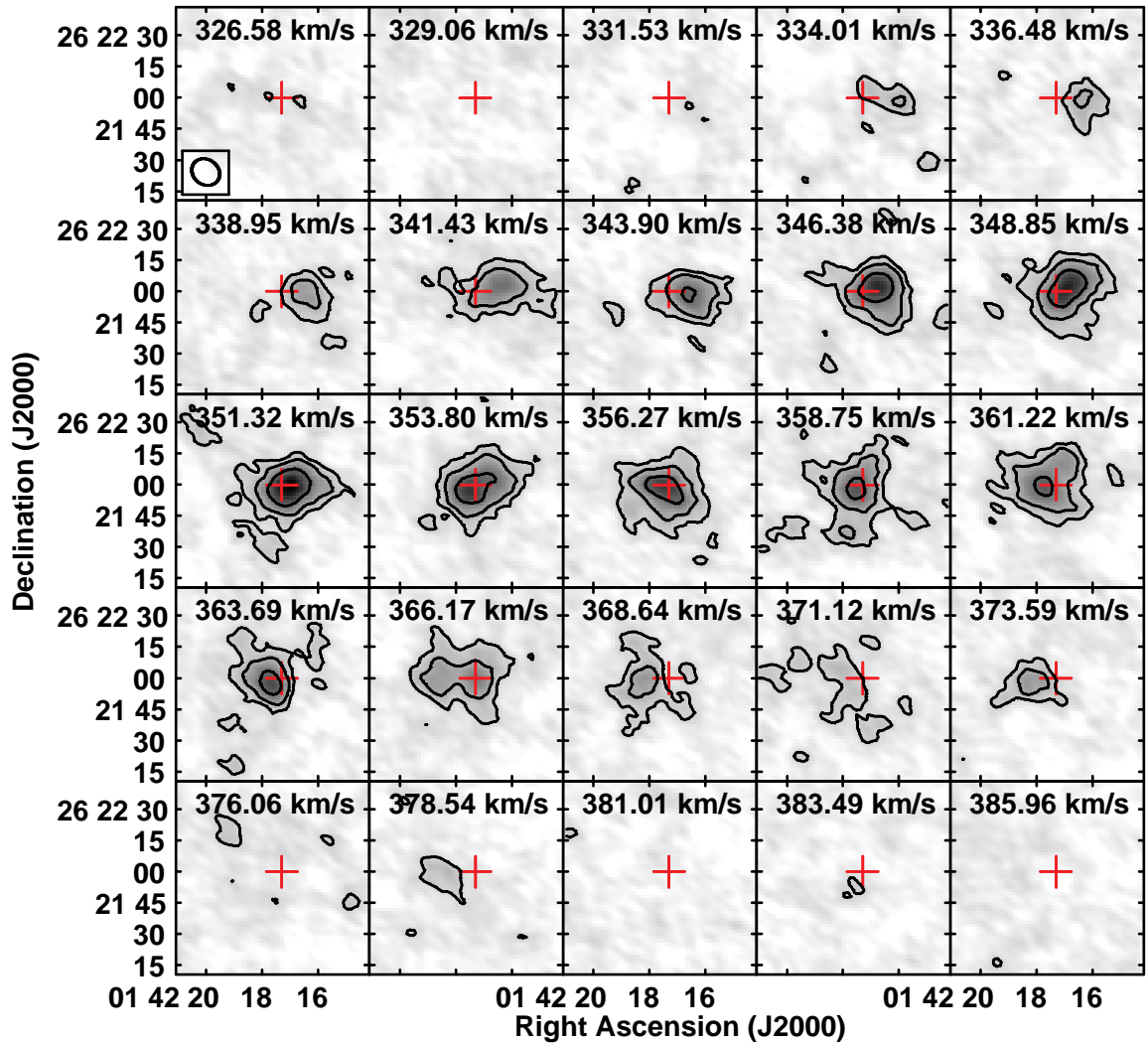}\vspace{-2 cm}
\caption{Channel map of the naturally-weighted, Hanning smoothed (by 3
  channels) data cube for AGC\,110482. The beam size is shown in the
  top left panel; the red crosshair is located at the identified
  dynamical center (see Table \ref{im.props}). The contours proceed in
  doubling intervals above 1$\times$10$^{20}$ atoms\,cm$^{-2}$.}
\label{110482.chmap}
\end{figure}\clearpage

\begin{figure}
\includegraphics[width=\textwidth]{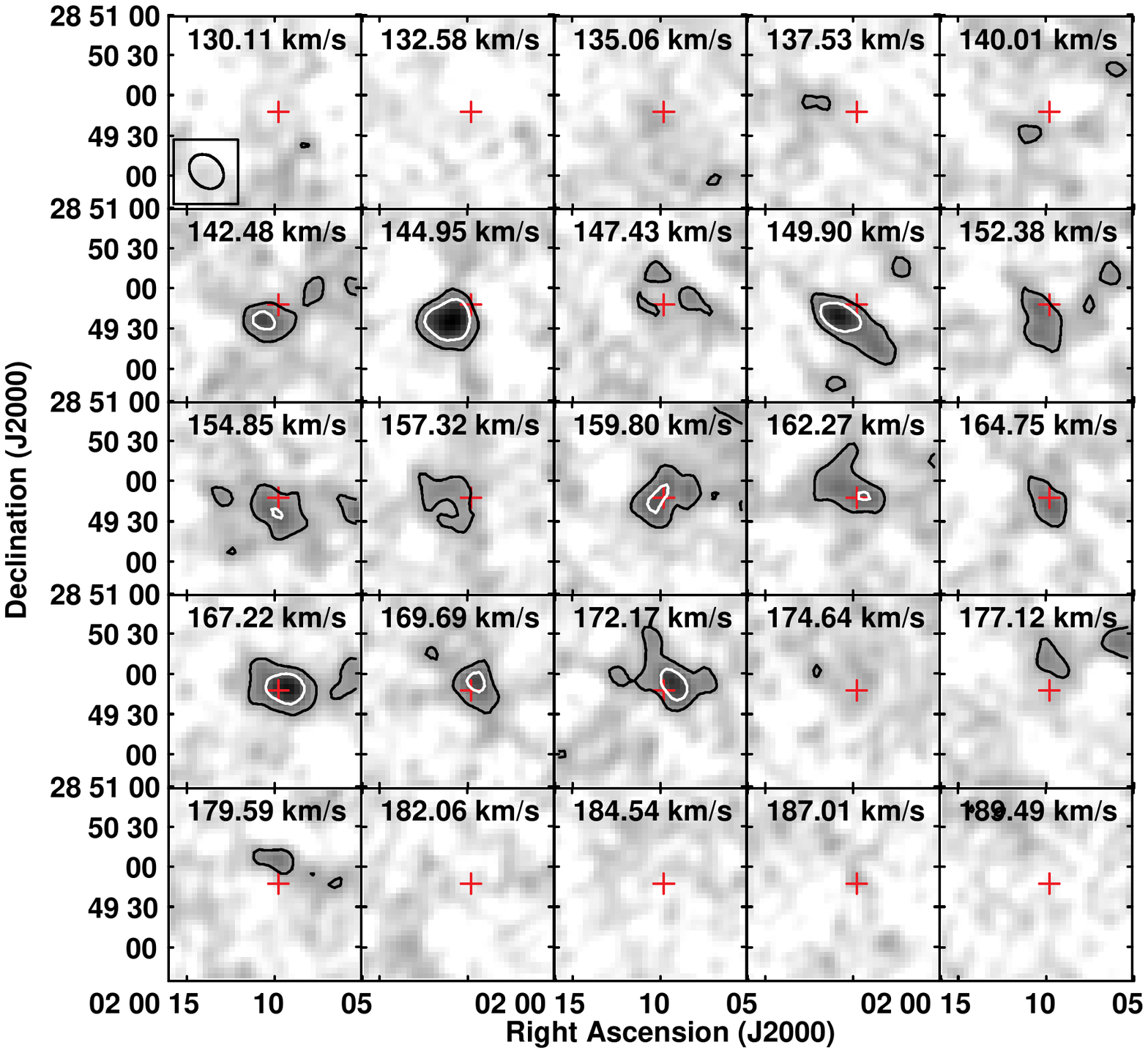}\vspace{-2 cm}
\caption{Same as Figure 33, for AGC\,111164.}
\label{111164.chmap}
\end{figure}\clearpage

\begin{figure}
\includegraphics[width=\textwidth]{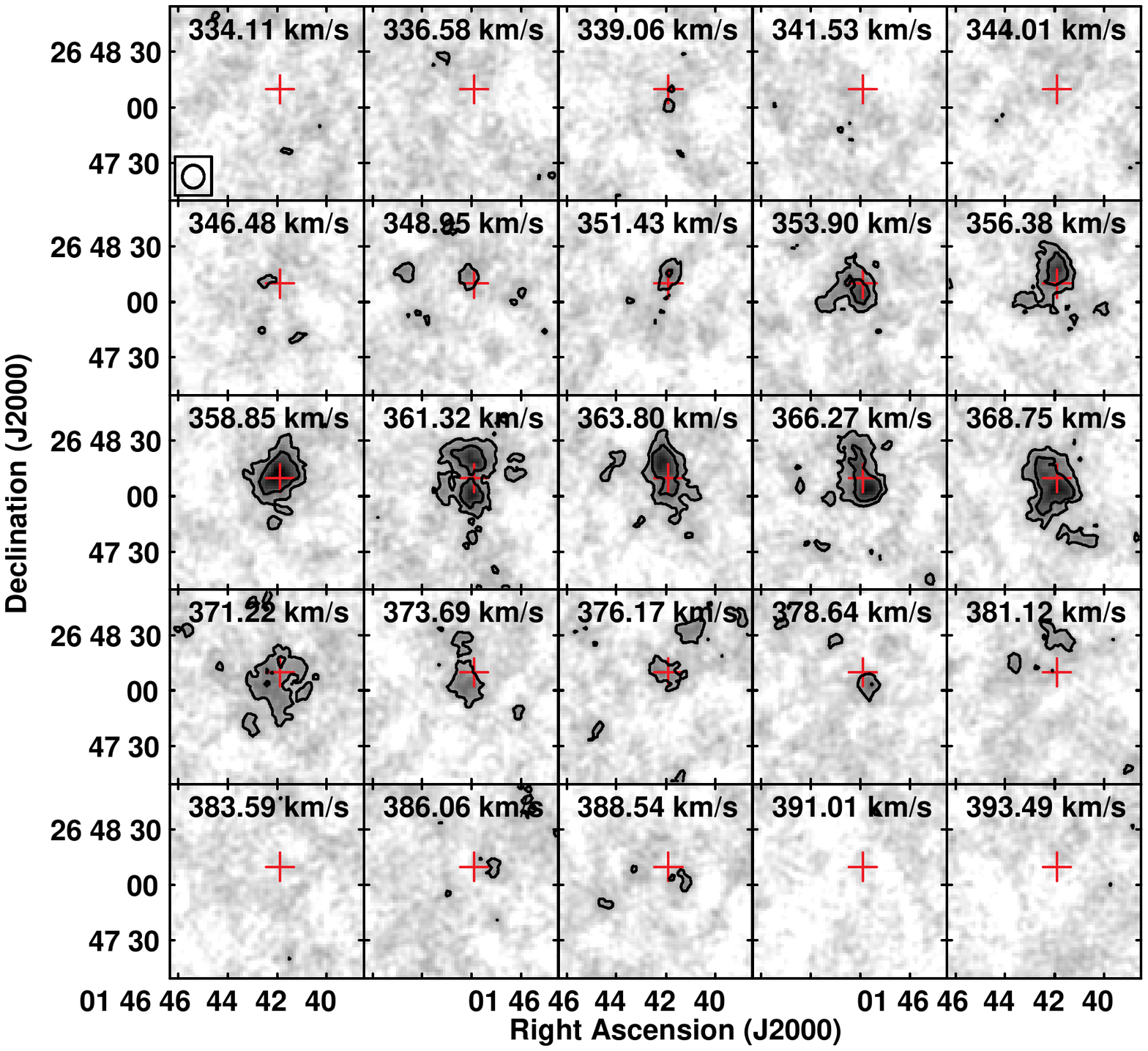}\vspace{-2 cm}
\caption{Same as Figure 33, for AGC\,111946.}
\label{111946.chmap}
\end{figure}\clearpage

\begin{figure}
\includegraphics[width=\textwidth]{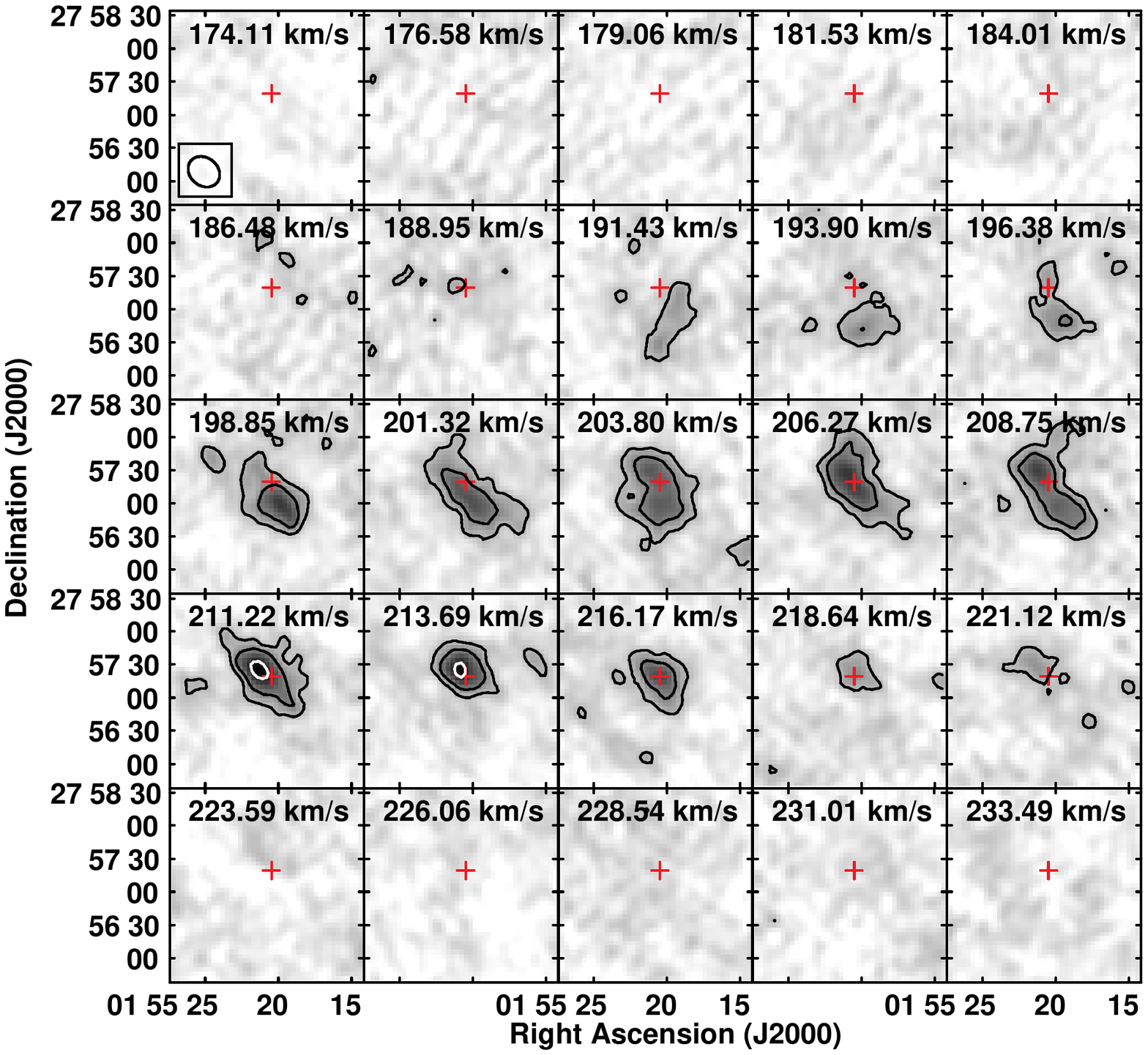}\vspace{-2 cm}
\caption{Same as Figure 33, for AGC\,111977.}
\label{111977.chmap}
\end{figure}\clearpage

\begin{figure}
\includegraphics[width=\textwidth]{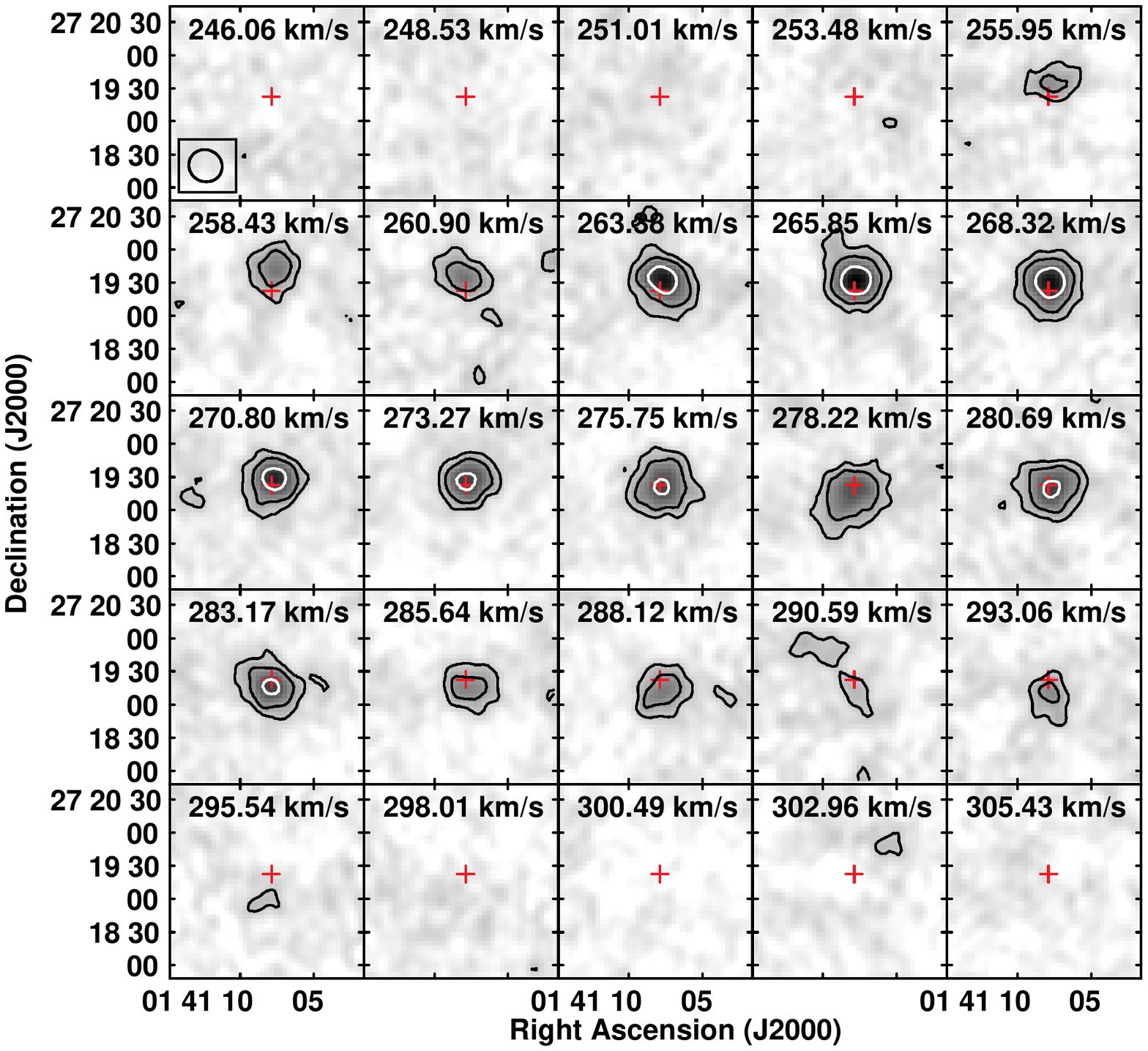}\vspace{-2 cm}
\caption{Same as Figure 33, for AGC\,112521.}
\label{112521.chmap}
\end{figure}\clearpage

\begin{figure}
\includegraphics[width=\textwidth]{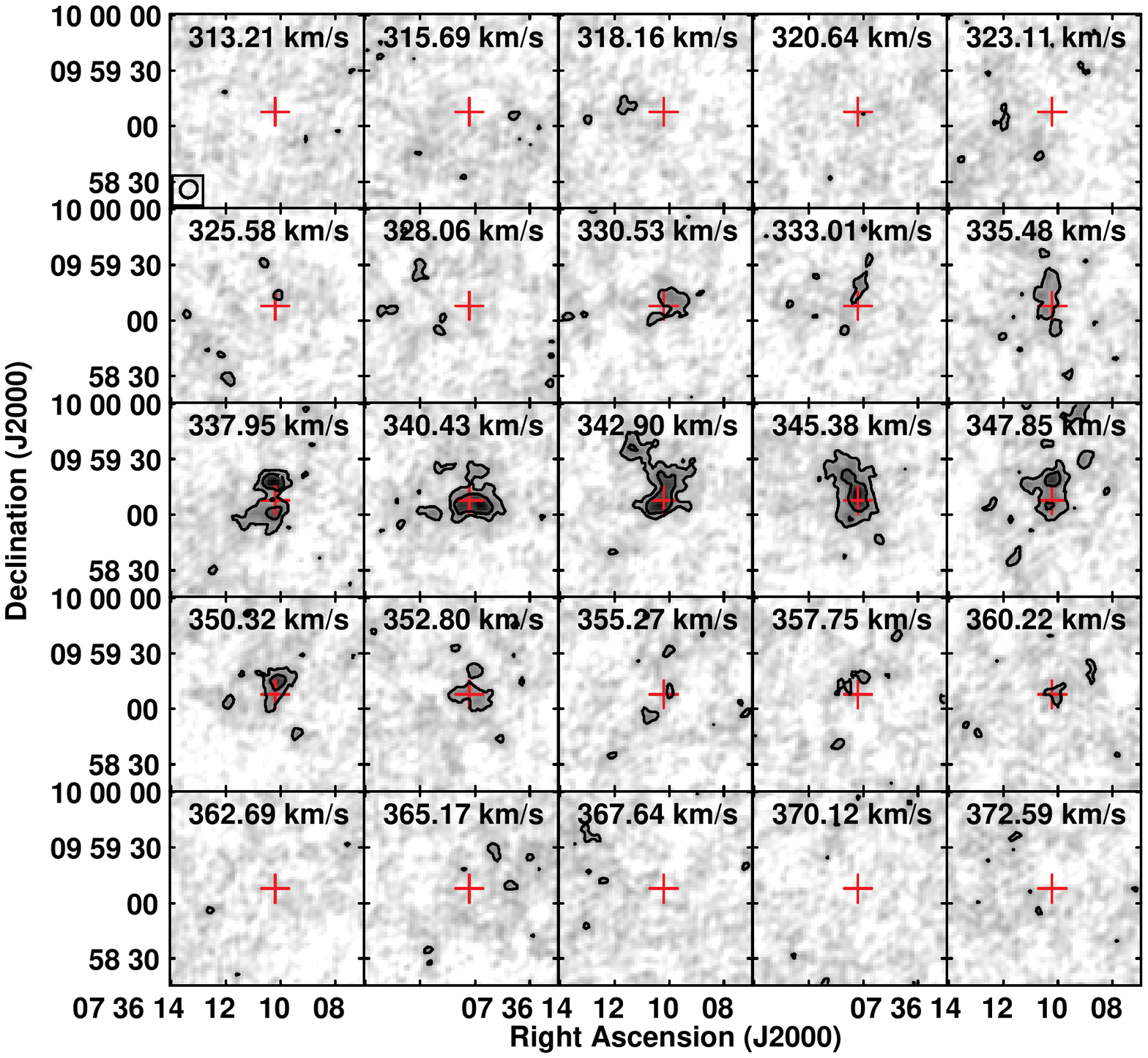}\vspace{-2 cm}
\caption{Same as Figure 33, for AGC\,174585.}
\label{174585.chmap}
\end{figure}\clearpage

\begin{figure}
\includegraphics[width=\textwidth]{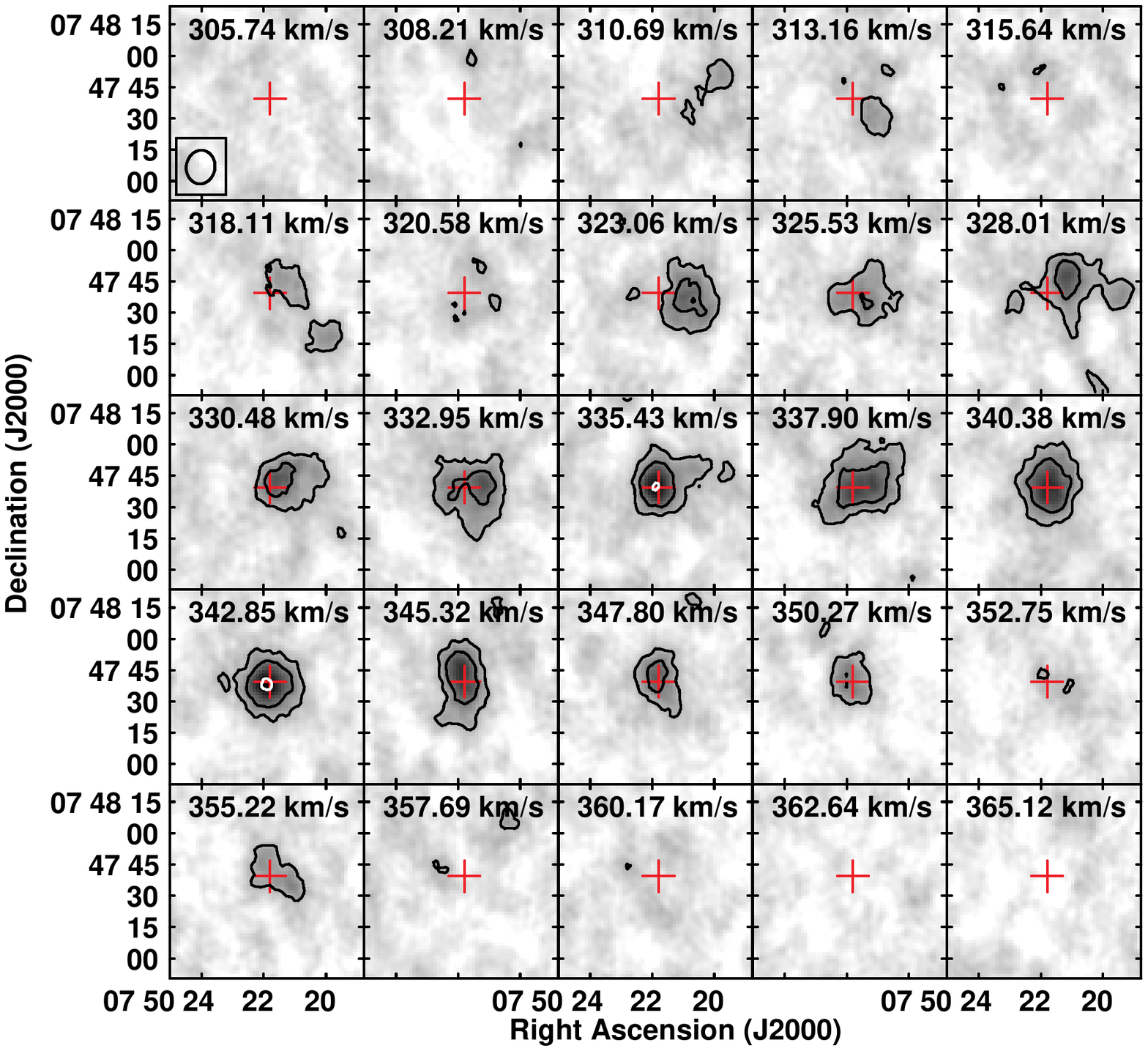}\vspace{-2 cm}
\caption{Same as Figure 33, for AGC\,174605.}
\label{174605.chmap}
\end{figure}\clearpage

\begin{figure}
\includegraphics[width=\textwidth]{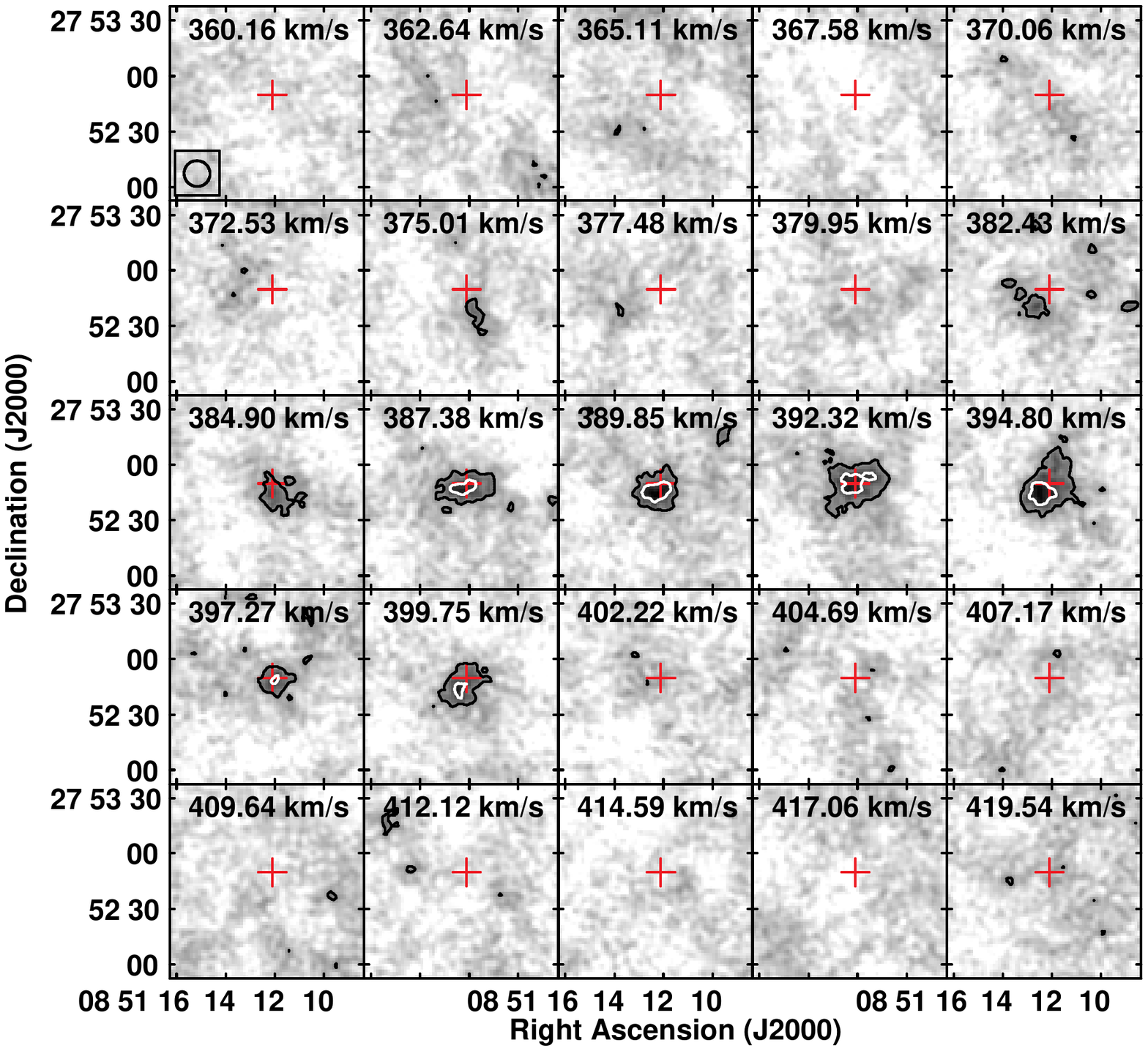}\vspace{-2 cm}
\caption{Same as Figure 33, for AGC\,182595.}
\label{182595.chmap}
\end{figure}\clearpage

\begin{figure}
\includegraphics[width=\textwidth]{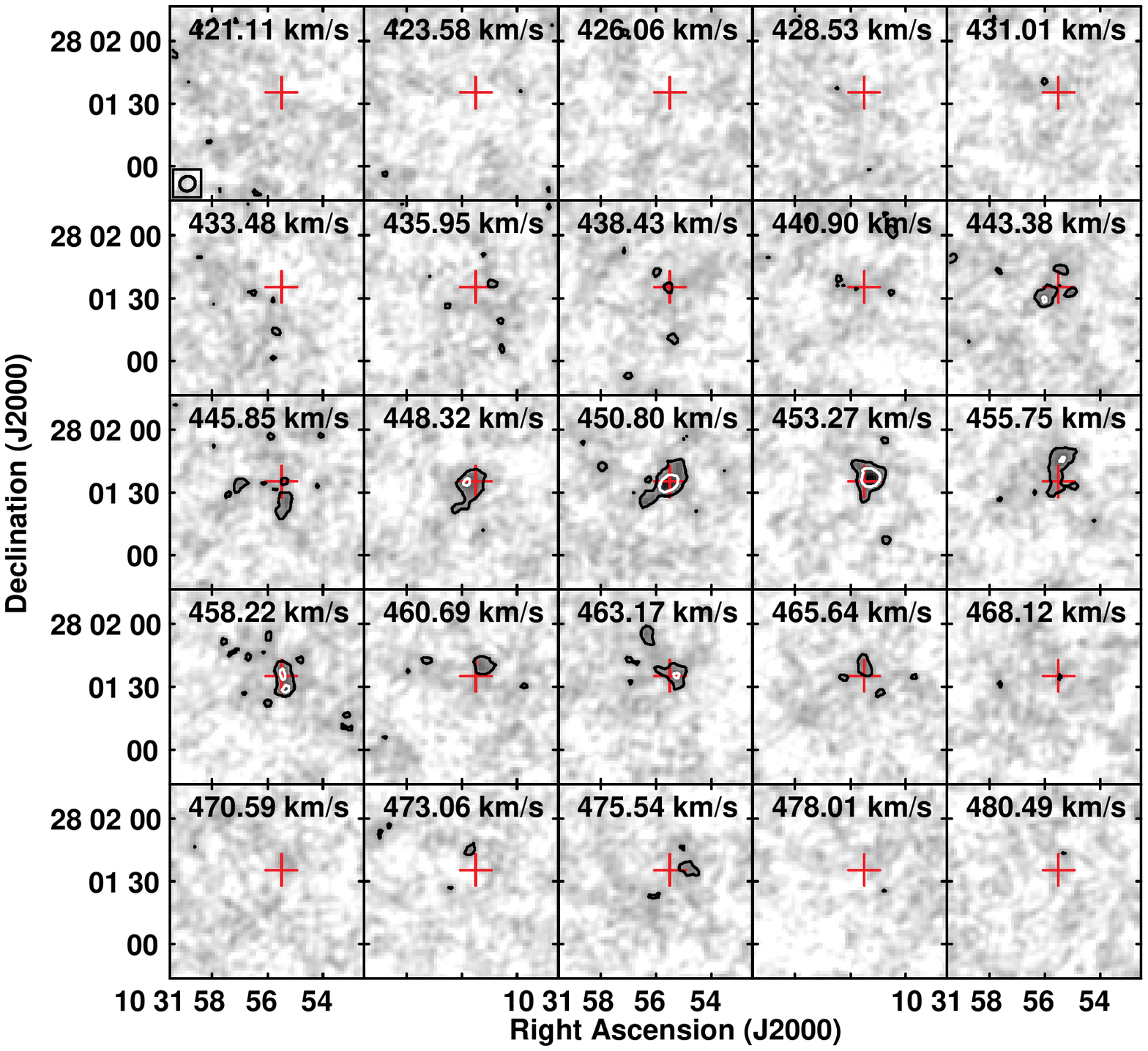}\vspace{-2 cm}
\caption{Same as Figure 33, for AGC\,731457.}
\label{731457.chmap}
\end{figure}\clearpage

\begin{figure}
\includegraphics[width=\textwidth]{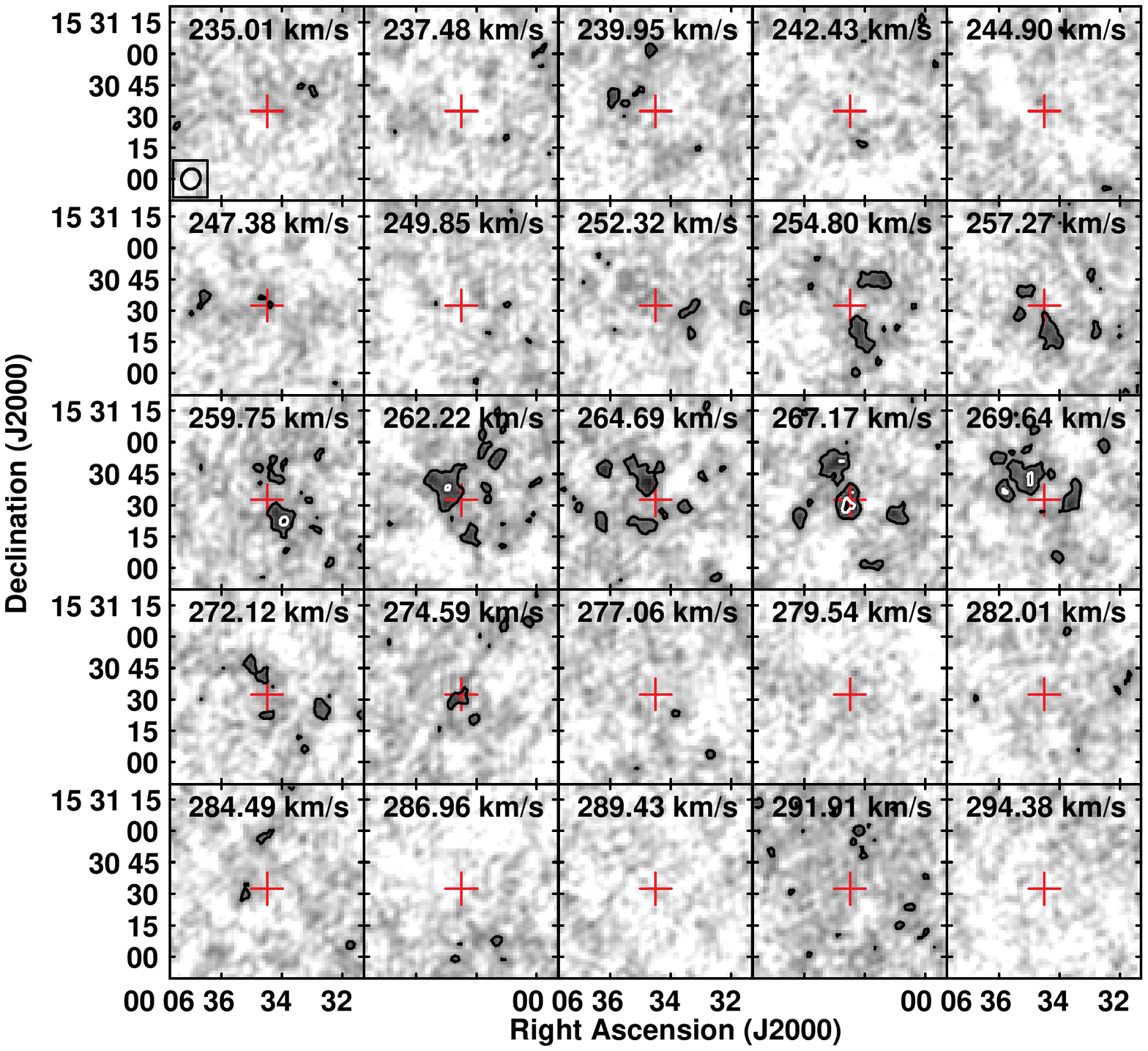}\vspace{-2 cm}
\caption{Same as Figure 33, for AGC\,748778.}
\label{748778.chmap}
\end{figure}\clearpage

\begin{figure}
\includegraphics[width=\textwidth]{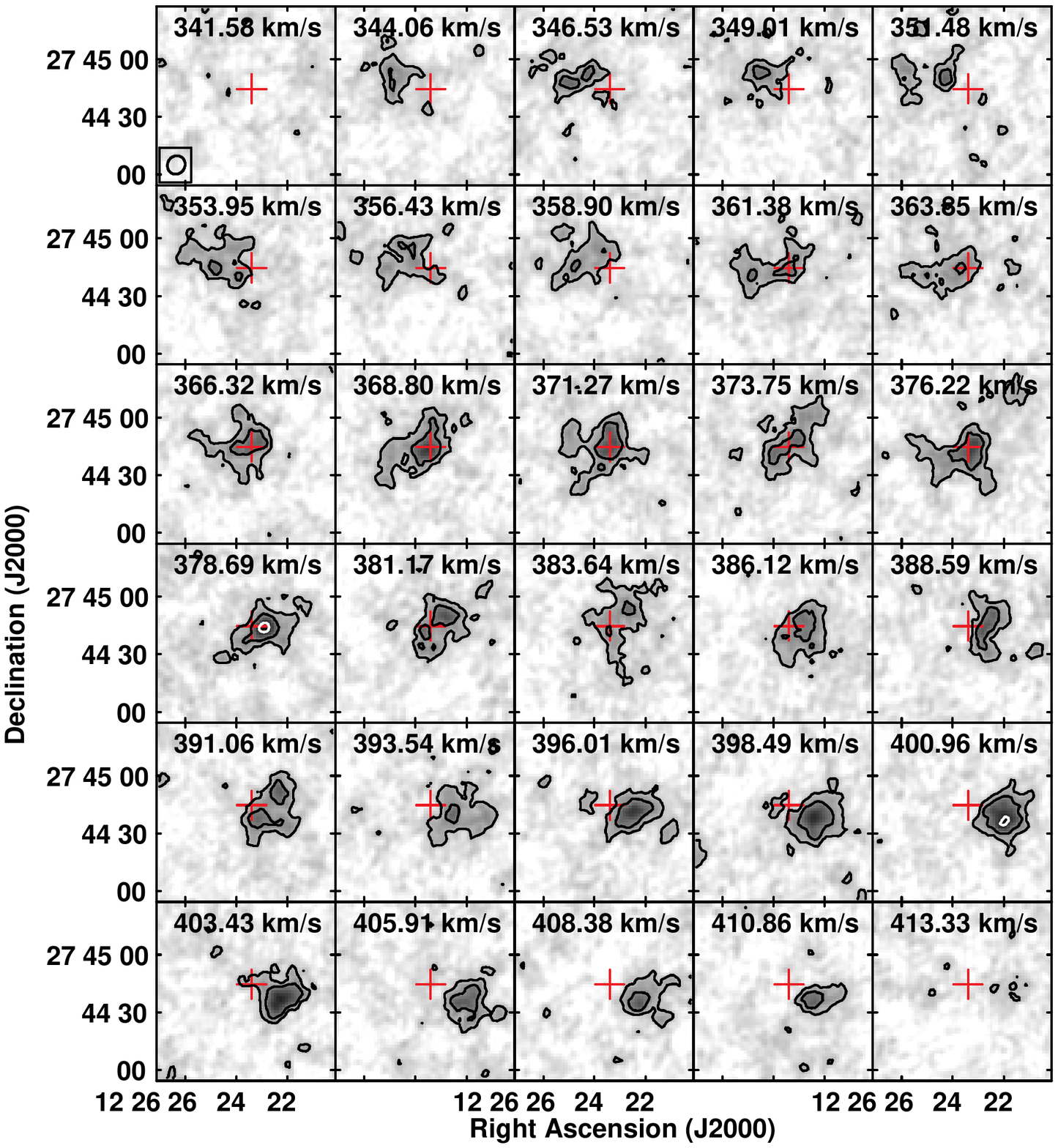}\vspace{-2 cm}
\caption{Same as Figure 33, for AGC\,749237.}
\label{749237.chmap}
\end{figure}\clearpage

\begin{figure}
\includegraphics[width=\textwidth]{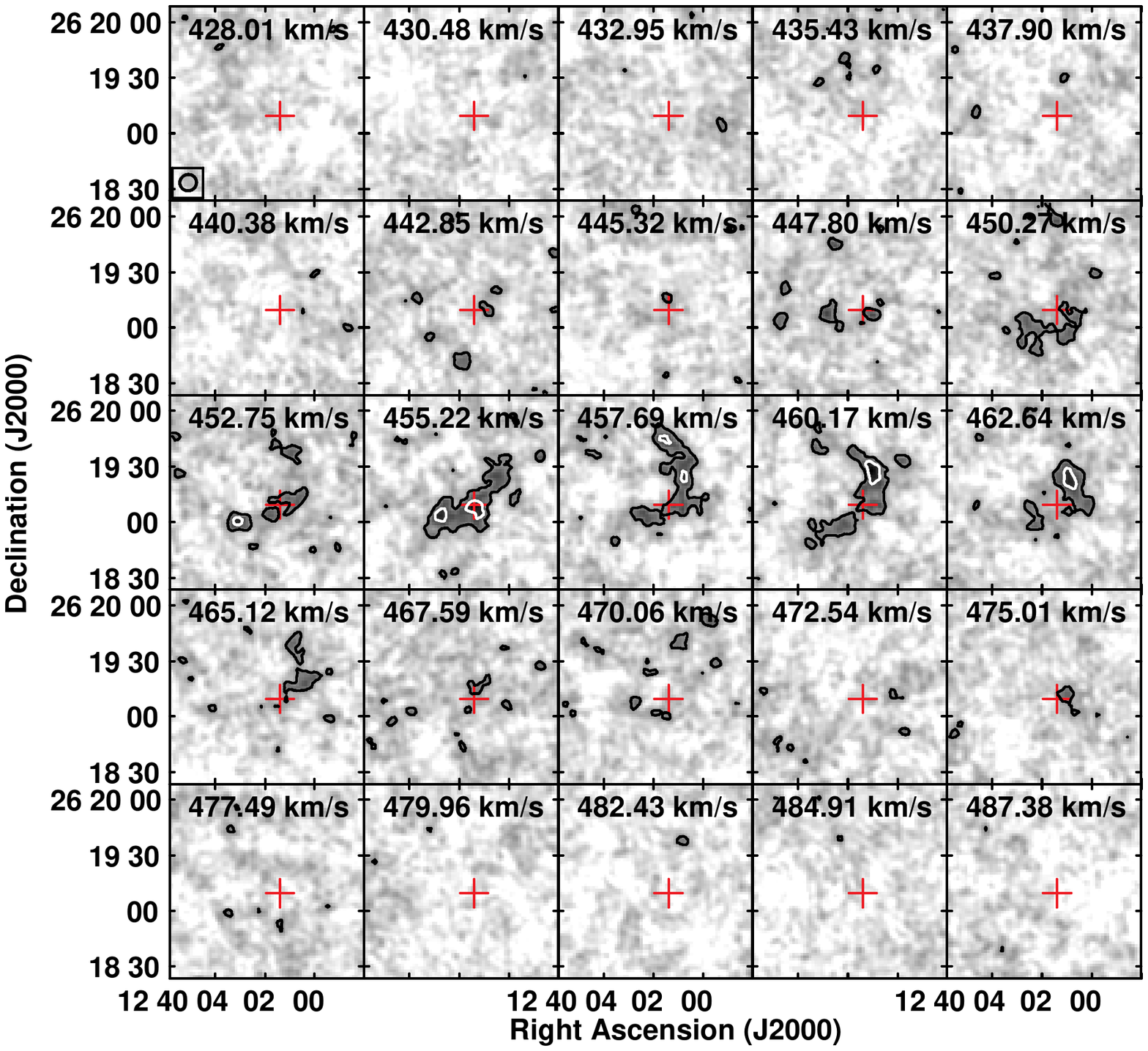}\vspace{-2 cm}
\caption{Same as Figure 33, for AGC\,749241.}
\label{749241.chmap}
\end{figure}\clearpage

\end{document}